\documentclass[usegraphicx,useAMS,usenatbib]{mn2e}

\usepackage{amsmath}
\usepackage{aas_macros}
\usepackage[usenames,dvipsnames,svgnames]{xcolor}

\usepackage{xspace}

\usepackage{todonotes}

\usetikzlibrary{arrows,decorations.pathmorphing}

\usepackage[T1]{fontenc}
\usepackage{ae,aecompl}
\usepackage{newtxtext,newtxmath}

\newcommand{\ateb}{\umdrahd{$\beta$}}
\newcommand{\msol}{M_{\odot}}

\newcommand{\Msol}{M_{\odot}}

\newcommand{\zehn}[1]{10^{#1}}
\newcommand{\zehnh}[2]{{#1} \times 10^{#2}}

\newcommand{\Alfven}{Alfv{\'e}n\xspace}

\newcommand{\pns}{proto neutron star\xspace}

\newcommand{\ms}{\textrm{ms}}

\newcommand{\km}{\textrm{km}}
\newcommand{\mtr}{\textrm{m}}
\newcommand{\cm}{\textrm{cm}}
\newcommand{\cms}{\textrm{cm s}^{-1}}

\newcommand{\erg}{\textrm{erg}}
\newcommand{\ergs}{\textrm{erg} \, \textrm{s}^{-1}}
\newcommand{\sek}{\textrm{s}}
\newcommand{\isek}{\textrm{s}^{-1}}

\newcommand{\gccm}{\textrm{g\,cm}^{-3}}

\newcommand{\Gauss}{\textrm{G}}
\newcommand{\grad}{^{\circ}}

\newcommand{\ie}{i.e.\xspace}
\newcommand{\cf}{cf.\xspace}
\newcommand{\eg}{e.g.\xspace}
\newcommand{\viz}{viz.\xspace}
\newcommand{\wrt}{w.r.t.\xspace}

\newcommand{\Eqref}[1]{Eq.\,(\ref{#1})}
\newcommand{\figref}[1]{Fig.\,\ref{#1}}
\newcommand{\tabref}[1]{Tab.\,\ref{#1}}
\newcommand{\secref}[1]{Sect.\,\ref{#1}}

\newcommand{\nth}[1]{${#1}^{\mathrm{th}}$}
\newcommand{\kbb}{k_{\mathrm{B}}}
\newcommand{\miRyC}{RYC2018-024938-I}
\newcommand{\jcop}{J.~Comput.~Phys.}

\newcommand{\modl}[1]{model \texttt{#1}}
\newcommand{\modls}[1]{models \texttt{#1}}
\newcommand{\Modl}[1]{Model \texttt{#1}}
\newcommand{\Modls}[1]{Models \texttt{#1}}
\newcommand{\modelname}[1]{\texttt{#1}}

\newcommand{\panel}[1]{\textit{(#1)}}
\newcommand{\banel}[1]{\textit{#1}}

\newcommand{\iPNS}{{\textsc{pns}}}

\newcommand{\Erot}{\mathcal{T}}
\newcommand{\Erotpns}{\Erot_\iPNS}

\newcommand{\tauAlf}{\tau_{\textrm{Alf}}}
\newcommand{\tauadv}{\tau_{\textrm{adv}}}
\newcommand{\tauhtg}{\tau_{\textrm{heat}}}
\newcommand{\tautau}{\tauadv / \tauhtg}
\newcommand{\tautaum}{\tauadv / \tauAlf}

\newcommand{\tpb}{t_{\textrm{pb}}}

\newcommand{\ashlm}[2]{a^{\textrm{sh}}_{#1,#2}}

\newcommand{\aPNSlm}[2]{a^{\iPNS}_{#1,#2}}
\newcommand{\rsh}{R_{\textrm{sh}}}
\newcommand{\rshM}{R_{\textrm{sh;max}}}
\newcommand{\eej}{E_{\textrm{ej}}}
\newcommand{\mej}{M_{\textrm{ej}}}
\newcommand{\vpb}{\varpi_{\textrm{b}}}

\newcommand{\Rw}{\modelname{W}\xspace}
\newcommand{\RO}{\modelname{O}\xspace}
\newcommand{\Rp}{\modelname{P}\xspace}
\newcommand{\Rs}{\modelname{S}\xspace}
\newcommand{\mRw}{\modl{W}\xspace}
\newcommand{\mRO}{\modl{O}\xspace}
\newcommand{\mRp}{\modl{P}\xspace}
\newcommand{\mRs}{\modl{S}\xspace}
\newcommand{\MRw}{\Modl{W}\xspace}
\newcommand{\MRO}{\Modl{O}\xspace}
\newcommand{\MRp}{\Modl{P}\xspace}
\newcommand{\MRs}{\Modl{S}\xspace}
\newcommand{\mnRw}{\modelname{W}\xspace}
\newcommand{\mnRO}{\modelname{O}\xspace}
\newcommand{\mnRp}{\modelname{P}\xspace}
\newcommand{\mnRs}{\modelname{S}\xspace}

\newcommand{\MPNS}{M^{\iPNS}}
\newcommand{\ErPNS}{\Erot^{\iPNS}}
\newcommand{\Emag}{\mathcal{B}}
\newcommand{\EmPNS}{\Emag^{\iPNS}}
\newcommand{\igain}{{\textrm{gain}}}

\renewcommand{\ateb}{\beta^{-1}}

\newcommand{\MpreSN}{M_{\mathrm{pre}\textsc{sn}}}

\newcommand{\Ye}{Y_{\rm e}}

\defcitealias{Obergaulinger_Aloy__2020__mnras__MagnetorotationalCoreCollapseofPossibleGRBProgenitorsIExplosionMechanisms}{Paper I}
\defcitealias{Aloy_Obergaulinger_2020__mnras_PaperII}{Paper II}

\begin{document}

\title[3D MHD core collapse] {Magnetorotational core collapse of
  possible GRB progenitors.  III. Three-dimensional models}

\author[Obergaulinger \& Aloy]{
  M.~Obergaulinger$^{1,2}$, M.\'A.~Aloy$^1$
  \\
  $^1$ Departament d{\'{}}Astronomia i Astrof{\'i}sica, Universitat de
  Val{\`e}ncia, \\ Edifici d{\'{}}Investigaci{\'o} Jeroni Munyoz, C/
  Dr.~Moliner, 50, E-46100 Burjassot (Val{\`e}ncia), Spain
  \\
  $^{2}$ Institut f{\"u}r Kernphysik, Technische Universit{\"a}t
  Darmstadt, Schlossgartenstra{\ss}e 2, 64289 Darmstadt, Germany
}

\maketitle

\begin{abstract}
  We explore the influence of non-axisymmetric modes on the dynamics
  of the collapsed core of rotating, magnetized high-mass stars in
  three-dimensional simulations of a rapidly rotating star with an
  initial mass of $M_{\textsc{zams}} = 35 \, \Msol$ endowed with four
  different pre-collapse configurations of the magnetic field, ranging
  from moderate to very strong field strength and including the field
  predicted by the stellar evolution model.  The model with the
  weakest magnetic field achieves shock revival due to neutrino
  heating in a gain layer characterized by a large-scale, hydrodynamic
  $m = 1$ spiral mode.  Later on, the growing magnetic field of the
  \pns launches weak outflows into the early ejecta.  Their
  orientation follows the evolution of the rotational axis of the
  \pns, which starts to tilt from the original orientation due to the
  asymmetric accretion flows impinging on its surface.  The models
  with stronger magnetization generate mildly relativistic,
  magnetically driven polar outflows propagating over a distance of
  $10^4$ \,km within a few $100 \, \ms$.  These jets are stabilized against
  disruptive non-axisymmetric instabilities by their fast propagation
  and by the shear of their toroidal magnetic field.  Within the
  simulation times of around $1 \, \sek$, the explosions reach
  moderate energies and the growth of the \pns masses ceases at values
  substantially below the threshold for black hole formation, which,
  in combination with the high rotational energies, might suggest a
  possible later proto-magnetar activity.
\end{abstract}

\begin{keywords}
  Supernovae: general - gamma-ray bursts: general
\end{keywords}

\section{Introduction}
\label{Sek:Intro}

Stars of masses in excess of about $8 \, \Msol$ experience the
collapse of their cores after the end of their hydrostatic burning
phases.  In most cases, the collapse and the subsequent phase in which
the shock wave launched at the formation of a \pns (PNS) stalls inside
the core leads to a core-collapse supernova (CCSN) explosion powered
by the standard neutrino mechanism
\citep{Bethe1985a,Janka__2012__ARNPS__ExplosionMechanismsofCore-CollapseSupernovae}.
In a small fraction of progenitor stars, however, fast rotation and
strong magnetic fields, only of minor importance in the neutrino
mechanism, may have a strong impact on the evolution.  

A large number of theoretical and numerical studies has been devoted
to addressing the conditions for rotation and magnetic fields to
affect core collapse, their possible dynamical consequences, and the
observational signatures and remnants of such events.  The numerical
costs of simulations of such a complex system are reflected in a
gradual increase of the degree of realism of the models in terms of
their resolution and dimensionality as well as their modelling of the
nuclear physics and the transport and interactions of neutrinos
\citep[see,
\eg][]{Bisnovatyi-Kogan_Popov_Samokhin__1976__APSS__MHD_SN,Mueller_Hillebrandt__1979__AA__MHD_SN,Symbalisty__1984__ApJ_MHD_SN,Akiyama_etal__2003__ApJ__MRI_SN,Kotake_etal__2004__Apj__SN-magrot-neutrino-emission,Thompson_Quataert_Burrows__2004__ApJ__Vis_Rot_SN,Moiseenko_et_al__2006__mnras__A_MR_CC_model_with_jets,Obergaulinger_2006A&A...457..209,Obergaulinger_et_al__2006__AA__MR_collapse_TOV,
  Dessart_et_al__2007__apj__MagneticallyDrivenExplosionsofRapidlyRotatingWhiteDwarfsFollowingAccretion-InducedCollapse,Burrows_etal__2007__ApJ__MHD-SN,Sawai_et_al__2013__apjl__GlobalSimulationsofMagnetorotationalInstabilityintheCollapsedCoreofaMassiveStar,Bugli__2020__mnras__TheImpactofNonDipolarMagneticFieldsinCoreCollapseSupernovae}.
The current state of the art is represented by simulations combining
three-dimensional (3D), (Newtonian or relativistic)
magnetohydrodynamics (MHD) and a leakage scheme
\citep{Winteler_et_al__2012__apjl__MagnetorotationallyDrivenSupernovaeastheOriginofEarlyGalaxyr-processElements,Mosta_et_al__2014__apjl__MagnetorotationalCore-collapseSupernovaeinThreeDimensions,Moesta_et_al__2015__nat__Alarge-scaledynamoandmagnetoturbulenceinrapidlyrotatingcore-collapsesupernovae}
or a two-moment transport scheme
\citep{Kuroda_et_al__2020__apj__MagnetorotationalExplosionofaMassiveStarSupportedbyNeutrinoHeatinginGeneralRelativisticThreeDimensionalSimulations}
for the neutrinos.  

The strongest impact of rotation and magnetic fields can be expected
if their energies reach, at least locally, equipartition with the
(non-rotational) kinetic or even the internal energies
\citep[\eg][]{Meier_etal__1976__ApJ__MHD_SN}.  Even accounting for
their growth due to the compression of the core and, for the magnetic
energy, due to mechanisms such as the ensuing differential rotation
and instabilities such as convection
\citep[\eg][]{Thompson_Duncan__1993__ApJ__NS-dynamo,Raynaud2020a} and
the standing-accretion shock instability \citep[SASI,
\eg][]{Endeve_et_al__2010__apj__Generation_of_Magnetic_Fields_By_the_SASI,Guilet_Foglizzo__2010__apj__Toward_a_MHD_Theory_of_the_SASI_Toy_Model_of_the_AAC_in_a_Magnetized_Flow}
or the magneto-rotational instability \cite[MRI,
\eg][]{Balbus_Hawley__1998__RMP__MRI,Akiyama_etal__2003__ApJ__MRI_SN,Obergaulinger_etal__2009__AA__Semi-global_MRI_CCSN,Moesta_et_al__2015__nat__Alarge-scaledynamoandmagnetoturbulenceinrapidlyrotatingcore-collapsesupernovae,Masada_et_al__2015__apjl__MagnetohydrodynamicTurbulencePoweredbyMagnetorotationalInstabilityinNascentProtoneutronStars, Rembiasz_et_al__2016__mnras__Onthemaximummagneticfieldamplificationbythemagnetorotationalinstabilityincore-collapsesupernovae, Guilet_et_al__2015__mnras__Neutrinoviscosityanddrag:impactonthemagnetorotationalinstabilityinprotoneutronstars, Rembiasz_et_al__2016__mnras__Terminationofthemagnetorotationalinstabilityviaparasiticinstabilitiesincore-collapsesupernovae},
this condition corresponds to rotational velocities and magnetic field
strengths that can be expected only in an, as yet undetermined, though
likely rather small, fraction of the progenitor stars
\cite[\eg][]{Heger_et_al__2005__apj__Presupernova_Evolution_of_Differentially_Rotating_Massive_Stars_Including_Magnetic_Fields,Woosley_Heger__2006__apj__TheProgenitorStarsofGamma-RayBursts,Aguilera-Dena_et_al__2018__apj__RelatedProgenitorModelsforLong-durationGamma-RayBurstsandTypeIcSuperluminousSupernovae}. 

Stars that meet these conditions may produce explosions powered by the
rotational energy magnetically extracted from the PNS rather than by
neutrino heating.  Such an explosion mechanism may lead to the
generation of very violent hypernova explosions
\citep{Iwamoto_1998Natur.395..672} with much higher energies than the
$10^{51} \, \erg$ typical for neutrino-driven standard CCSNe and the
production of very fast collimated outflows.  The most extreme cases
within this spectrum includes gamma-ray bursts (GRBs) driven by the
spin down of rapidly rotating PNSs with very strong magnetic fields
\citep[proto-magnetars,
PMs][]{Metzger_et_al__2011__mnras__Theprotomagnetarmodelforgamma-raybursts},
and superluminous supernovae (SLSNe)
\citep{GalYam__2019__araa__TheMostLuminousSupernovae}.  This last
possibility, commonly supported in two-dimensional (2d) axisymmetric
models
\citep[\eg][]{Obergaulinger__2004__Dipl__MHD_collapse,Obergaulinger_2006A&A...457..209,Obergaulinger_et_al__2006__AA__MR_collapse_TOV,
  Obergaulinger_Aloy__2017__mnras__Protomagnetarandblackholeformationinhigh-massstars},
has been questioned based on some 3Dsimulations in which
non-axisymmetric instabilities disrupt jets briefly after their
formation \citep[][but note that other 3d, low-resolution models do
not fully support this claim, \eg
\citealt{Obergaulinger_Aloy__2020__mnras__MagnetorotationalCoreCollapseofPossibleGRBProgenitorsIExplosionMechanisms,Aloy_Obergaulinger_2020__mnras_PaperII};
hereinafter
\citetalias{Obergaulinger_Aloy__2020__mnras__MagnetorotationalCoreCollapseofPossibleGRBProgenitorsIExplosionMechanisms}
and \citetalias{Aloy_Obergaulinger_2020__mnras_PaperII},
respectively]{Mosta_et_al__2014__apjl__MagnetorotationalCore-collapseSupernovaeinThreeDimensions,Kuroda_et_al__2020__apj__MagnetorotationalExplosionofaMassiveStarSupportedbyNeutrinoHeatinginGeneralRelativisticThreeDimensionalSimulations}.
Before reaching a radius of around 1000 km, the jets are quenched and
turn into wide lobes expanding in a less collimated geometry.  Such an
evolution might still be consistent with the subsequent generation of
GRBs in the collapsar scenario
\citep{MacFadyen_Woosley__1999__ApJ__Collapsar}, if ongoing accretion
causes the PNS to collapse to a black hole (BH) and sufficient
rotational energy permits the formation of an accretion disk
\citepalias[though see][for the possibility of forming a Type-III
collapsar without a well developed accretion
disc]{Aloy_Obergaulinger_2020__mnras_PaperII}.

In previous studies
(\citealt{Obergaulinger_Aloy__2017__mnras__Protomagnetarandblackholeformationinhigh-massstars};
\citetalias{Obergaulinger_Aloy__2020__mnras__MagnetorotationalCoreCollapseofPossibleGRBProgenitorsIExplosionMechanisms,
  Aloy_Obergaulinger_2020__mnras_PaperII}) we have investigated the
collapse of stars considered potential progenitors of the class of
very violent explosions outlined above.  In order to cover the large
parameter space adequately and achieve long simulation times, we had
reduced the computational costs by restricting the majority of our
simulations to axisymmetry (only two low-resolution 3D models were
included in the previous work to support, to some extent, some of our
findings in axial symmetry).  We had confirmed the aforementioned
evolutionary paths of mostly bipolar explosions driven by a
combination of neutrino heating and magneto-rotational stresses,
depending on the pre-collapse values of rotational frequency and
magnetic field strength.  In this article, we extend our work to 3d
models.  Due to the higher computational effort per model, we select
only a small subset of four of the models investigated in axisymmetry,
varying the magnetic field of a zero-age main-sequence mass
$M_{\textsc{zams}} = 35 \, \Msol$
\citep{Woosley_Heger__2006__apj__TheProgenitorStarsofGamma-RayBursts}
around the predictions made by the stellar evolution calculation for
this star. 

Specific questions to be addressed here are:
\begin{itemize}
\item What is the explosion mechanism, how is it affected by the
  presence of a strong field, and how do these results differ from
  the axisymmetric case?
\item Are MHD-driven jets destroyed by strong 3D instabilities or do
  they remain collimated over a long time?
\item How does the PNS evolve, does it show non-axisymmetric modes,
  and do 3D effects alter the likelihood of BH formation?
\end{itemize}

The paper is organised as follows.  We discuss the methodology and
initial data of our models in \secref{Sek:NumInit}, describe our
results in \secref{Sek:Res}, and present a summary and conclusions
from the results in \secref{Sek:Concl}.

\section{Numerical method and simulation setup}
\label{Sek:NumInit}

As a continuation of our previous work, the present study uses the
same input physics and numerical method as
\cite{Obergaulinger_Aloy__2017__mnras__Protomagnetarandblackholeformationinhigh-massstars},\citetalias{Obergaulinger_Aloy__2020__mnras__MagnetorotationalCoreCollapseofPossibleGRBProgenitorsIExplosionMechanisms},
and \citetalias{Aloy_Obergaulinger_2020__mnras_PaperII}.  We refer to
these publications for a detailed description of these aspects.

Our 3D grid is formed by $(n_r,n_\theta,n_\phi)=(300,64,128)$
numerical zones in the $r$-, $\theta$- and $\phi$-directions; to be
compared with the 2D grid used in previous papers
$(n_r,n_\theta)=(300,128)$.  The radial zones are spaced
logarithmically up to an outer radius of
$r_{\mathrm{max}} = \zehnh{5}{11} \, \cm$.  The relatively coarse grid
is compensated for by the high-resolution methods used in solving the
MHD and transport equations with a spatial reconstruction in \nth{5}
order \citep{Suresh_Huynh__1997__JCP__MP-schemes}.  In energy space,
we used $n_{\epsilon} = 10$ logarithmically distributed bins in the
interval $[3\,\text{MeV}, 240\,\text{MeV}]$.

We selected four of our axisymmetric models for resimulation in 3D
(see \tabref{Tab:mods}). 
They are all based on the stellar model 35OC for a
$M_{\textsc{zams}} = 35 \, \Msol$ star
\citep{Woosley_Heger__2006__apj__TheProgenitorStarsofGamma-RayBursts}.
The spherically symmetric stellar evolution model includes the effects
of rotation and magnetic fields according to the prescription for
magnetic instabilities, a dynamo, and the redistribution of angular
momentum of \cite{Spruit__2002__AA__Dynamo}.  At the time of collapse,
the star has a mass of $\MpreSN \approx 28.1 \, \Msol$ and a large Fe
core of $M_{\mathrm{Fe}} \approx 2.02 \, \Msol$.  Its centre rotates
with an angular velocity of
$\Omega_{\mathrm{c}} \approx 2.0 \, \sek^{-1}$.  The model contains a
dominantly toroidal magnetic field in radiative zones of the star with
a maximum field strength of
$B_{\mathrm{max}} \approx \zehnh{1.2}{12} \, \Gauss$. Convective
regions are not magnetised in the pre-SN model 35OC.

\begin{table*}
  \centering
  \begin{tabular}{|cc|l|cc|cc|cccccc| }
    \hline
    name & 2D name& field
    & $t_{\mathrm{f}} $
    & fate
    & $M_{\rm exp}$& $E_{\rm exp,51}$
    & $\MPNS$ &  $\ErPNS_{51}$             &$\EmPNS_{51}$
    & $\bar{\Omega}$
    & $B^{\mathrm{pol}}_{\mathrm{surf}}$ & $B^{\mathrm{tor}}_{\mathrm{surf}}$
    \\
    &                &        & $[s]$  &      &$[\Msol]$
    &
    &    $[\Msol]$   &  & & $[10^3 \, \sek^{-1}]$
    & $[10^{14} \mathrm{G}]$ & $[10^{14} \mathrm{G}]$
    \\
    \hline
    \Rw & \modelname{35OC-Rw}
    & $a(10,10)$ & 0.85 & $\nu$-$\Omega$ & 0.58 & 0.50
    & $2.16$ & $15$  & $0.53$ & $1.8$
    & 2.8 & 0.98 \\
    \RO &\modelname{35OC-RO}
    & Or ($1.1\times^{11} \, \Gauss$) & 0.81 & MR & 0.21 & 0.62
    & $2.03$  & $19$ &  $0.23$ & $2.6$
    & 1.3 & 0.25 \\
    \Rp & \modelname{35OC-Rp3}
    & $3{\rm p},1{\rm t}$  ($1.5\times^{11} \, \Gauss$) & 1.5 & MR & 0.16 & 1.7
    & $1.88$ & $11$  & $0.048$ & $2.3$
    & 0.73 & 0.72 
    \\
    \Rs & \modelname{35OC-Rs}
    & $a(12,12)$ & 1.15 & MR & 1.7 & 13
    & $1.75$& $3.3$ & $0.078$ & $1.3$
    & 1.4 & 1.9 \\
    \hline
  \end{tabular}
  \caption{
    List of our models (first column) and their corresponding axysymmetric versions (second column). 
    The third column shows the type of magnetic field:
    ``Or'' indicates the magnetic field profile of the original
    stellar evolution model, $x\mathrm{p},y\mathrm{t}$ means that the
    original poloidal and toroidal fields have been multiplied by
    factors $x$ and $y$, respectively, and $\mathrm{a}(x,y)$ stands
    for an artificial dipolar field with maximum poloidal and toroidal field
    components of $10^x$ and $10^y$ G, respectively.  
    For models \modelname{O} and \modelname{P}, we
      add the values of the field strength at $r = 0$ in parenthesis.
    The fourth column gives the final time of the simulation,
    $t_{\mathrm{f}}$.
    The column ``fate'' gives a brief indication of the evolution of
    the model: $\nu$ means a standard neutrino-driven shock revival,
    $\nu$-$\Omega$ one strongly affected by rotation, and MR a
    magneto-rotational explosion.  The last two columns provide proxy
    values for the explosion mass ($M_{\rm exp}$) and for the
    explosion energy $E_{\rm exp}$ (in units of $\zehn{51}\,$erg) at
    the end of our simulations.
    The next five columns present the PNS mass, rotational energy,
    magnetic energy (both in units of $\zehn{51}\,$erg), average
    rotational frequency $\bar{\Omega}=J^\iPNS/I^\iPNS$ and poloidal
    and toroidal field strength at the end of
    the computed time.
  }
  \label{Tab:mods}
\end{table*}

Instead of using several progenitors, we explored the influence of
variations of the magnetic field on the dynamics.  A thorough
justification of the modifications included in the magnetic field
topology and strength can be found in
\citetalias{Aloy_Obergaulinger_2020__mnras_PaperII}.  In the present
work, we use the same rotational profile for all models, namely, the
one provided by the stellar evolution model, but initialise our models
with four different distributions of the magnetic fields, related to
four of the axisymmetric models:
\begin{description}
\item[\MRO] is a 3D version of the eponymous axisymmetric model.  It
  uses the original magnetic field predicted by the stellar evolution
  model.  In axisymmetry, the model explodes in a bipolar way
  $\tpb\sim 0.18\,\sek$ after bounce ($\tpb=t-t_{\rm b}$, where
  $t_{\rm b}\simeq 0.4\,\sek$ is the bounce time) predominantly driven
  by magnetic stresses and produces a BH within about $3.2\,\sek$.
\item[\MRp] corresponds to the axisymmetric model \modelname{35OC-Rp3}
  and is defined by an enhancement of the poloidal field component by
  a factor 3 with respect to the pre-SN progenitor, while the toroidal
  component remains unchanged.  The axisymmetric model produces a
  magnetically driven jet-like explosion in the first $80\,\ms$ after
  bounce.  Accretion onto the PNS is weaker than in model
  \modelname{35OC-RO} and, instead of collapsing to a BH, the model is
  a possible candidate for a PM-driven GRB.
\item[\MRs] starts  with a very strong magnetic field set up following
  \cite{Suwa_etal__2007__pasj__Magnetorotational_Collapse_of_PopIII_Stars}
  with both toroidal and poloidal components normalized to central
  values of $10^{12} \, \Gauss$.  The axisymmetric model
  \modelname{35OC-Rs} exhibits the strongest explosion of all of the
  2D models, setting in without a delay immediately after bounce.
  Like model \modelname{35OC-Rp3}, it does not lead to a BH collapse.
\item[\MRw] employs the same prescription as \mRs, but reducing the
  field strength by two orders of magnitude.  Its 2D equivalent, model
  \modelname{35OC-Rw}, develops a bipolar neutrino-driven explosion
  aided by the large rotational energy reservoir of the collapsed core
  at $t_{\rm exp}=378\,$ms.  At its
    termination, the model developed a very massive PNS close to the
    threshold of BH formation.
\end{description}
We computed the pre-bounce evolution of all models in axisymmetry
and mapped to the 3D grid at the time of bounce.

\section{Results}
\label{Sek:Res}

\subsection{General overview}
\label{sSek:GenOvi}

\begin{figure}
  \centering
   \begin{tikzpicture}
   \pgftext{%
     \includegraphics[width=\linewidth]{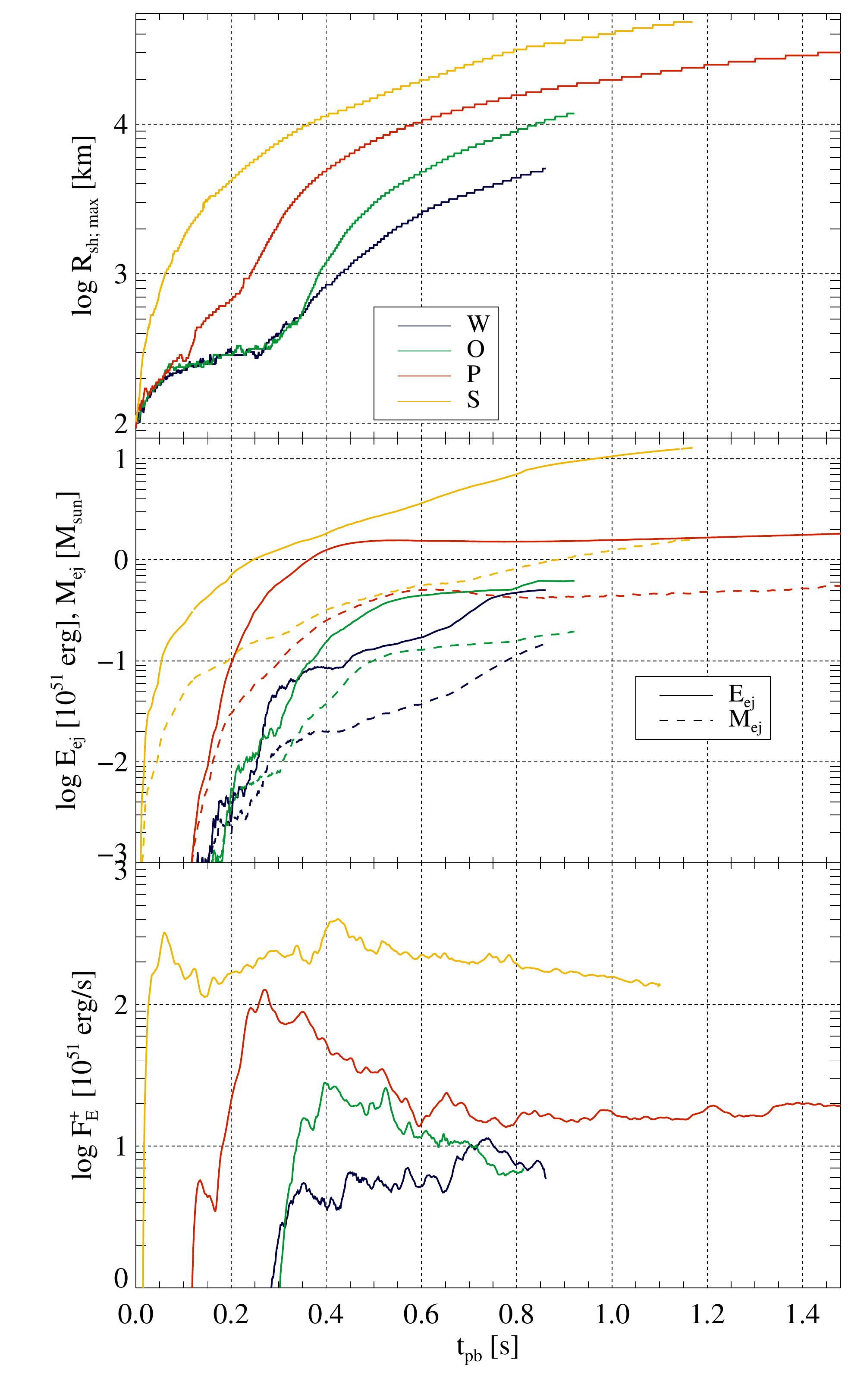}
   }%
   \node at (-3.95,+2.8) {\large (a)};
   \node at (-3.95,-1.3) {\large (b)};
   \node at (-3.95,-5.7) {\large (c)};
  \end{tikzpicture}
  \caption{
    Properties of the explosion.  The top, middle, and bottom panels show the time
    evolution of the maximum shock radius, the mass and energy of
    the ejecta, and the total outward flux of energy measured
      at a radius of $r = 300 \, \km$, respectively.
  }
  \label{Fig:globvars2}
\end{figure}

\begin{figure}
  \centering
  \begin{tikzpicture}
   \pgftext{%
  \includegraphics[width=\linewidth]{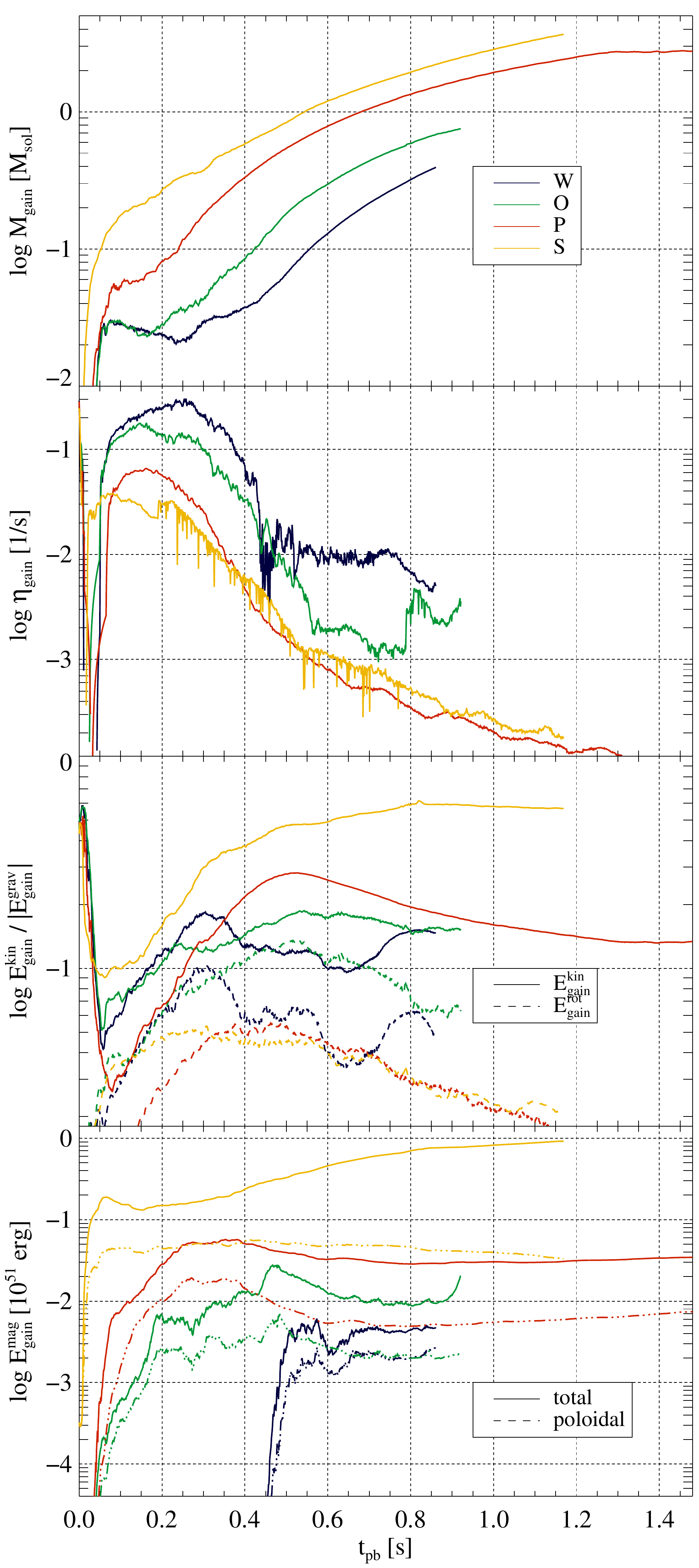}
          }%
   \node at (-3.95,+5.0) {\large (a)};
   \node at (-3.95,+0.55) {\large (b)};
   \node at (-3.95,-4.0) {\large (c)};
   \node at (-3.95,-8.4) {\large (d)};
  \end{tikzpicture}
  \caption{
    Properties of the gain layer.  From top to bottom, the panels
    display the mass contained in the gain layer, the heating
    efficiency, the ratio between (total and rotational) kinetic
    energy and the gravitational energy, and the total and poloidal
    magnetic energy.
  }
  \label{Fig:globvars3}
\end{figure}

\begin{figure}
  \centering
    \begin{tikzpicture}
   \pgftext{%
  \includegraphics[width=0.99\linewidth]{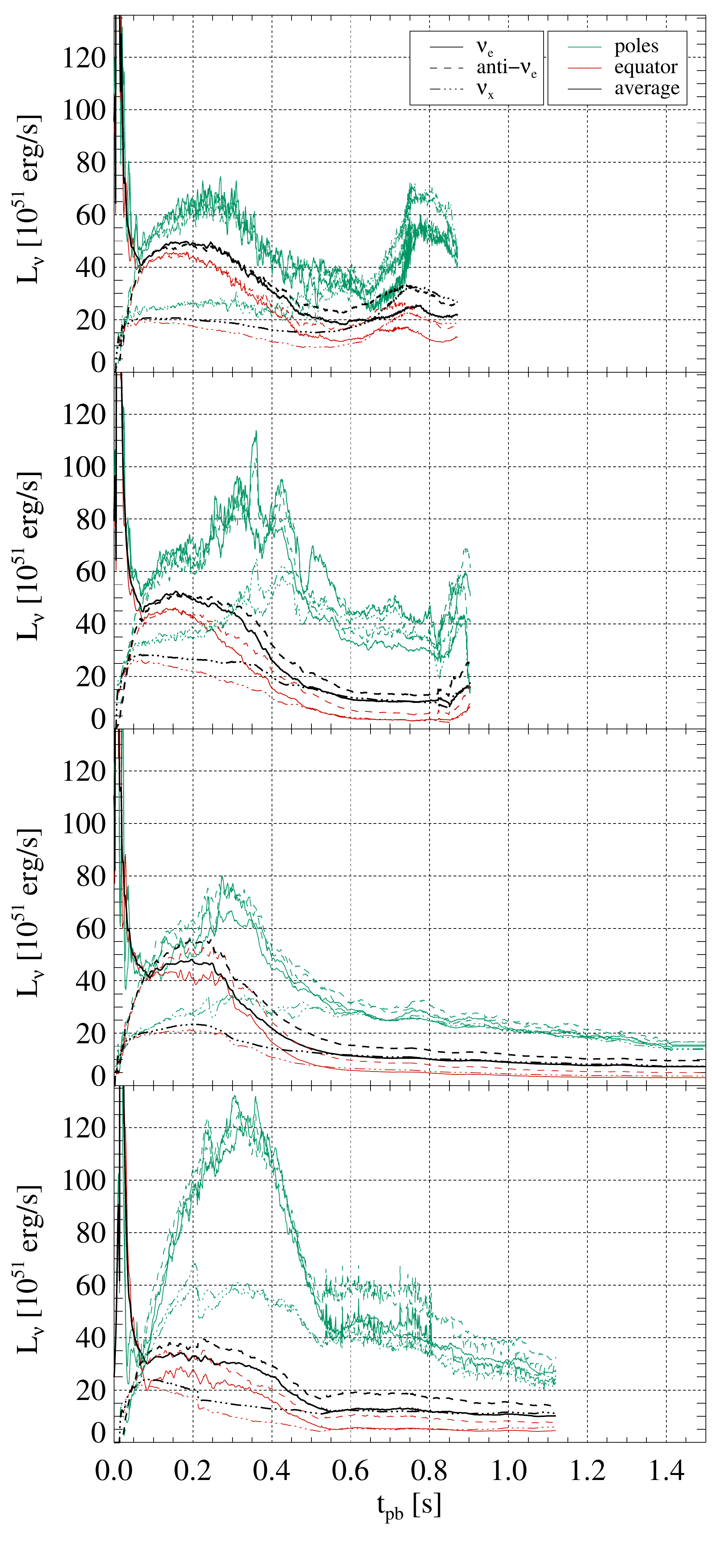}
          }%
   \node at (-3.75,+5.1) {\large (a)};
   \node at (-3.75,+0.95) {\large (b)};
   \node at (-3.75,-3.4) {\large (c)};
   \node at (-3.75,-7.6) {\large (d)};
   \node [fill=white, opacity=1, text opacity=1] at (-2.35,+8.75) {\large \textcolor{MidnightBlue}{\Rw}};
   \node [fill=white, opacity=1, text opacity=1] at (-2.35,+4.58) {\large \textcolor{PineGreen}{\RO}};
   \node [fill=white, opacity=1, text opacity=1] at (-2.35,0.4) {\large \textcolor{red}{\Rp}};
   \node [fill=white, opacity=1, text opacity=1] at (-2.35,-3.8) {\large \textcolor{YellowOrange}{\Rs}};
  \end{tikzpicture}
  \caption{
    Neutrino luminosities of models \Rw, \RO, \Rp, \Rs (top to
    bottom).  We show the time evolution of the luminosities of all
    three flavours (distinguished by line styles as indicated in the
    legend).  Colours show different geometries: green lines display
    the emission along the north and south polar
    directions (green lines), red lines show the emission in the
    equator, averaged over longitudes, and black lines represent the
    average over all solid angles. 
  }
  \label{Fig:nulums}
\end{figure}

\begin{figure}
  \centering
      \begin{tikzpicture}
   \pgftext{%
     \includegraphics[width=\linewidth]{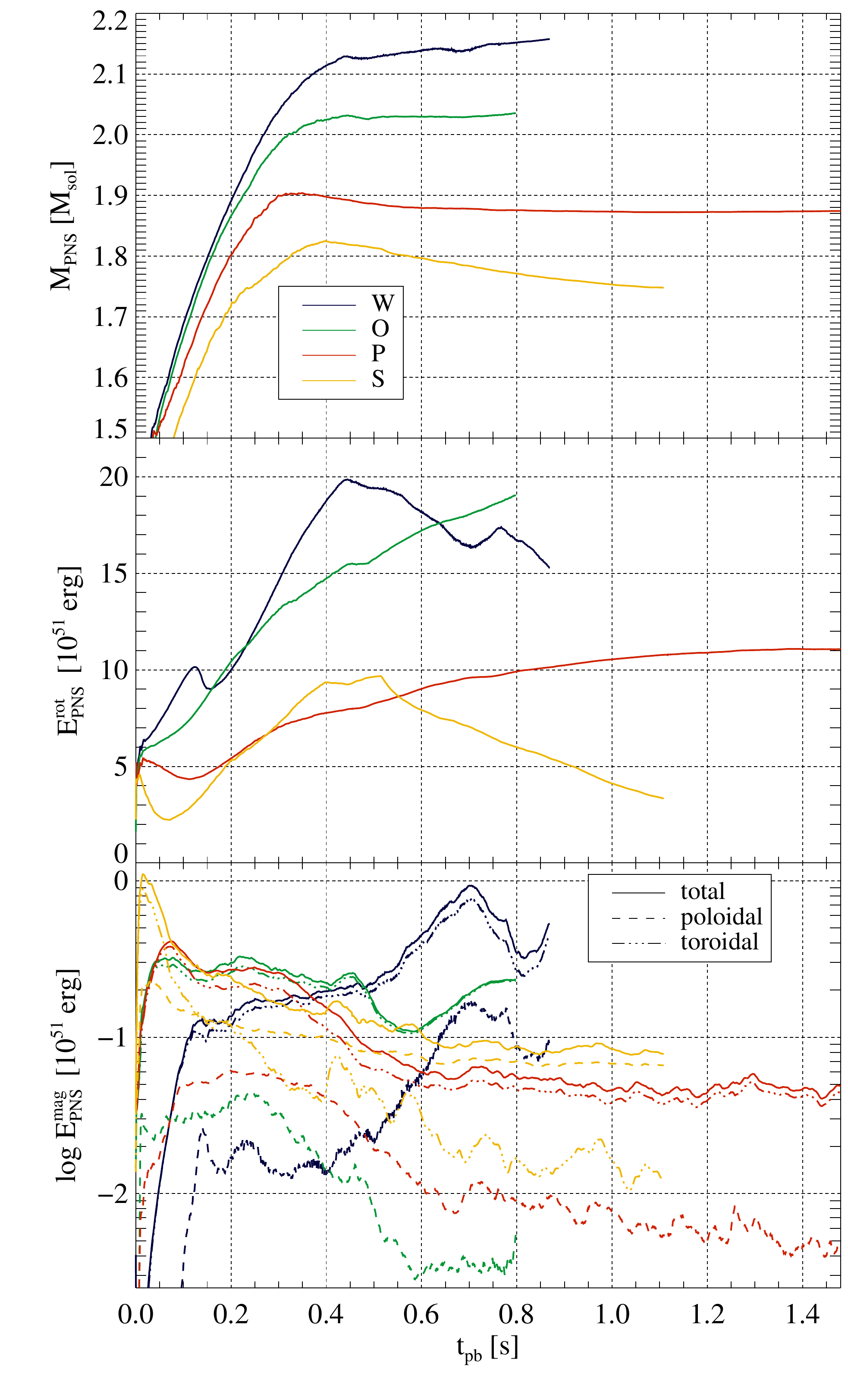}
   }%
   \node at (-3.99,+2.75) {\large (a)};
   \node at (-3.99,-1.65) {\large (b)};
   \node at (-3.99,-6.20) {\large (c)};
   \node[fill=white, opacity=1, text opacity=1, rotate=90] at (-3.6,4.80) {$\MPNS [\Msol]$};
   \node[fill=white, opacity=1, text opacity=1, rotate=90] at
   (-3.6,+0.50) {$\ErPNS \, [10^{51} \, \erg]$};
   \node[fill=white, opacity=1, text opacity=1, rotate=90] at
   (-3.6,-3.8) {$\log\EmPNS \, [10^{51} \, \erg]$};
 \end{tikzpicture}
 \caption{
   Evolution of the PNSs: from top to bottom, the mass, the rotational
   energy, and the total, poloidal, and toroidal magnetic
   energies are shown.
 }
 \label{Fig:globvars4}
\end{figure}

All our models develop supernova explosions within the first $\sim
0.25\,\sek$ after core
collapse. Figures~\ref{Fig:globvars2}--\ref{Fig:globvars4} summarize
selected global quantities of the evolution of our four models: the
run-away of the maximum shock radii, $\rshM$,
(\figref{Fig:globvars2}\panel{a}), the increase of the diagnostic
explosion energy, $\eej$, the mass, $\mej$, of the ejecta
(\figref{Fig:globvars2}\panel{b}), and the integral of the
  outward directed flux of energy over a surface of radius $r = 300 \,
  \km$, $F^{+}_{\mathrm{E}}$ (\figref{Fig:globvars2}\panel{c}). As we shall
see in the following, these properties are ordered according to the
poloidal field strength in the initial model. The stronger the
poloidal field, the more energetic the explosions are. Similarly, the
ejected mass and the shock radius at times sufficiently separated from
core bounce (say, $\tpb>0.35\,\sek$) grow with larger pre-SN poloidal
field strengths. In this regard, 2D and 3D models are qualitatively
similar \citepalias[see][]{Aloy_Obergaulinger_2020__mnras_PaperII}.

The two models \mnRw and \mnRO initiate the shock expansion almost at
the same time, after a phase of about 250 ms during which the shock
stagnates at around 200 km (\figref{Fig:globvars2}\panel{a}).
Immediately thereafter, \mRO exhibits a rise of the shock radius to
$\rshM \approx \zehn{4} \, \km$ after 850\,ms.  The ejecta energy and
mass (\figref{Fig:globvars2}\panel{b}) increase continuously and very
rapidly within the first 100\,ms after shock revival, then more
gradually.  The final state of the simulation with
$\eej \approx \zehnh{5}{50} \, \erg$ and $\mej \approx 0.15 \, \msol$
suggests a moderately violent supernova.  The somewhat stronger
magnetised \mRO reaches similar final ejecta masses and energies,
while the maximum shock radius is about twice that of \mRw.  Despite
the similar global evolution, a detailed comparison of the two models
reveals important differences regarding the explosion mechanism and
outflow properties (see below).  While the rise of the diagnostic
explosion energy (panel \panel{b}) becomes more gradual towards the
end of the simulation, we find that both models keep injecting energy
into the ejecta at a rate close to
$F^{+}_{\mathrm{E}} \lesssim \zehn{51} \, \ergs$ (panel \panel{c}).
Despite the lower growth rate of $\eej$, this continued energy
injection indicates that it is the simulations would have to be run
for a longer time in order to determine final asymptotic explosion
energies and masses.

We find much stronger explosions for the two models with enhanced
pre-collapse magnetic fields.  \MRp explodes after about 100\,ms (\ie
at about the same time as its axisymmetric counterpart
\modl{35OC-Rp3}; Paper I) with a rapidly growing shock radius.  By the
end of the simulation, \ie at $t \approx 1.6 \, \sek$, the shock
achieves a maximum radius of $\rshM \approx \zehnh{3}{4} \, \km$.
Within less than half a second after the onset of the explosion,
energy and mass of the ejecta show a fast early rise to
$\eej \approx \zehnh{1.15}{51} \, \erg$ and
$\mej \gtrsim 0.5 \, \msol$, respectively.  The early rise of the
explosion energy concindes with a peak of
$F^{+}_{\mathrm{E}} \gtrsim \zehn{52} \, \ergs$.  As the energy flux
relaxes to a level of
$F^{+}_{\mathrm{E}} \approx \zehnh{2}{51} \, \ergs$, maintained
throughout the entire evolution, the explosion energy first declines
by a small amount, then stabilizes and, finally, starts to grow at a
comparably low rate.  The ejecta mass follows a similar trend.  This
behaciour is in contrast to the ongoing growth of both quantities
displayed by the axisymmetric version of this model.  Nevertheless, as
in the cases shown above, it is too early to determine the final
explosion energies.  As in 2D, the most extremely magnetised model,
\Rs, directly explodes without any shock stagnation.  Its shock wave
expands the fastest ($\rshM \approx \zehnh{5}{4} \, \km$ at
$t \approx 1.15 \, \sek$) and its final explosion energy
($\eej \approx \zehnh{1.3}{52} \, \erg$) and mass
($\mej \approx 1.7 \, \msol$) are by far the highest of all models and
keep increasing when we had to terminate the simulation.  The high
explosion energies, putting \mRs into the range of potential
hypernovae, are powered by the most intense energy fluxes of all
models in excess of $F^{+}_{\mathrm{E}} > \zehn{52} \, \ergs$ at all
time after the initial rise, though a gradual decline is observable
after $t \approx 0.4 \, \sec$.

Similarly to the variables characterizing the shock wave and the
ejecta, the properties of the gain layer (\figref{Fig:globvars3}) as
well as the neutrino emission (\figref{Fig:nulums}) exhibit an
ordering with the initial (poloidal) magnetic field.  It should
be noted that an analysis of the gain layer is relevant mostly before
shock revival, after which its growth parallels that of the expanding
shock.

In the case of \mRO, the mass in the gain layer ($M_{\igain}$) starts
to rise already before the begin of the increase of the maximum shock
radius, while it continues to gradually decrease for a little longer
in \mRw.  During shock stagnation, models \Rw and \RO launch their
explosions out of a gain layer of $M_{\igain} \gtrsim 0.02 \, \msol$.
\MRw emits neutrinos at the highest rate reaching
$L_{\nu_e} \approx \zehnh{6.6}{52} \, \ergs$ after the end of the
neutrino burst.  Consequently, its neutrino heating is strongest of
all models with an specific heating rate,
$\eta_{\igain} = Q_{\nu} / ( M_{\igain} c^2)$, where $Q_\nu$ is the
energy deposition due to neutrinos averaged over the entire volume of
the gain layer (see Eq.\,22 of Paper I), peaking at
$\eta_{\igain} \approx 0.3 \, \isek$ close to the start of the shock
runaway.  \MRO achieves shock revival at the same time and somewhat
more violently despite lower neutrino luminosities (about $10 \, \%$
less than \Rw) and heating rate
($\eta_{\igain} \lesssim 0.2 \, \isek$) smaller than in \mRw.

In both models, the total and rotational kinetic energies in the gain
layer grow at a similar rate until the onset of the explosion.  The
magnetic energy, dominated by the toroidal component, reaches up to
$10 \, \%$ of the rotational energy in the gain layer of \mRO.  For
\mRw, it is insignificant in the phase leading to shock revival.  Its
sharp increase thereafter is the result of the ejection of
magnetised, hot matter (with entropy per baryon $>20\,\kbb$) from
above the north polar region of the PNS (see Sect.\,\ref{sSek:Rw}).

Compared to models \Rw and \RO, \Rp shows a gain layer of a higher
mass exposed to slightly lower neutrino luminosities.  Consequently,
the specific heating rates are low.  Despite the higher gain mass,
\MRp begins its post-bounce evolution with kinetic and rotational
energies comparable to models \Rw and \RO.  Its magnetic energies,
both for the poloidal and the toroidal components, exceed those of
\mRO by a factor of a few (\figref{Fig:globvars3}\panel{d}),
consistent with the enhanced poloidal field in the initial conditions,
and those of \Rw by several orders of magnitude.  The corresponding
values of gain mass and kinetic and magnetic energies of \mRs are
still higher than in \mRp, which reflects the lack of a phase of shock
stagnation and the fact that the gain layer is almost identical to the
expanding post-shock region at all times.

Important properties characterizing the PNS show a similar ordering
with the initial magnetic fields as can be seen in
\figref{Fig:globvars4}.  Accretion increases their masses
(\figref{Fig:globvars4}\panel{a}) for a few 100 ms to maximum values
of $\MPNS \approx 2.16 \, \msol,\, 2.03 \, \msol,\, 1.88 \, \msol, $
and $1.75 \, \msol$ for models \Rw, \RO, \Rp, and \Rs, respectively.
The PNS mass levels off or, for stronger fields, even enters a phase
of decrease showing a qualitative agreement with the behaviour
displayed by the equivalent axisymmetric models (see Paper II).  PNSs
of models with weaker fields possess very high rotational energies
($\ErPNS \lesssim \zehnh{2}{52} \, \erg$;
\figref{Fig:globvars4}\panel{b}), while stronger fields reduce the
rotational energy by a factor of about 2.  Even in these cases, the
average spin frequencies are above $\Omega \gtrsim 10^{3} \, \isek$.
Right after bounce, the ratio of magnetic to rotational energy of the
PNS of \MRs is $\EmPNS / \ErPNS \sim 0.2$, i.e., about one order of
magnitude higher than in \modls{W} and \Rs.  The strong field causes
the rotation to slow down after 500\,ms in \modls{W} and \Rs, while
\modls{O} and \Rp show a growing trend for $\ErPNS$. The behaviour of
the rotational energy of the 3D models is qualitatively similar to
their 2D counterparts. However, the values of $\ErPNS$ are
systematically larger in axisymmetry.  The modification of the
rotational profile by the magnetic field explains the behaviour of the
magnetic energy of the PNS in all models.  As a result, the models
with weaker initial fields end up with higher final magnetic energies
of around $\EmPNS \sim \zehnh{(2\ldots5)}{50}\,\erg$
(\figref{Fig:globvars4}\panel{c}).

At late times, the evolution of the magnetic energies of models \Rw
and \RO is more variable than in the case of the other two models.  We
find, in particular, a strong increase to
$\EmPNS \approx \zehn{51} \, \erg$ at $t \approx 0.7 \, \sek$ and
subsequent decline for \mRw and a somewhat weaker intermediate minimum
of the magnetic energy at $t \approx 0.55 \, \sek$ in \RO.  The
typical timescales of the variation are of the order of $100 \, \ms$,
\ie, much slower than the dynamical timescales of the PNSs.  we do not
have a definite explanation for this behaviour, but will present a
likely interpretation when discussing the models below.

Except for \mRw, which, as will be shown below, follows a different
dynamical path than the other three models, the evolution of the
explosion energies and energy fluxes correlates with that of the
poloidal field energy of the PNS.  Among them, the one with the most
violent explosion, \mRs, possesses the strongest field and, in
particular, a predominantly poloidal field geometry.  The ordering of
the poloidal field energies of \Rp and \RO the same as that of the
energy fluxes.  Furthermore, the time evolution of
$F^{+}_{\mathrm{E}}$ is parallel to that of the poloidal field energy.

In the following discussion of individual models, we quantify the
deformations of surfaces such as the shock wave or the outer boundary
of the PNS by expanding their radii as a function of the angular
coordinates, $R (\theta,\phi)$, into the spherical harmonic components
(Superscripts ${}^{\mathrm{sh}}$ and ${}^{\mathrm{PNS}}$ will be used
for the shock wave and the PNS surface, respectively.).  Following
\cite{Burrows_et_al__2012__apj__AnInvestigationintotheCharacterofPre-explosionCore-collapseSupernovaShockMotion},
we define the amplitudes
\begin{equation}
  \label{Gl:r-acomp}
  a_{l,m} =
  \frac{( -1)^{|m|}}{\sqrt{4 \upi ( 2 l + 1 ) } }
  \int \mathrm{d} \Omega \,
  R (\theta, \phi ) Y^m_l ( \theta, \phi)
\end{equation}
in terms of the spherical harmonics, $Y^m_l$, which in turn depend on
the associated Legendre polynomials, $P^m_l (\cos \theta)$,
\begin{eqnarray}
  \label{Gl:Ylm}
  Y^m_l ( \theta, \phi )
  & = &
        \left\{
        \begin{array}{ll}
          N^0_l P^0_l (\cos \theta ) & m = 0,
          \\
          \sqrt{2} N^{|m|}_l P^{|m|}_l (\cos \theta ) \cos |m| \phi & m \neq 0
        \end{array}
                                                                      \right.,
  \\
  \label{Gl:Nlm}
  N^m_l
  & = &
        \sqrt{
        \frac{2 l + 1}{4 \upi}
        \frac{( l - m )!}{(l + m)!}
        }.                                                                      
\end{eqnarray}
The lower order coefficients have direct physical meaning: $a_{0,0}$
is the average radius, the dipole coefficients $a_{1,m}$ represent the
average displacement of the $R(\theta,\phi)$-surface from the origin
in the three coordinate directions, and the quadrupole amplitude
$a_{2,0}$ can be used to quantify a prolate or oblate deformation of
the surface.  For an analysis of the latter property, we furthermore
determine the radii along the polar axis and the maximum, minimum, and
average equatorial radii of the $R(\theta,\phi)$-surface.

We will employ several variables commonly used in the analysis of core
collapse such as the mass and (total) energy in the gain layer or the
ratio between the timescales for advection of gas through the gain
layer and for heating by neutrinos.  These variables can be defined in
different ways.  We will use the definitions of the timescales for
advection, $\tauadv$, neutrino heating, $\tauhtg$, and the propagation
of \Alfven waves through the gain layer, $\tauAlf$, put forward in
\citetalias{Obergaulinger_Aloy__2020__mnras__MagnetorotationalCoreCollapseofPossibleGRBProgenitorsIExplosionMechanisms}
-see their Eqs.\,24, 25, and 28.  In these quantities, we can account
for their dependence on the angular coordinates by computing radial
integrals or averages on radial rays at fixed $(\theta, \phi)$.

\subsection{\MRw: neutrino-driven explosion}
\label{sSek:Rw}

\begin{figure*}
  \centering
  \begin{tikzpicture}
    \pgftext{
      \hbox{ 
        \includegraphics[width=0.49\linewidth]{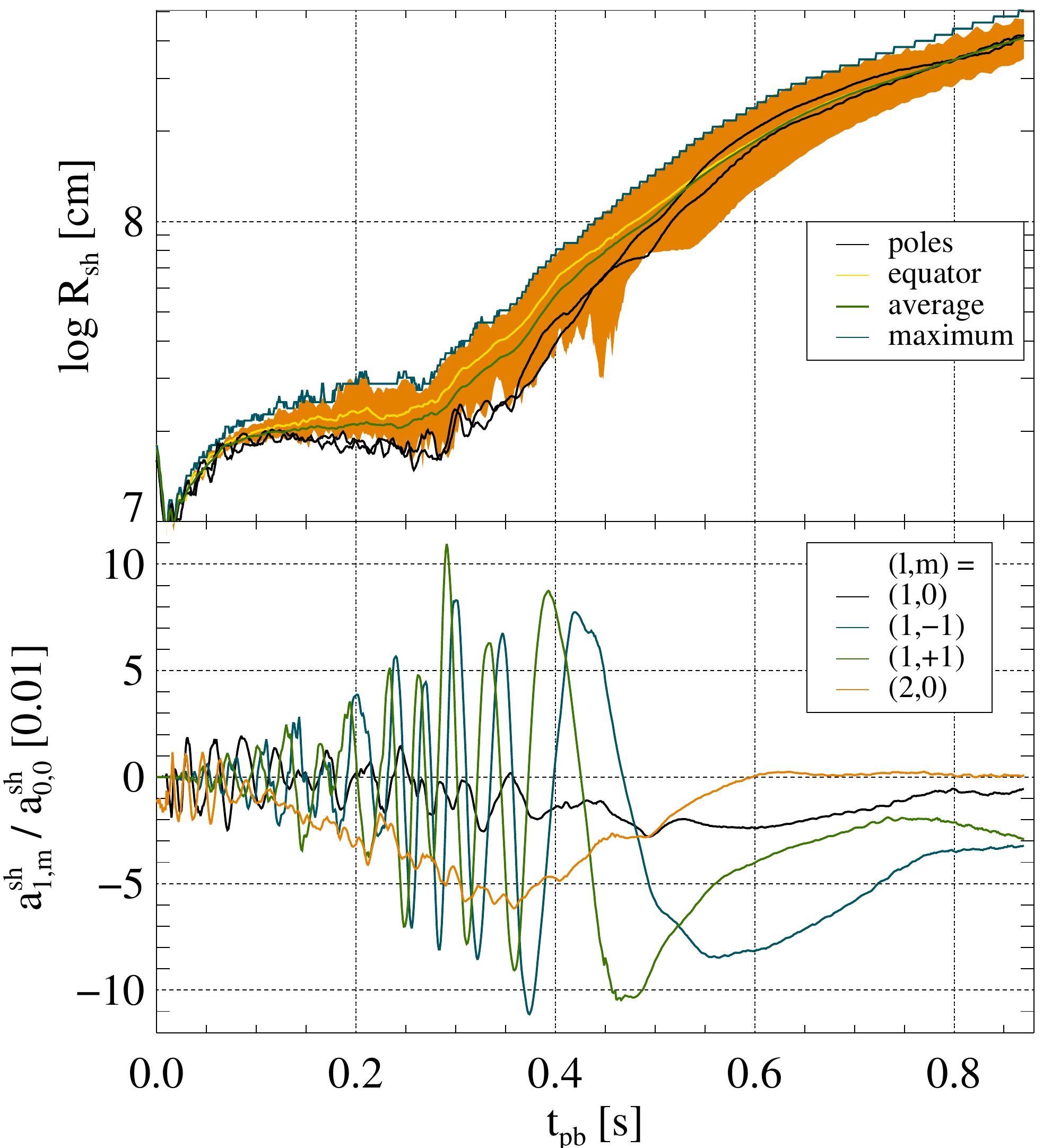}
        \includegraphics[width=0.49\linewidth]{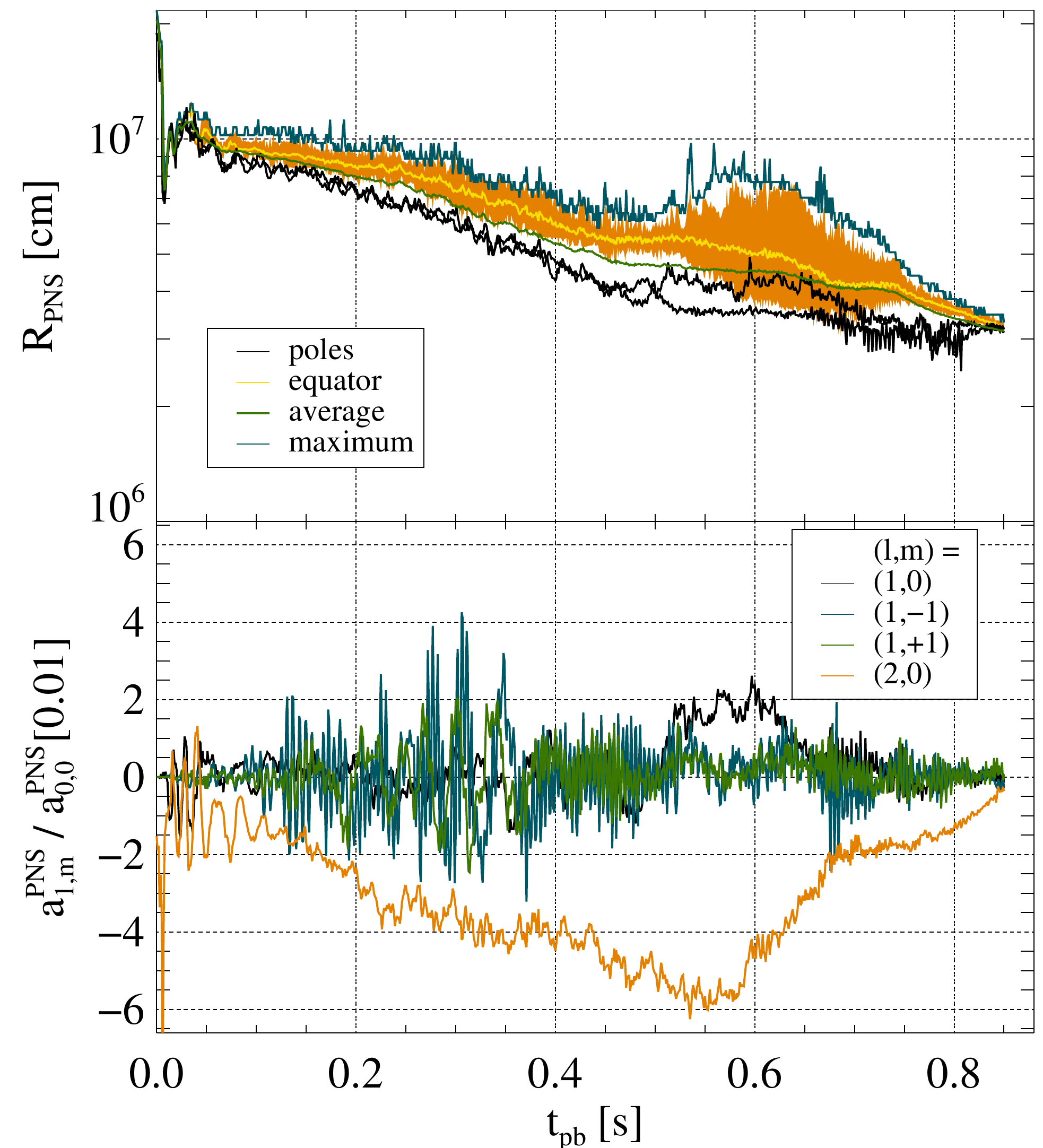}
      }
    }%
    \node[fill=white, opacity=1, text opacity=1] at (-8.25,+4.49) {\large (a)};
    \node[fill=white, opacity=1, text opacity=1] at (0.54,+4.49) {\large (b)};
    \node[fill=white, opacity=0, text opacity=1] at (-8.25,-3.99) {\large (c)};
    \node[fill=white, opacity=0, text opacity=1] at (0.54,-3.99) {\large (d)};	
  \end{tikzpicture}
  \caption{
    Evolution of shock (\banel{left}) and PNS (\banel{right} panel)
    radii of \mRw.  In the top panels, the radii measured along the
    polar directions and the maximum radii as well as the average over
    all longitudes in the equatorial plane and over the entire surface
    are shown by lines as indicated in the legend, and the range of
    radii in the equatorial plane is represented by the salmon band.
    The \banel{bottom} panels show the evolution of the three
    normalised multipole amplitudes of the shock/PNS
    surface, $a^{\mathrm{sh/PNS}}_{1,m} / a^{\mathrm{sh/PNS}}_{0,0}$
    ($m = 0, \pm 1$), and of one of the normalised quadrupole
    amplitudes,
    $a^{\mathrm{sh/PNS}}_{2,0} /
    a^{\mathrm{sh/PNS}}_{0,0}$.
  }
  \label{Fig:35OC-Rw-rad}
\end{figure*}

\begin{figure*}
  \centering
  \begin{tikzpicture}
    \pgftext{\vbox{
        \hbox{ 
          \includegraphics[width=0.05\linewidth]{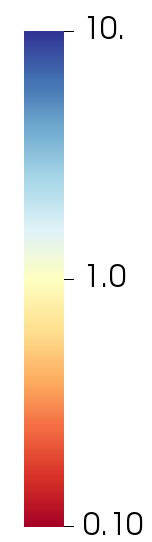}
          \includegraphics[width=0.23\linewidth]{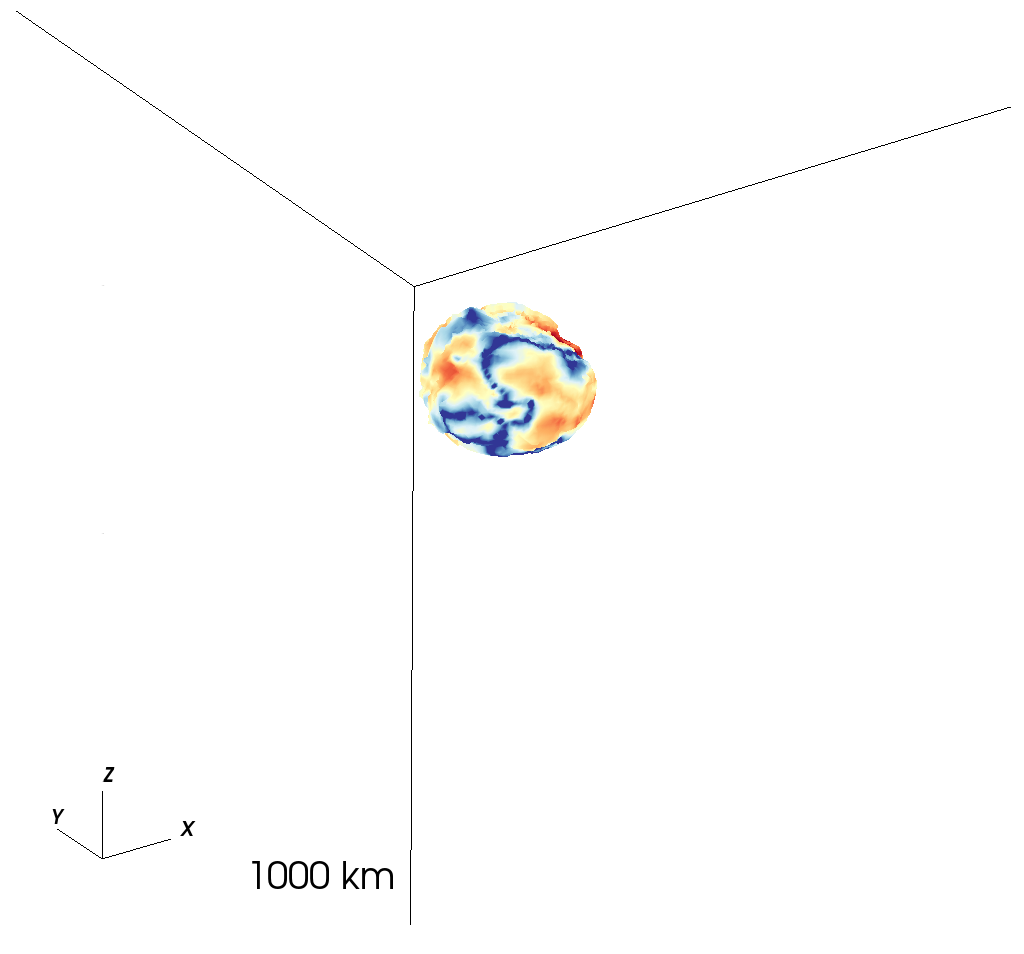}
          \includegraphics[width=0.23\linewidth]{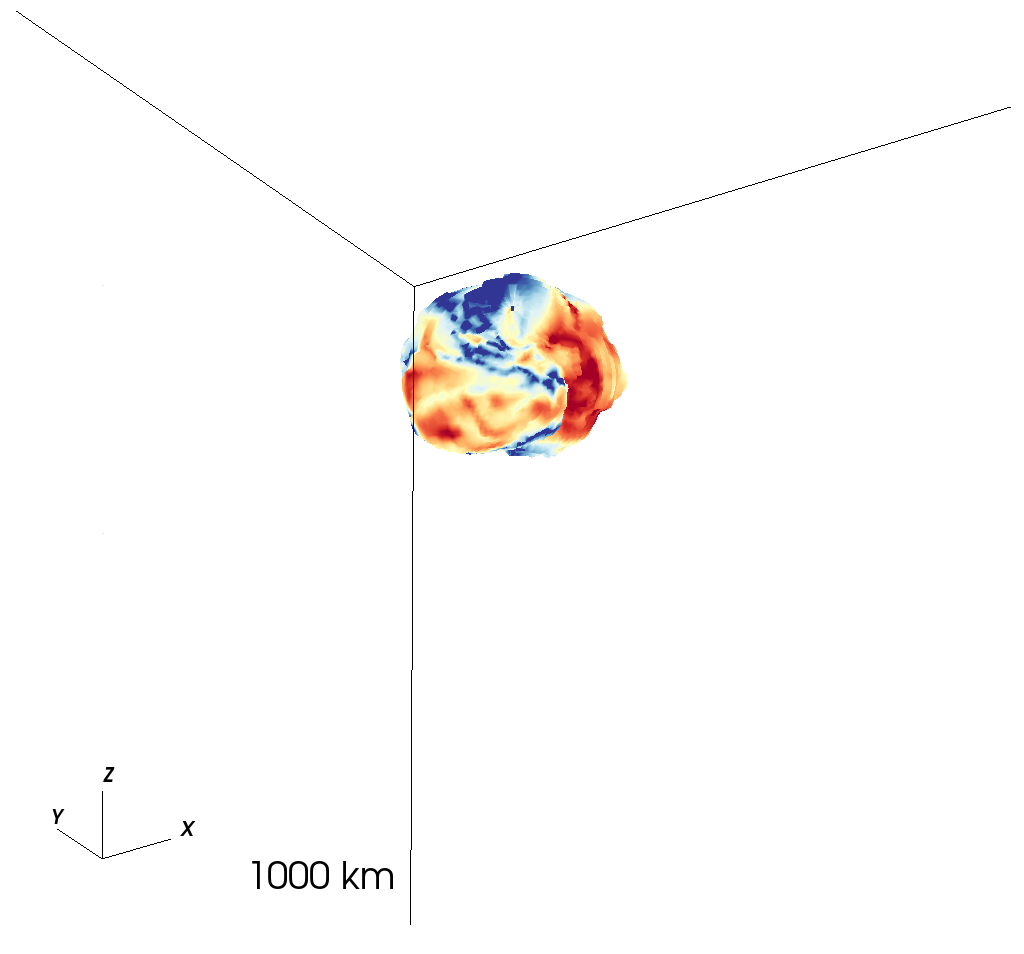}
          \includegraphics[width=0.23\linewidth]{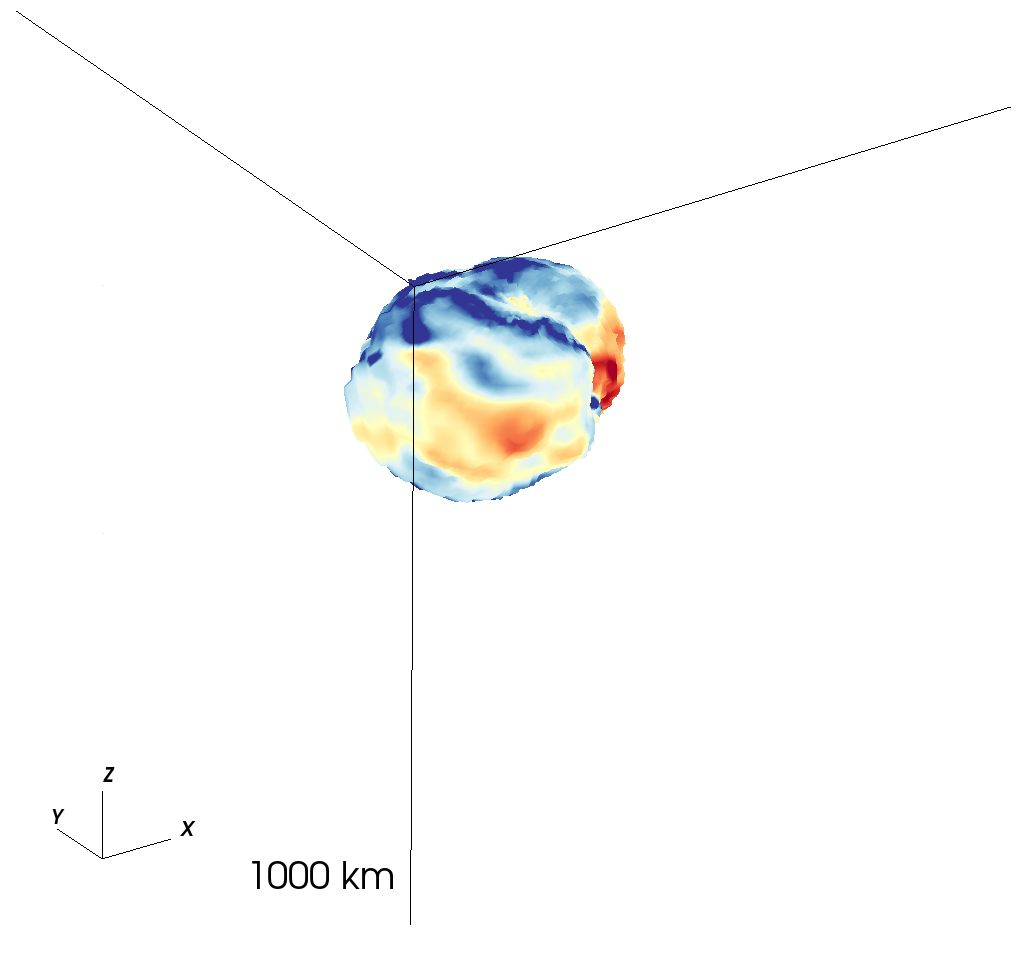}
          \includegraphics[width=0.23\linewidth]{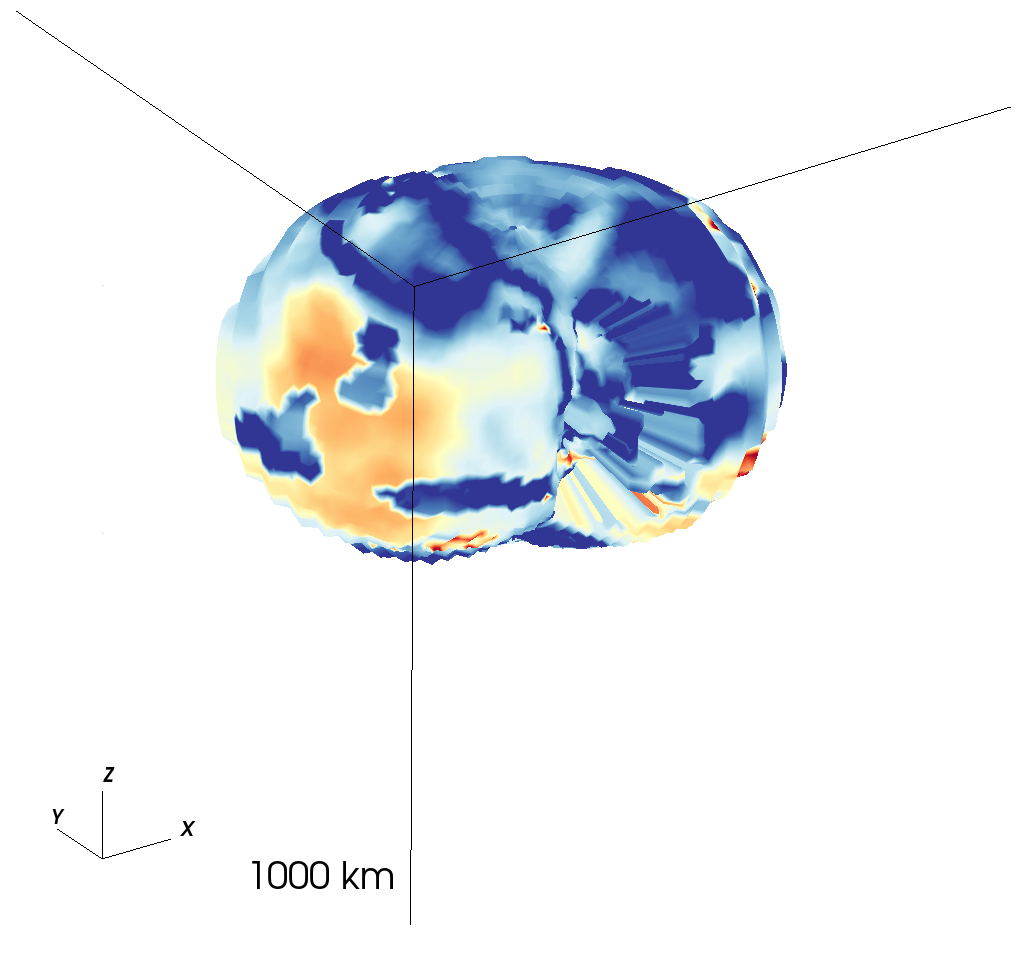}
        }
        \hbox{
          \includegraphics[width=0.05\linewidth]{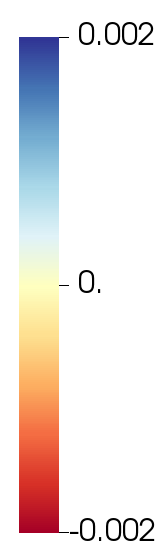}
          \includegraphics[width=0.23\linewidth]{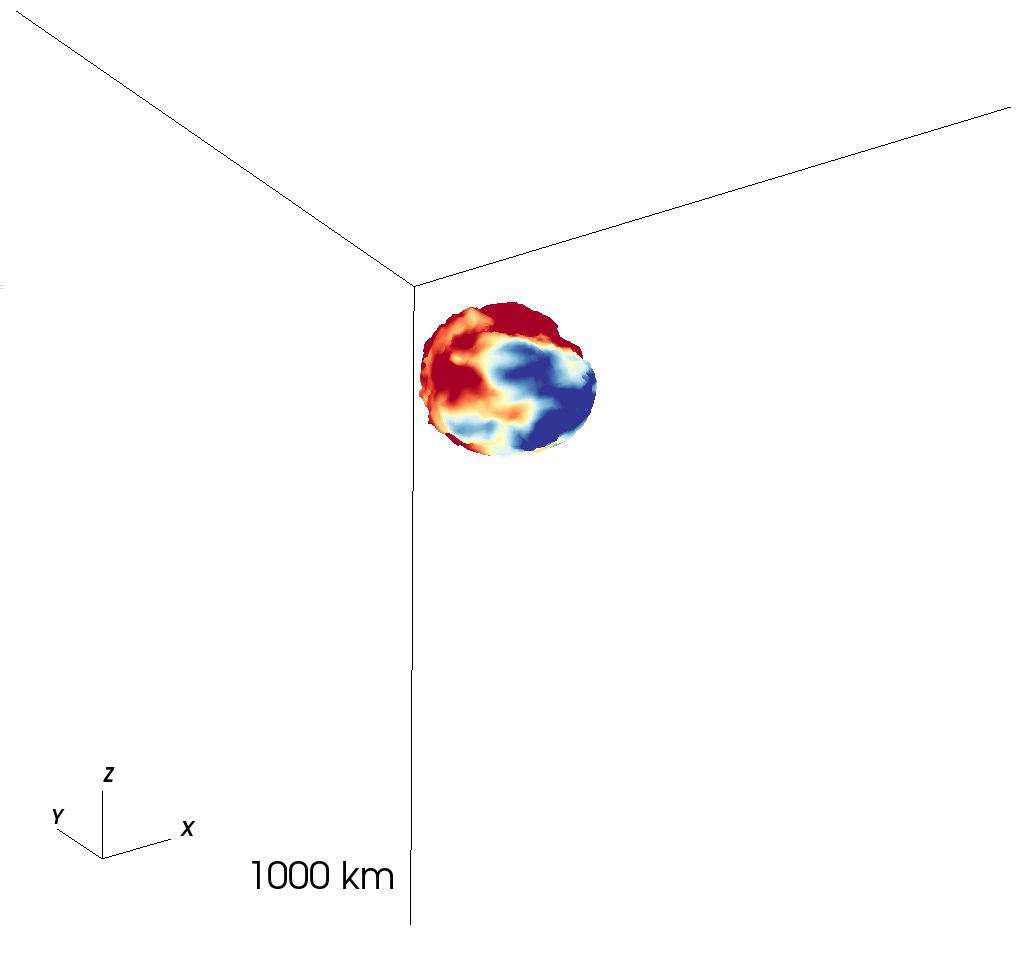}
          \includegraphics[width=0.23\linewidth]{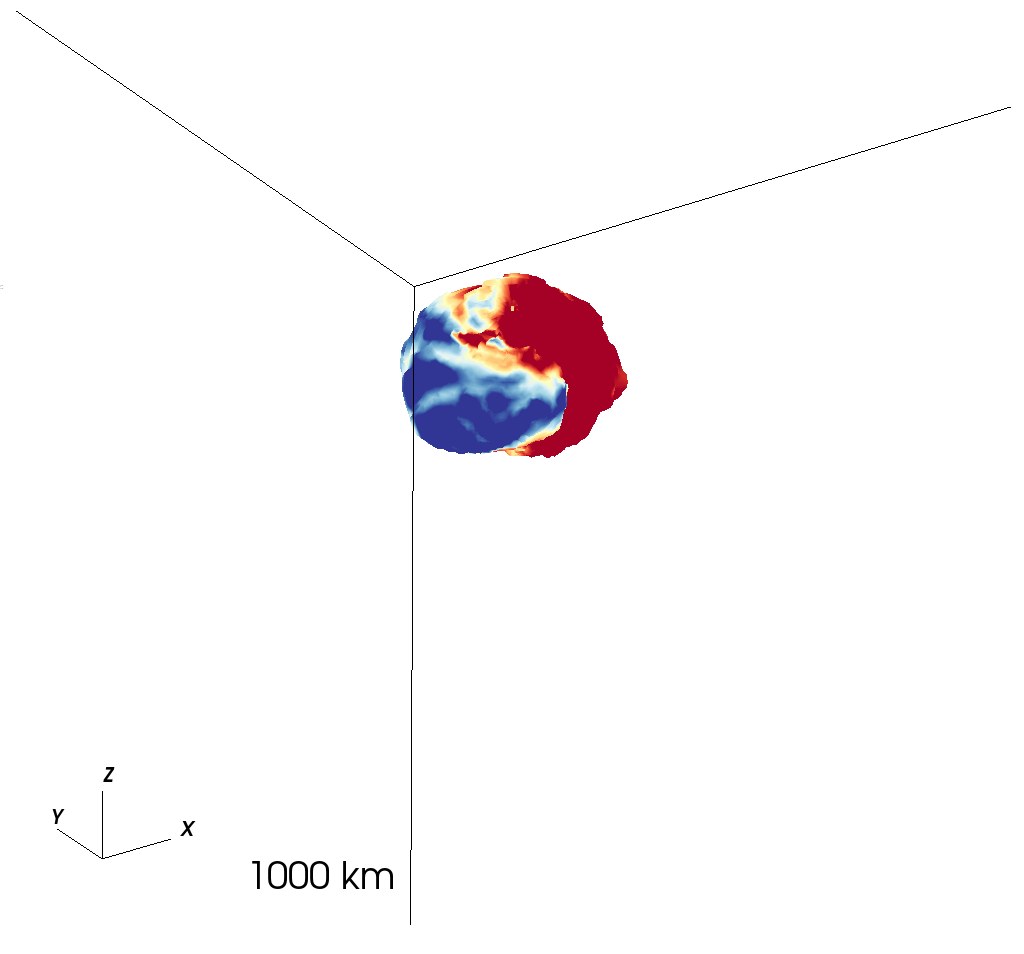}
          \includegraphics[width=0.23\linewidth]{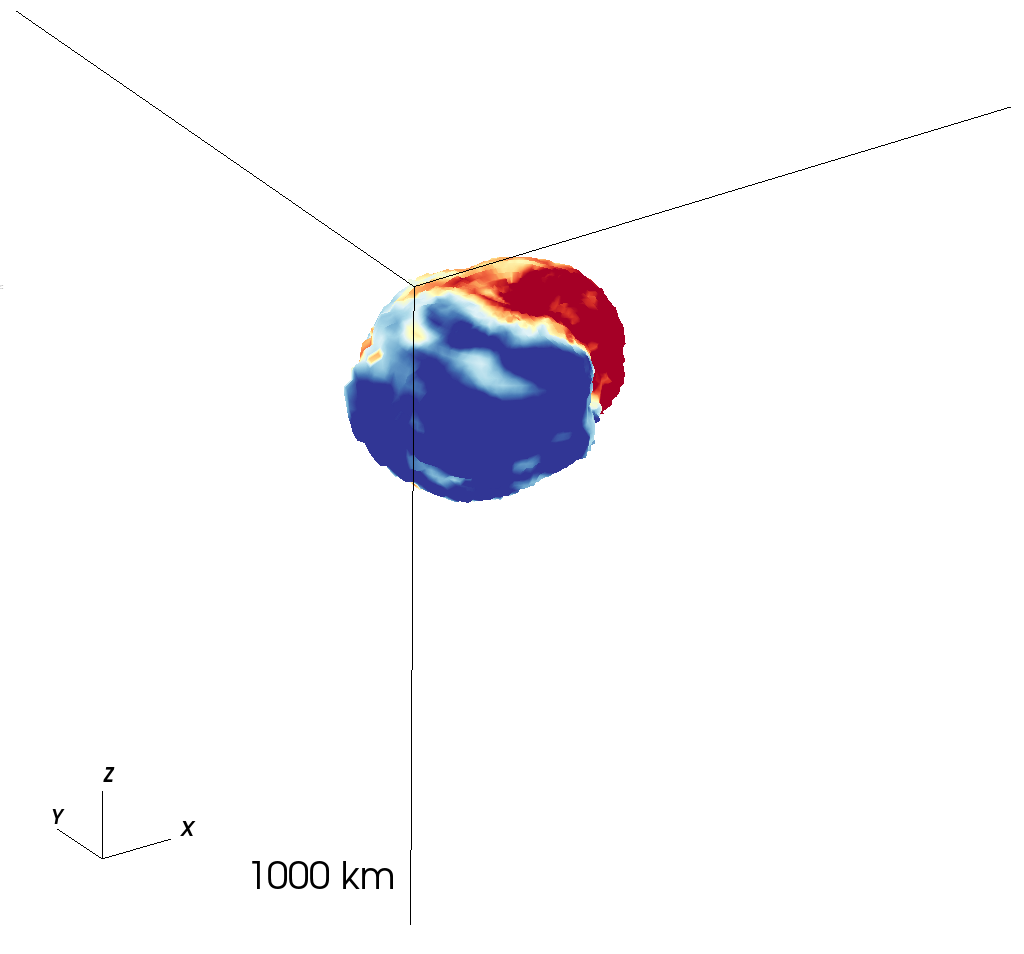}
          \includegraphics[width=0.23\linewidth]{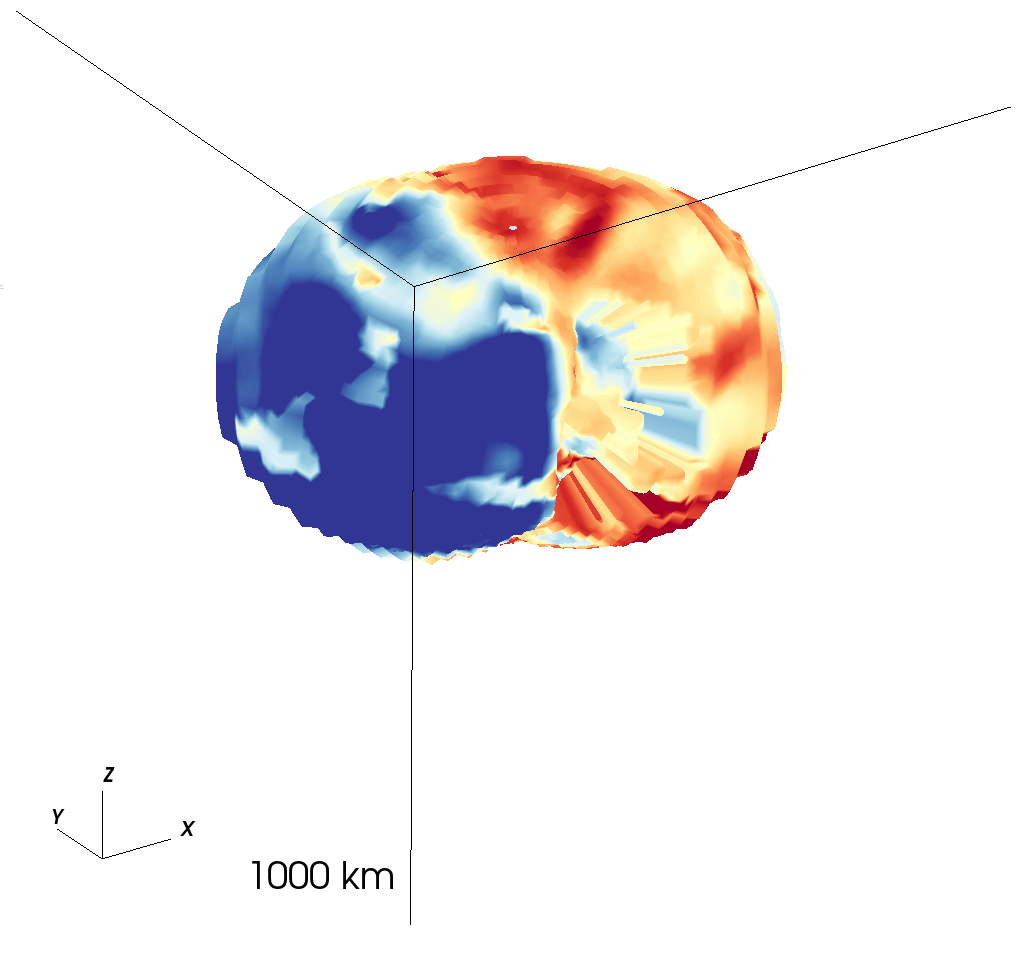}
        }
      }
    }
    \node[fill=white, opacity=0, text opacity=1] at (-8.2,+3.3) {$\tautau$};
    \node[fill=white, opacity=1, text opacity=1] at (-5.9,0.3) {(a)};
    \node[fill=white, opacity=1, text opacity=1] at (-1.7,0.3) {(b)};
    \node[fill=white, opacity=1, text opacity=1] at (2.4,0.3) {(c)};	
    \node[fill=white, opacity=1, text opacity=1] at (6.6,0.3) {(d)};
    \node[fill=white, opacity=0, text opacity=1] at (-8.2,-0.6) {$e [c^2]$};
    \node[fill=white, opacity=0, text opacity=1] at (-5.9,-3.5) {(e)};
    \node[fill=white, opacity=1, text opacity=1] at (-1.70,-3.5) {(f)};
    \node[fill=white, opacity=0, text opacity=1] at (2.4,-3.5) {(g)};	
    \node[fill=white, opacity=1, text opacity=1] at (6.6,-3.5) {(h)};
    \draw [->,thick] (7.64, -3.29) node[below] {\sffamily triple point} -- (6.8,-2.19) ;
  \end{tikzpicture}  
  \caption{
    Shock surface of \mRw at, from left to right, $t =
    0.265, 0.29, 0.34, 0.41 \, \sek$.  The 
    shock is represented by a surface whose colours 
    display quantities characterizing the explosion conditions:
    in the \banel{top} and \banel{bottom} panels, the ratio of heating
    and advection time scales (blue/red:
    heating faster/slower than advection), and the specific
    total energy (blue/red: unbound/bound matter), respectively.
  }
  \label{Fig:Rw-3dtauratio}
\end{figure*}

In the roughly 250\,ms between core bounce and the onset of the
explosion, the strong centrifugal force of the rapidly rotating matter
leads to a decidedly oblate shape of the shock wave.  While its polar
radii shrink, the shock wave continuously expands in the equatorial
plane, as shown in \figref{Fig:35OC-Rw-rad}.  Consequently, the
pole-to-equator axis ratio (compare black lines to yellow line and
orange band in \figref{Fig:35OC-Rw-rad}\panel{a}) decreases to values
around $3:2$ after $t \approx 200 \, \ms$.  This tendency is reflected
in the negative values of $\ashlm{2}{0} / \ashlm{0}{0}$
(\figref{Fig:35OC-Rw-rad}\panel{c}).  The dipole coefficients
$\ashlm{1}{\pm 1} / \ashlm{0}{0}$ show quasi-periodic oscillations
with an increasing amplitude and a phase shift of $\approx \upi/2$
between each other.  Such an evolution is indicative of a growing
$m = 1$ spiral mode rotating around the PNS.

This mode is visible in \figref{Fig:Rw-3dtauratio} showing the shape
of the shock wave at different times,
$t = 0.265, 0.29, 0.34, 0.41 \, \sek$.  The panels follow the
expansion of the shock surface.  Their orientation is such that the
rotation of the gas and the spiral pattern around the $z$-axis
proceeds in a counterclockwise sense.  In the \banel{top} and
\banel{bottom} rows of panels, the colours of the surface encode the
logarithm of the ratio between the advection and heating timescales,
$\tautau$: red and blue shades correspond in the \banel{top} panels to
regions where heating is slower and faster than advection and in the
bottom panel to regions with negative and positive total energies in
the gain layer, respectively.  Already at an early stage of its
growth, the spiral mode shows a close correlation to the pattern of
the neutrino heating with the patter of fast/slow cooling and
increasingly positive total energy (blue shades in the \banel{bottom}
row) developing behind the triple point of the $m = 1$ mode.

This geometry with the most favourable conditions for the explosion
($\tautau > 1$, positive total energy) found near the equator
contrasts with the emission geometry of the neutrinos.  The
luminosities of all flavours are higher at high latitudes than close
to the equator; in \figref{Fig:nulums}, compare the green lines (north
and south poles) to the red ones (equator).  As a consequence, both
$\tauadv$ and $\tauhtg$ decrease above the poles, with the net effect
being a rise of $\tautau$ (as in axial symmetry).  During the entire
pre-explosion phase, $\eta_{\igain}$ is consistently larger by a
factor of at least 4 at $\theta = 0, 180 \, \grad$ than at
$\theta \approx 90 \, \grad$.  This enhanced efficiency, however, does
not translate into a polar explosion (like in axial symmetry) because
the poles are also the locations of the fastest accretion through the
gain layer, while the advection time increases in the growing $m = 1$
mode near the equator, thus favouring an oblate explosion.

When the shock wave enters the explosion phase ($\tpb = 0.34 $ and
$0.41 \, \sek$; \figref{Fig:Rw-3dtauratio}\panel{g} and \panel{h}),
the hot bubble has grown such as to encompass almost the entire
post-shock layer, with the most notable exception being the north and
south poles.  As the shock continues to expand, its shape partially
loses the oblate character and the polar, equatorial, and average
radii grow in parallel.  The dipole and quadrupole coefficients of the
shock surface evolve slowly.  They maintain a small, but
non-zero, value until the end of the simulation, indicating a
continuing moderate asphericity of the ejecta.

\begin{figure*}
  \centering
  \begin{tikzpicture}
    \pgftext{\hbox{ 
        \includegraphics[width=0.245\linewidth]{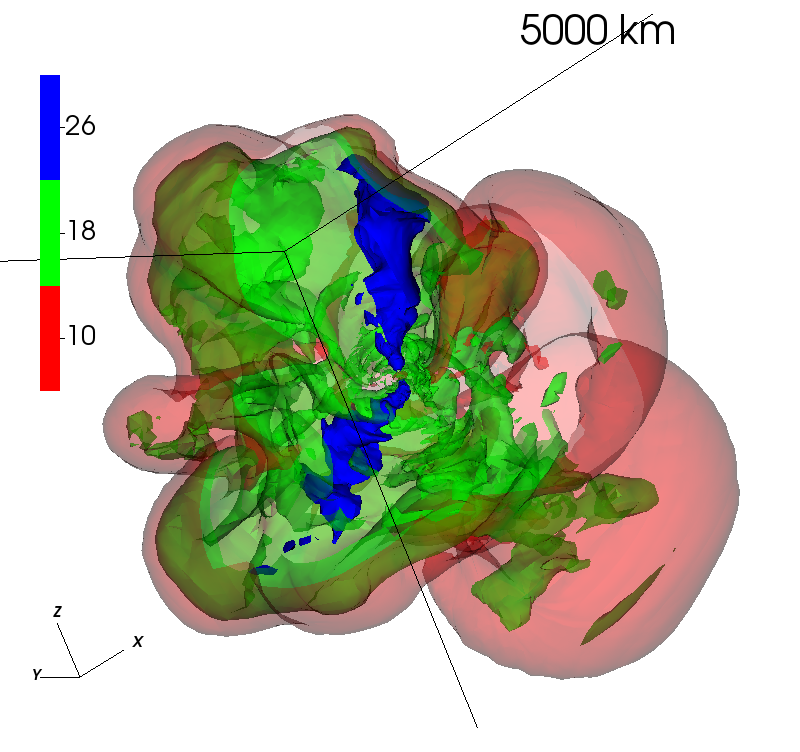}
        \includegraphics[width=0.245\linewidth]{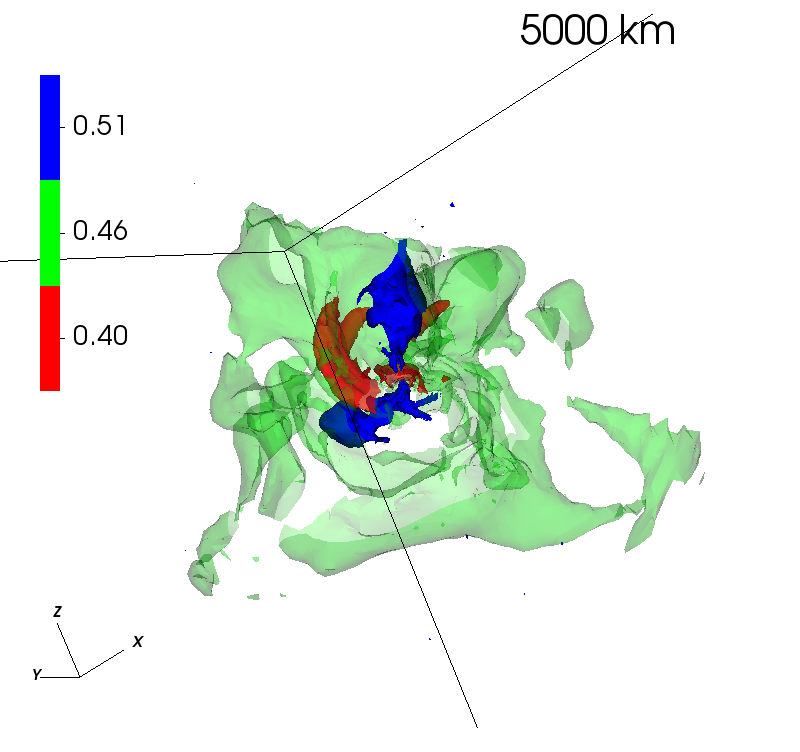}
        \includegraphics[width=0.245\linewidth]{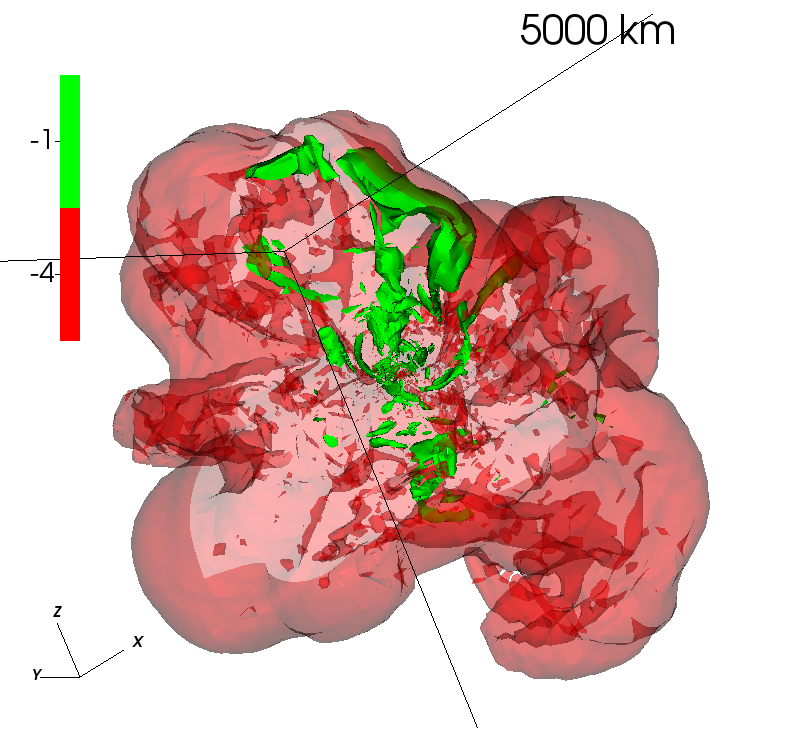}
        \includegraphics[width=0.245\linewidth]{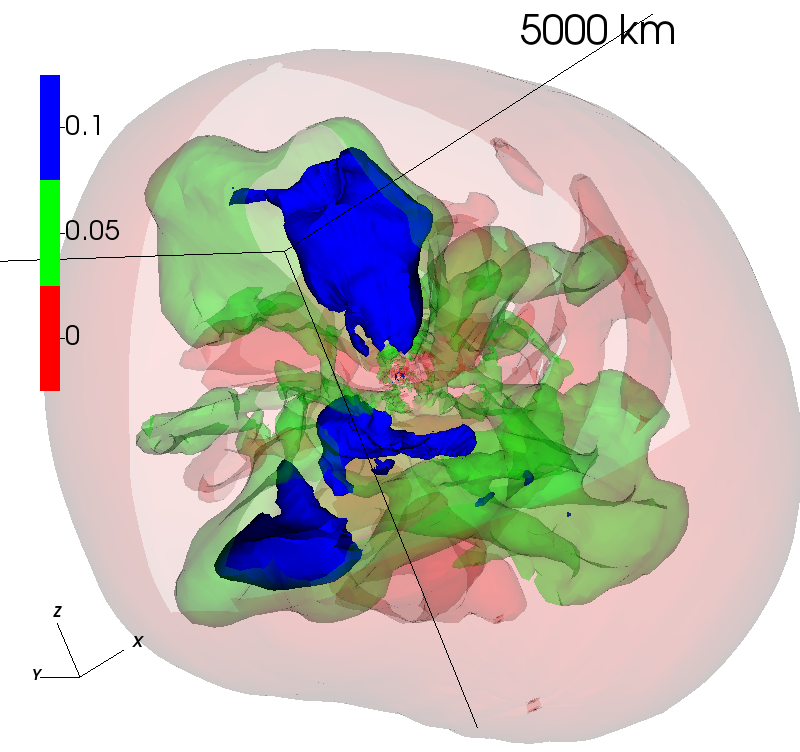}
      }
    }%
    \node[fill=white, opacity=0, text opacity=1] at (-8.75,-1.85) {(a)};
    \node[fill=white, opacity=0, text opacity=1] at (-4.30,-1.85) {(b)};
    \node[fill=white, opacity=0, text opacity=1] at (0.10,-1.85) {(c)};	
    \node[fill=white, opacity=0, text opacity=1] at (4.54,-1.85) {(d)};
    \node[fill=white, opacity=1, text opacity=1] at (-8.45,1.80) {\sffamily\scriptsize $s\,[k_{\textsc{b}}/\text{baryon}]$}; 
    \node[fill=white, opacity=1, text opacity=1] at (-4.07,1.80) {\sffamily\scriptsize $Y_{\rm e}$};	
    \node[fill=white, opacity=1, text opacity=1] at (0.40,1.85) {\sffamily\scriptsize $\log\beta^{-1}$};	
    \node[fill=white, opacity=1, text opacity=1] at (4.6,1.7) {\sffamily\scriptsize energy flux [$c$]};
    \node[fill=white, opacity=0, text opacity=1] at (4.6,1.9) {\sffamily\scriptsize specific};
  \end{tikzpicture}
  \caption{
    Structure of \mRw near the end of the simulation ($\tpb\approx 0.8
    \, \sek$).  Left to right
    panels show isosurfaces of the specific entropy, the electron
    fraction, the inverse plasma-$\beta$, and the specific energy
    flux.
  }
  \label{Fig:Rw-3dplots}
\end{figure*}

The final stages of the simulation (\figref{Fig:Rw-3dplots}) exhibit
ejecta propagating in the form of several large bubbles.  Inside the
shock wave, the gas possesses a highly asymmetric distribution of
entropy (\figref{Fig:Rw-3dplots}\panel{a}) and electron fraction
(\figref{Fig:Rw-3dplots}\panel{b}).  Two of the bubbles at the highest
radii are filled by relatively cool gas ($s < 18 \, \kbb$ to the right
of the plot).  Hotter gas with $18 \, \kbb < s < 26 \, \kbb$ expands
inside the shock in the form of blobs of a wide range of sizes.  The
largest fraction of the ejecta is slightly neutron-rich with
$0.4 < Y_e < 0.455$ (in the figure, the gas contained by the green
surface, but outside the red one) and $0.455 < Y_e < 0.51$ (between
the green and blue surfaces).  In addition, a more neutron-rich
structure is expanding at low to intermediate latitudes (inside the
red surface).  At very late times, a pair of fast outflows with a
relatively narrow angle, filled by hot gas consisting of symmetric
matter (the blue structure with $s > 26 \, \kbb$, \panel{a}, and
$Y_e > 0.51$, \panel{b} panel) is launched into the post-shock region
from a region near the PNS.  The generation of these outflows begins
after a rise of the magnetic energy in the PNS and the gain layer.
While the ejecta are mostly only very weakly magnetised, the polar
outflows are characterized by $\ateb \gtrsim 0.1$.  The two outflows
form an angle of about $135 \grad$ and are fairly asymmetric and not
perfectly aligned along the rotation axis.  Instead, their propagation
direction varies substantially with time, partly following the motion
of the PNS (see below). Thus, the change in the propagation direction
is not totally random as in the most standard jittering jets model
\citep[JJM][but see also \citealt{Sternberg_Soker_2008MNRAS.384.1327}
for variants of the model that may follow the precession of the
PNS]{Papish_Soker__2011__mnras__Explodingcorecollapsesupernovaewithjitteringjets}.
Furthermore, they tend to widen at higher radii.  By the time of the
end of our simulation, they have not managed to reach the shock
surface.  Whether they maintain their coherence for longer time and
are able to penetrated beyond the more roundish shock wave remains
uncertain.  Even if that is not the case, they constitute an
additional mechanism for transporting energy from the centre of the
core to the surrounding regions and heating the ejecta. Also regarding
jets as a mean to transport energy away from the PNS but below the
shock radius, the computed evolution shares qualitative similarities
with the JJM.  We define the specific radial energy flux,
$f^{e} := \rho^{-1} \left[(e_{\star} + P_{\star}) v^r - \vec b \vec v
  v^r\right] $, where $e_{\star}$ and $P_{\star}$ are the total energy
and pressure, including the magnetic contributions, and show it in
\figref{Fig:Rw-3dplots}\panel{d}.  The blue surface in the northern
hemisphere aligned with the stronger one of the two outflows indicates
the most efficient magnetohydrodynamic energy transfer emanating from
the centre.

As evidenced by the radii in \figref{Fig:35OC-Rw-rad}\panel{b}, the
PNS is gradually contracting.  Around $\tpb \approx 0.5 \, \sek$, this
contraction is interrupted by an expansion in the equatorial region.
This phenomenon will be explored below.  The shape of the PNS is
rather aspherical.  At early times, we see oscillations similar to the
$m=1$ shock modes in the decomposition of the PNS surface in spherical
harmonics (\figref{Fig:35OC-Rw-rad}\panel{d}), albeit at lower
amplitude and higher frequency.

As a result of a relatively limited time of mass accretion ending
after $\tpb \gtrsim 0.45 \, \sek$, the model develops a massive,
rapidly rotating and strongly magnetised PNS with values at the end of
the computed time of $\MPNS \approx 2.16 \, \Msol$,
$\ErPNS \approx \zehnh{2.16}{52} \, \erg$ and
$\EmPNS \approx \zehnh{5.3}{50} \, \erg$.  While the rotational energy
grows parallel to the mass accretion, the magnetic energy experiences
its strongest increase after the mass and rotational energy have
achieved their maximum values.  Its growth between
$\tpb \approx 0.5 \, \sek$ and $\tpb \approx 0.7 \, \sek$ is
exponential, though with an $e$-folding time of about 200\,ms much
slower than typical timescales of the PNS such as the rotational
period or the crossing times of the flow speeds in its interior.

\begin{figure*}
  \centering
  \begin{tikzpicture}
    \pgftext{\hbox{ 
        \includegraphics[width=0.33\linewidth]{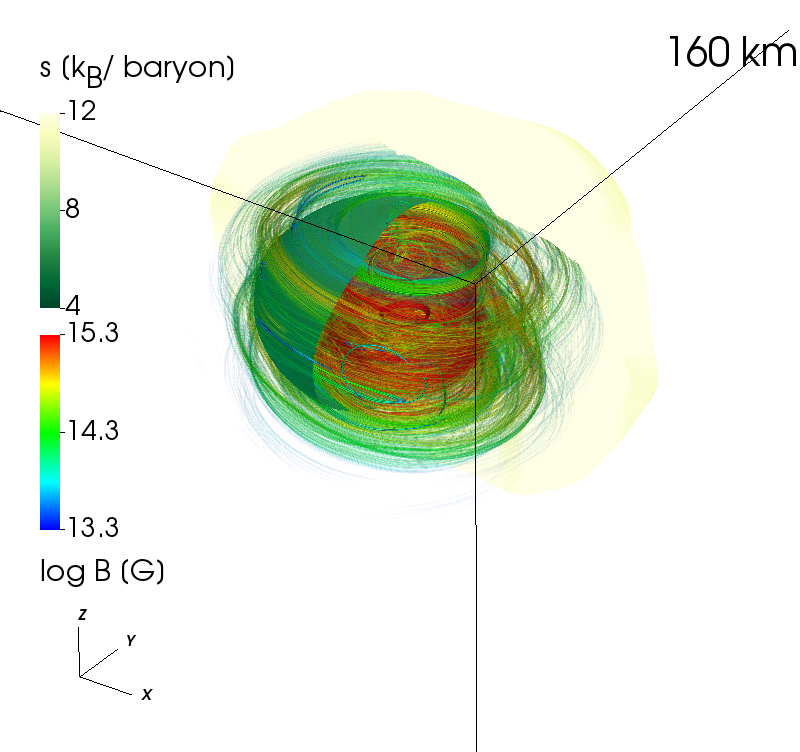}
        \includegraphics[width=0.33\linewidth]{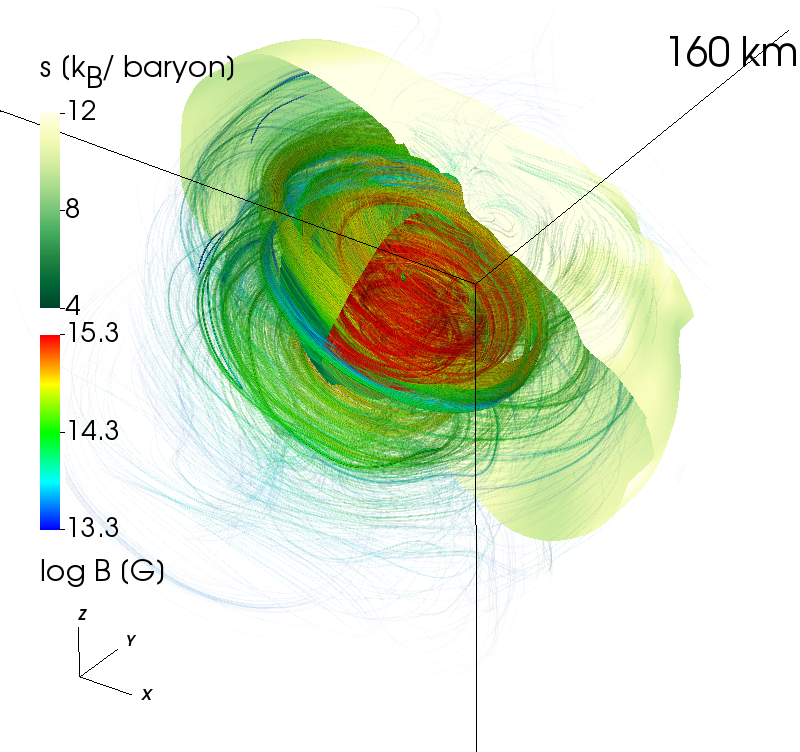}
        \includegraphics[width=0.33\linewidth]{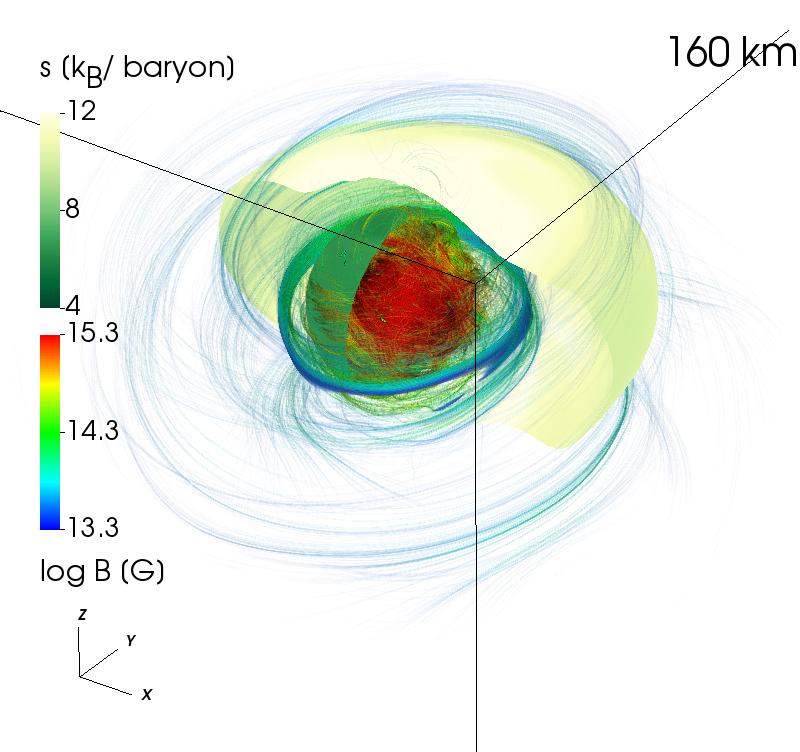}
      }
    }
    \node[fill=white, opacity=0, text opacity=1] at (-8.40,-2.55) {(a)};
    \node[fill=white, opacity=0, text opacity=1] at (-2.55,-2.55) {(b)};
    \node[fill=white, opacity=0, text opacity=1] at (3.34,-2.55) {(c)};
  \end{tikzpicture}
  \caption{
    Structure of the PNS and its magnetic field in \mRw at times $\tpb
    \approx 0.4, 0.6, 0.8 \, \sek$ (left to right).  The green
    surfaces, cut open in one hemisphere,
    represent the specific entropy on iso-density surfaces of
    $\rho = \zehn{10}$ and $ \zehn{12} \, \gccm$ and the colours on the field lines
    display the magnetic field strength.
  }
  \label{Fig:Rw-PNS3d}
\end{figure*}

This very low growth rate is, however, in rough agreement with a
different dynamical mode present in the PNS during the same interval
of time.  We show the evolution of the shape of the PNS and its
magnetic field in \figref{Fig:Rw-PNS3d}.  At $\tpb = 0.4 \, \sek$
(\figref{Fig:Rw-PNS3d}\panel{a}), the PNS has the form of a moderately
flattened ellipsoid with a pole-to-equator axis ratio around $8:5$
rotating about the $z-$axis.  The magnetic field is dominated by field
lines circling the axis.  The asymmetry of the surrounding layers,
modulated by the strong $m = 1$ mode, translates into a strong
asymmetry of the downflows feeding the PNS and a partial tilting of
its rotational axis.  At $\tpb = 0.6 \, \sek$, the outer layers rotate
about an axis forming an angle of $\sim 30 \, \grad$ \wrt the
$z-$axis, whereas the inner regions
($\rho \gtrsim \zehn{12} \, \gccm$) initially maintain the original
axis of rotation (\figref{Fig:Rw-PNS3d}\panel{b}).  Following this
transition, the axis of the magnetic field loops near the PNS surface
is also tilted.  The tilt slowly increases from
$\tpb \approx 0.45 \, \sek$ to $\tpb \approx 0.65 \, \sek$.

\begin{figure}
  \centering
  \includegraphics[width=\linewidth]{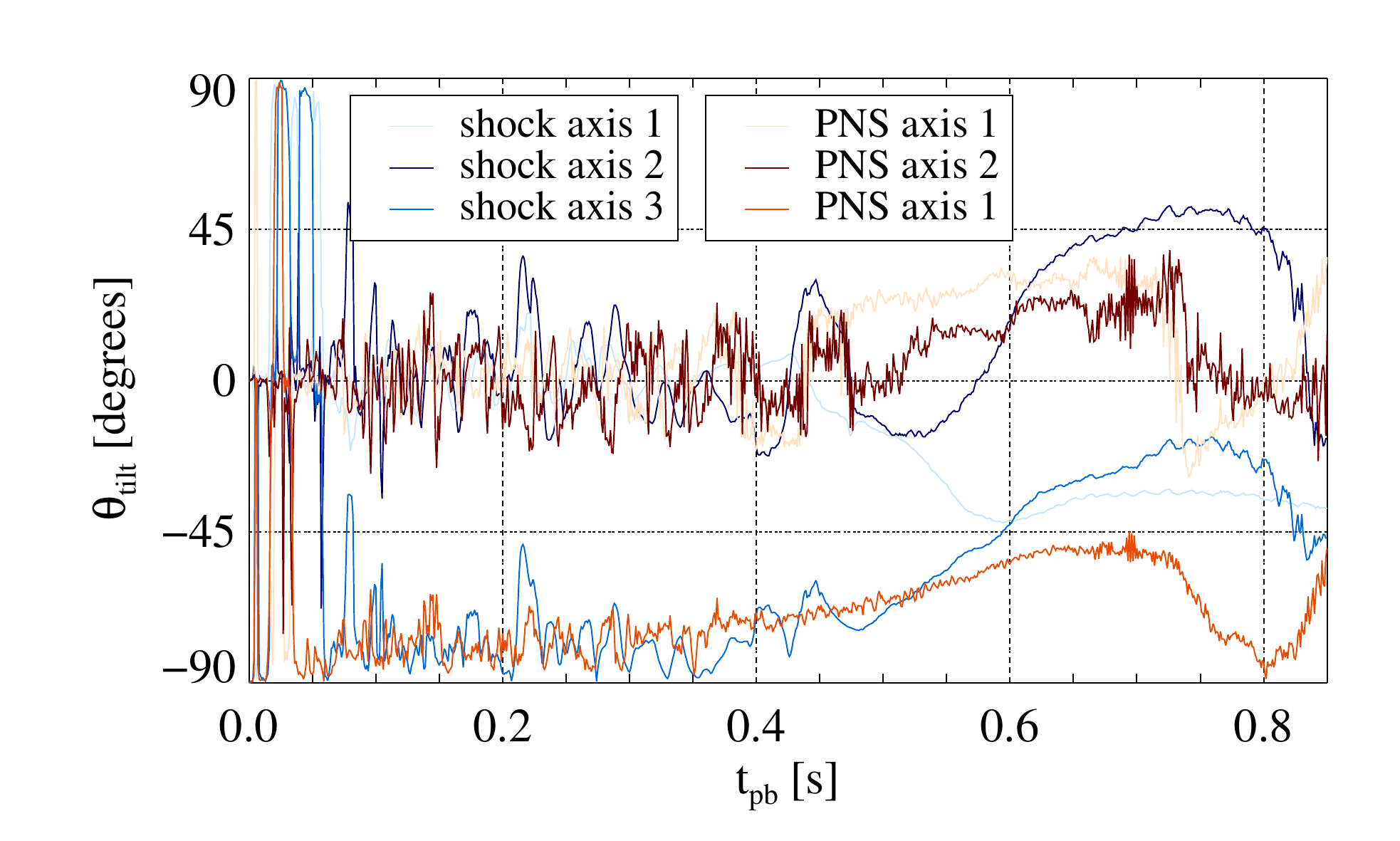}
  \caption{
    Time evolution of the orientation of the PNS and the shock surfaces of \mRw.  Red and blue
    lines show the latitudinal direction of the three major
    axes of the PNS and the shock, respectively.
  }
  \label{Fig:Rw-pnsshockangles}
\end{figure}

In order to quantify more precisely the evolution of the PNS tilting
angle, $\theta_{\rm tilt}^\iPNS$, we express it in terms of the major
axes of the PNS.  To this end, we approximate the PNS as well as the
shock surface by the ellipsoid determined by the principal axes of the
$3\times 3$-tensor formed from the quadrupole spherical harmonic
decomposition \citep[i.e. using Eq.\,\eqref{Gl:r-acomp} with $l=2$ and
$m=0,1,2$; see][\S\,4.1]{Jackson_Book_1962__Edyn} of this approximate
surface.  In both cases, the result is a pair of axes originally in
the equatorial plane and one axis along the $z$-axis.  The direction
of these axes will trace the changing orientation of the PNS or shock.
These changes are shown in the time evolution of the latitude of the
axes in \figref{Fig:Rw-pnsshockangles}.  During the first tens of
milliseconds, the PNS and the shock are close to spherical and the
angles vary in a random manner.  When the PNS starts to tilt and the
shock develops its large-scale deformations, the polar PNS axes drifts
from its original orientation, parallel to the $z$-axis, to latitudes
of $\theta = -50 \grad$.  The other two main axes experience a similar
change.  Temporarily, they return to their original orientations
($\tpb \sim 0.8\,\sek$).  The increase of the deviation by the end of
the simulation suggests that the PNS might continue to oscillate in a
similar way during the subsequent evolution.  We note that the shock
surface changes its orientation by a similar amount.

The tilt manifests itself in the growth of the quadrupole
coefficients, of which $\aPNSlm{2}{0}$ is shown in
\figref{Fig:35OC-Rw-rad}\panel{d}.  At times, the rather violent
changes of the PNS surface also show up in the large and quick
fluctuations of the dipole coefficient $\aPNSlm{1}{m}$.  The tilt is
also responsible for the interruption of the contraction of the PNS
(\cf \figref{Fig:35OC-Rw-rad}\panel{b}).  With decreasing mass
accretion rate, $\theta_{\rm tilt}^\iPNS$ saturates and later on
slowly decreases.  At $\tpb = 0.8 \, \sek$, the PNS shows a dichotomy
between a roughly spherical bulge with
$\rho \gtrsim \zehn{12} \, \gccm$ and a flat, disk-like component
whose rotational axis is mostly aligned with the original one
(\figref{Fig:Rw-PNS3d}\panel{c}).  The magnetic field in the bulge
exceeds the kinetic energy of the gas, which contributes to pinning
its rotational axis to the original orientation, whereas the weaker
field in the surrounding layers allows for the reorientation of their
rotational axis.  The magnetic field couples the entire PNS and leads
to an exchange of the magnetic axes of inner and outer layers such
that they are oblique to both each other and to the $z-$axis.  Most
likely, this configuration is not the final state and the factors at
work up to that point, external ones such as accretion and, possibly
later, fall-back or the expulsion of gas from the PNS surface as well
as internal ones such as the magnetic coupling will continue to modify
the structure and the magnetic field.

Finally, comparing Figs.~\ref{Fig:globvars4}\panel{c} and
\ref{Fig:35OC-Rw-rad}\panel{d}, we find that the increase of $\EmPNS$
and its subsequent relaxation to a lower value correspond to the
growth and reduction of $\aPNSlm{1}{0}$.  Furthermore, the region
where the magnetic energy varies the most coincides with the interface
between the inner core in which the orientation of the rotation
remains constant and the surrounding tilting layers.  This correlation
suggests that the amplification and relaxation of the field are
related to the change of orientation of the PNS rotation.  As a
tentative interpretation, we suggest that the change of the
large-scale structure of the PNS stretches the magnetic field lines
and thus increases the field energy, whereas the subsequent partial
return to the initial configuration relaxes the stretched field lines
and releases magnetic energy.

\subsection{\MRO: stronger field}
\label{sSek:RO}

\begin{figure*}
  \centering
  \begin{tikzpicture}
    \pgftext{
      \hbox{ 
        \includegraphics[width=0.49\linewidth]{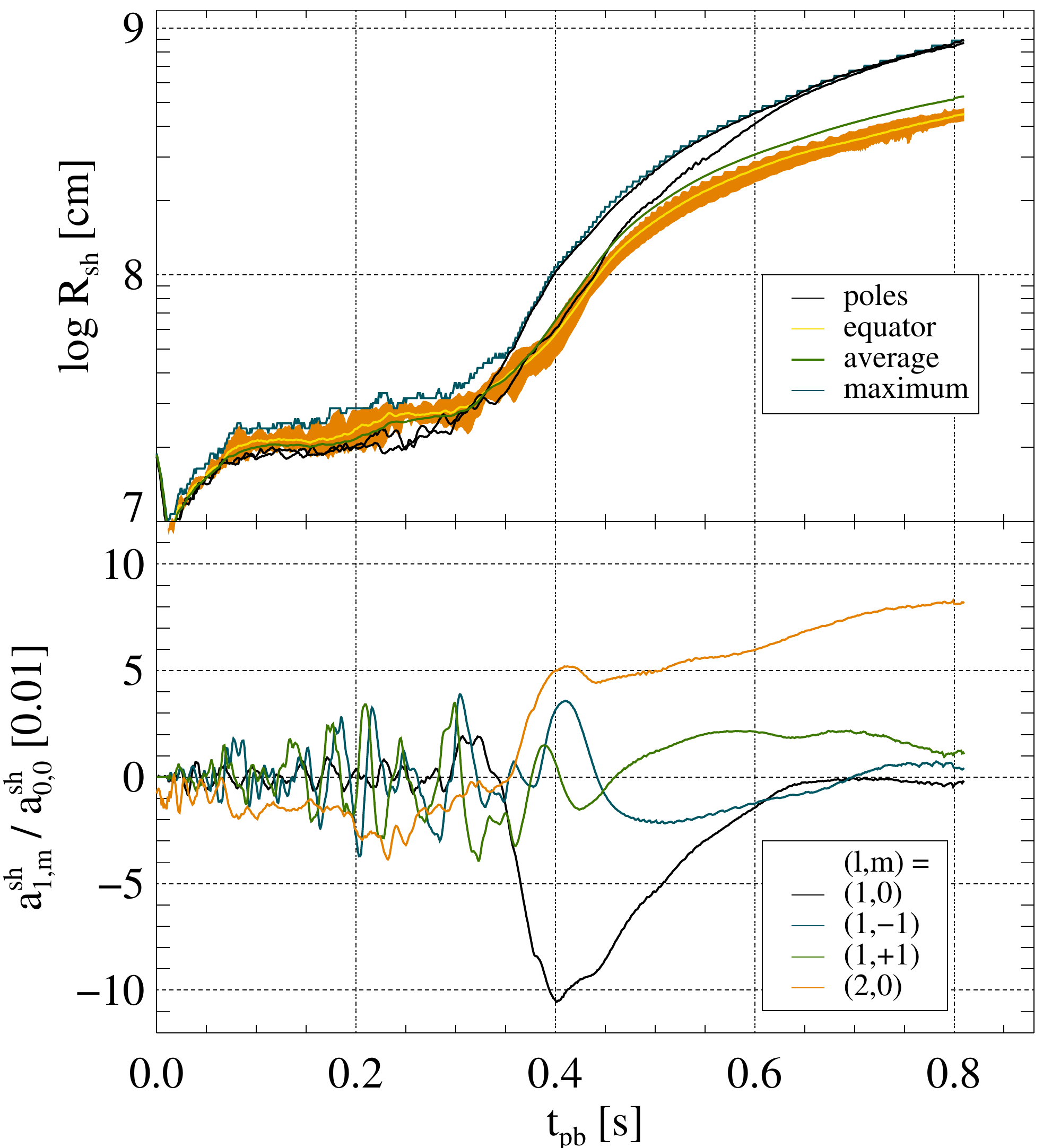}
        \includegraphics[width=0.49\linewidth]{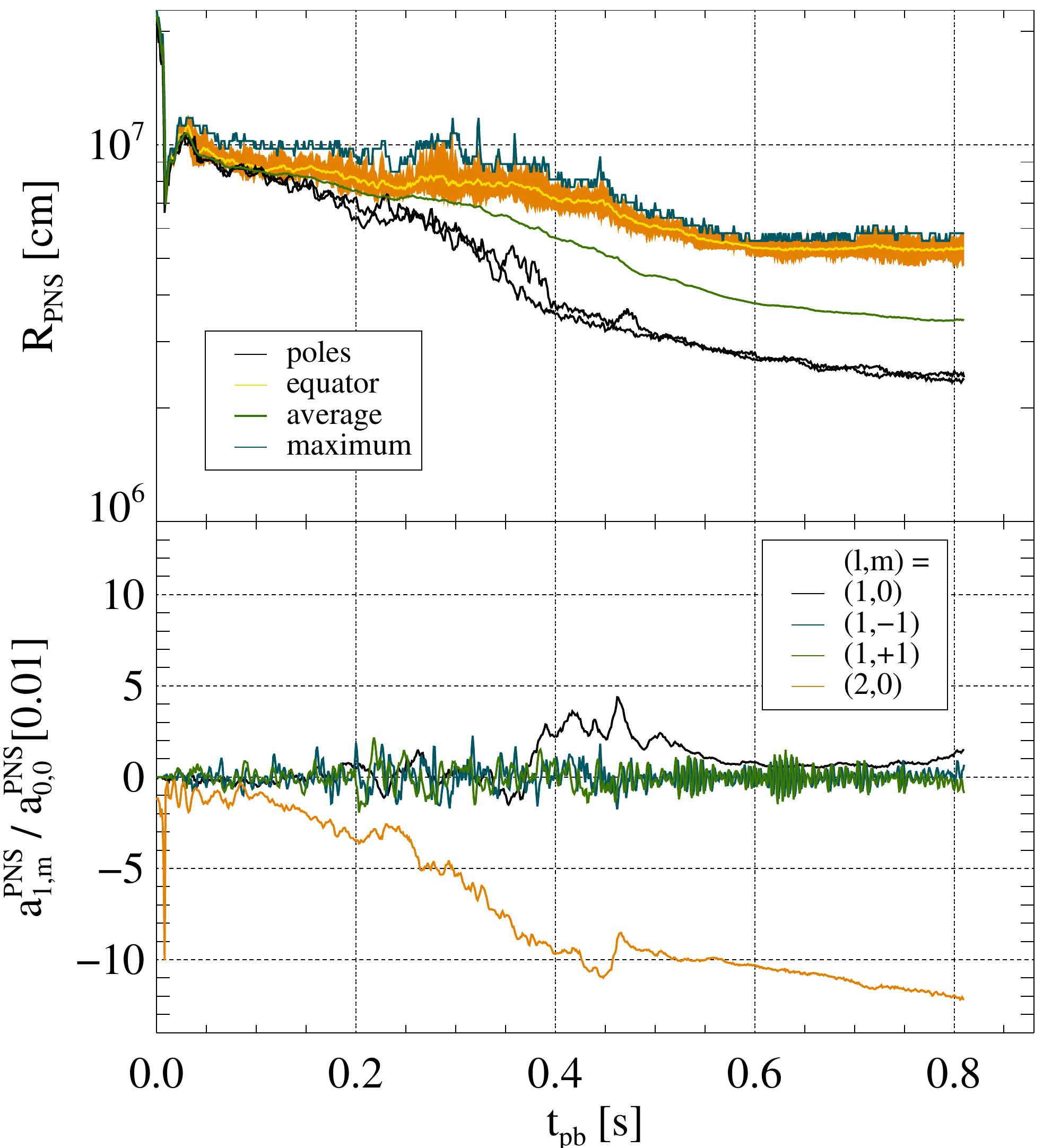}
      }
    }%
    \node[fill=white, opacity=1, text opacity=1] at (-8.25,+4.49) {\large (a)};
    \node[fill=white, opacity=1, text opacity=1] at (0.54,+4.49) {\large (b)};
    \node[fill=white, opacity=0, text opacity=1] at (-8.25,-3.99) {\large (c)};
    \node[fill=white, opacity=0, text opacity=1] at (0.54,-3.99) {\large (d)};	
  \end{tikzpicture}
  \caption{
    Same as \figref{Fig:35OC-Rw-rad}, but for \mRO.
  }
  \label{Fig:35OC-RO-rad}
\end{figure*}

\begin{figure*}
  \centering
  \begin{tikzpicture}
    \pgftext{\vbox{
        \hbox{ 
          \includegraphics[width=0.05\linewidth]{./W-taurat-colbar.png}
          \includegraphics[width=0.23\linewidth]{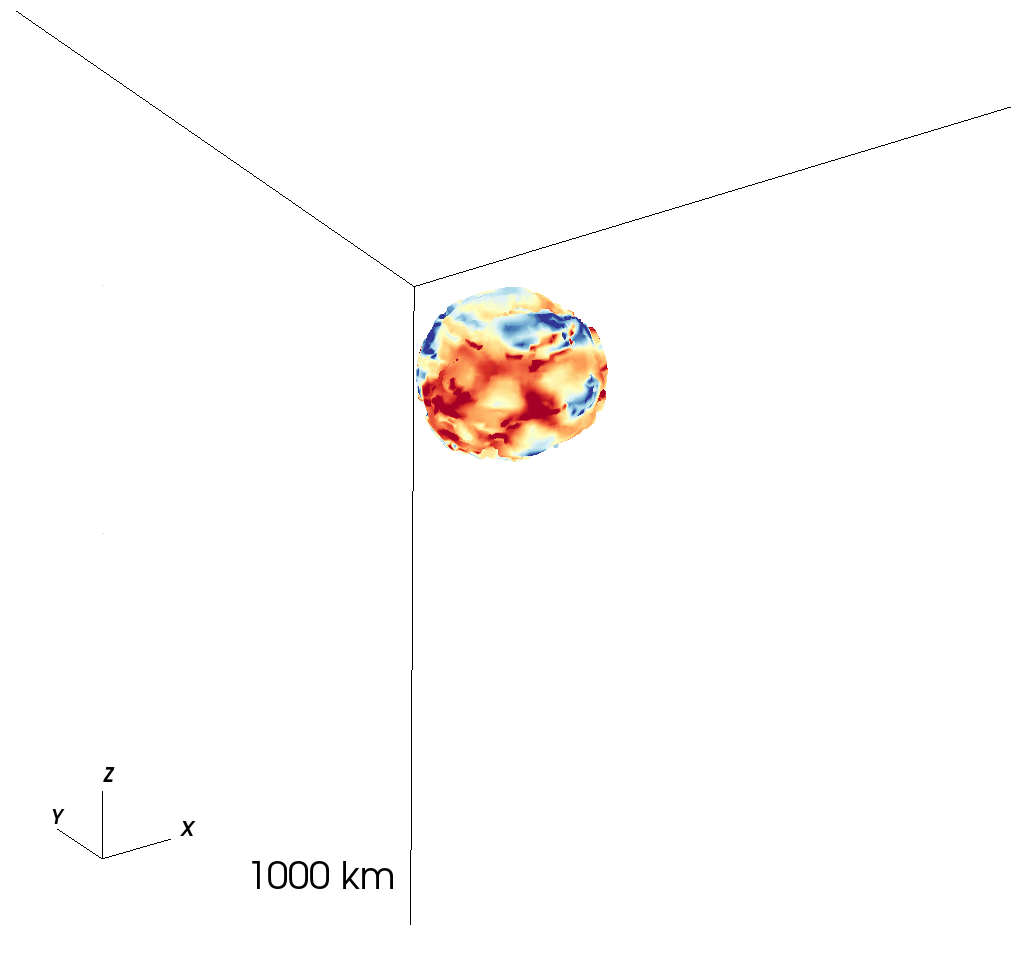}
          \includegraphics[width=0.23\linewidth]{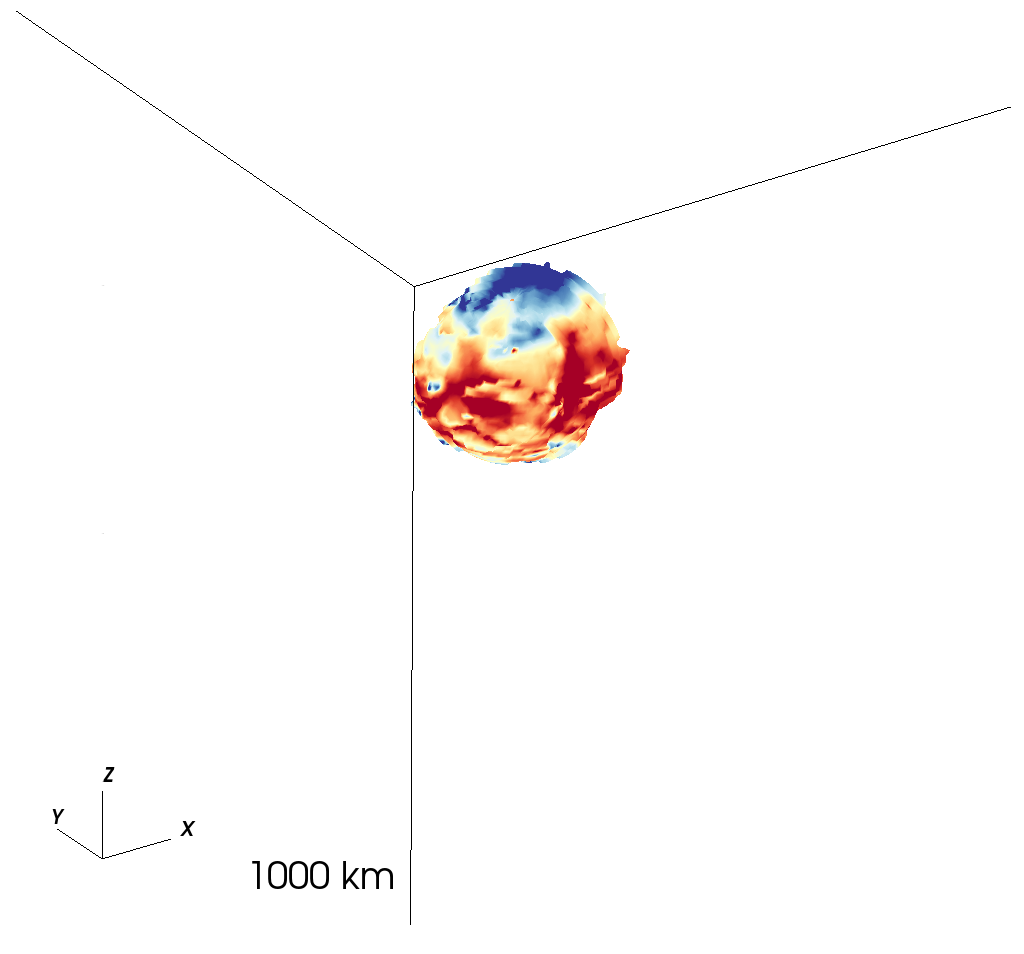}
          \includegraphics[width=0.23\linewidth]{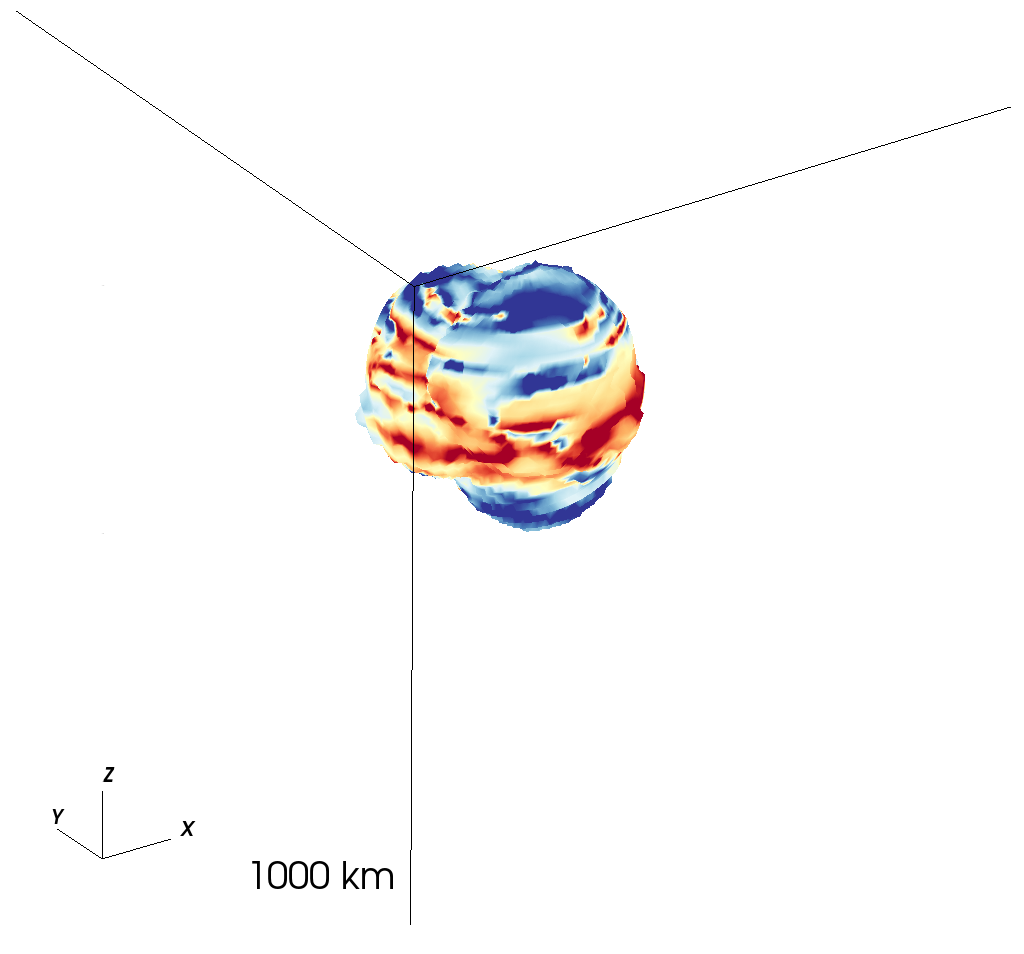}
          \includegraphics[width=0.23\linewidth]{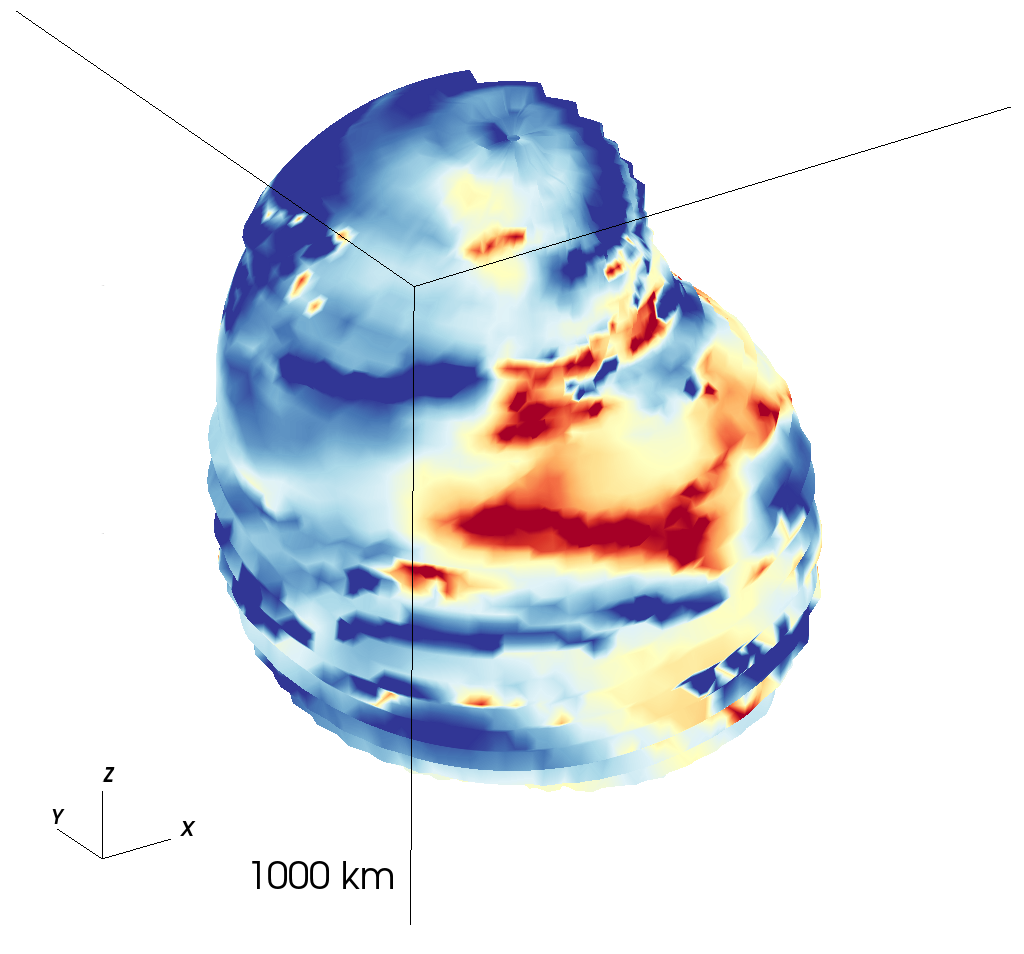}
        }
        \hbox{
          \includegraphics[width=0.05\linewidth]{./W-espec-colbar.png}
          \includegraphics[width=0.23\linewidth]{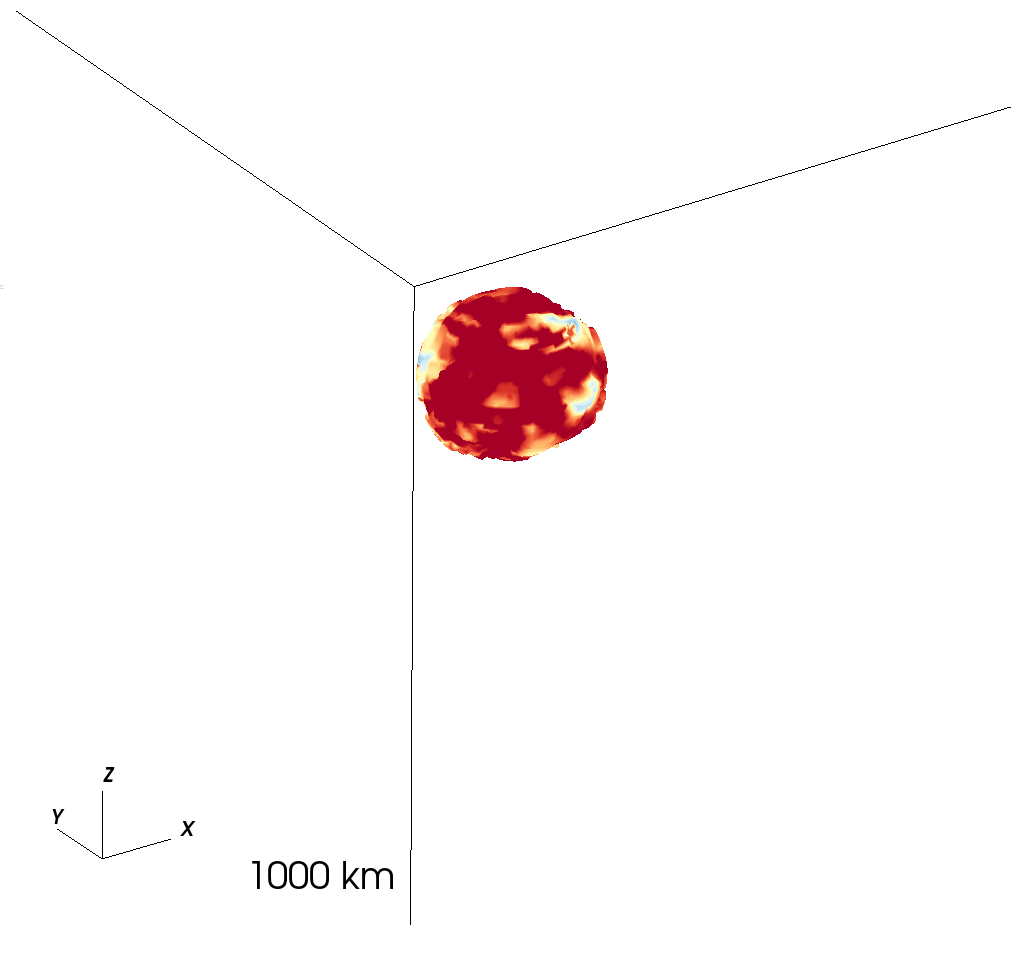}
          \includegraphics[width=0.23\linewidth]{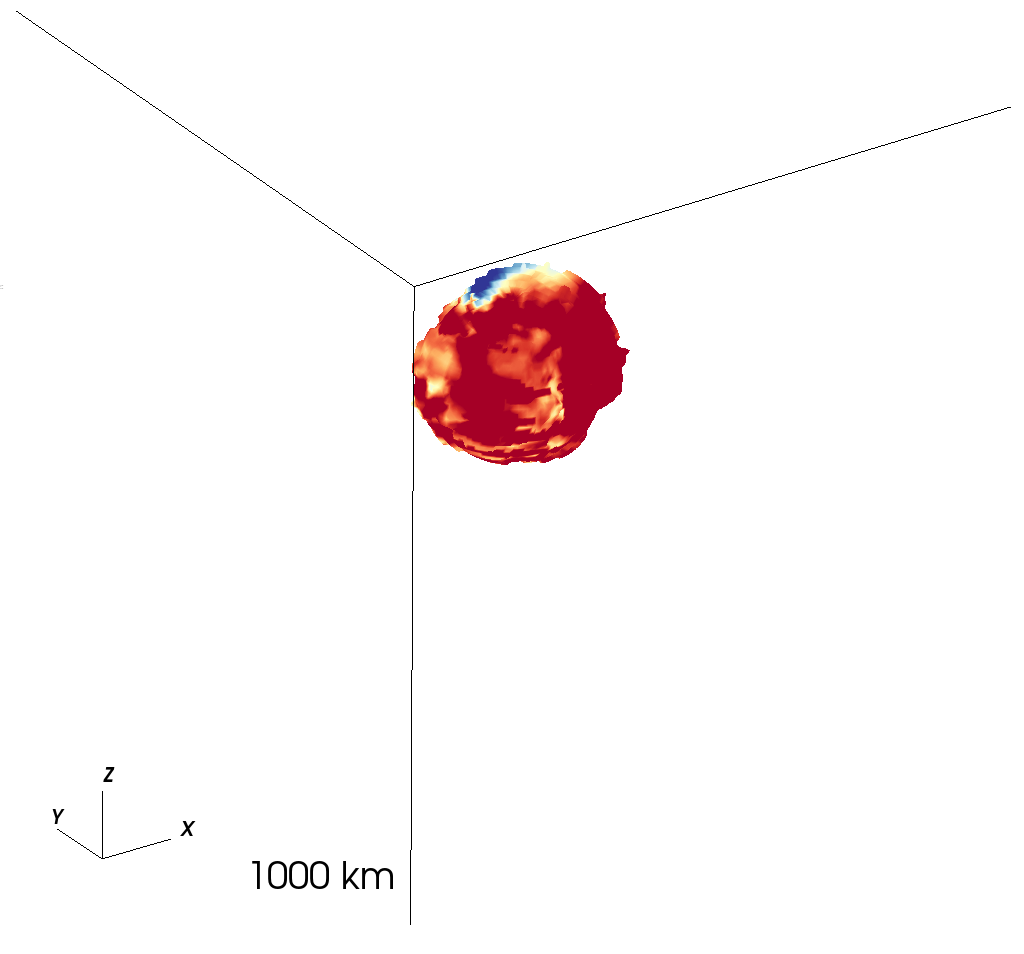}
          \includegraphics[width=0.23\linewidth]{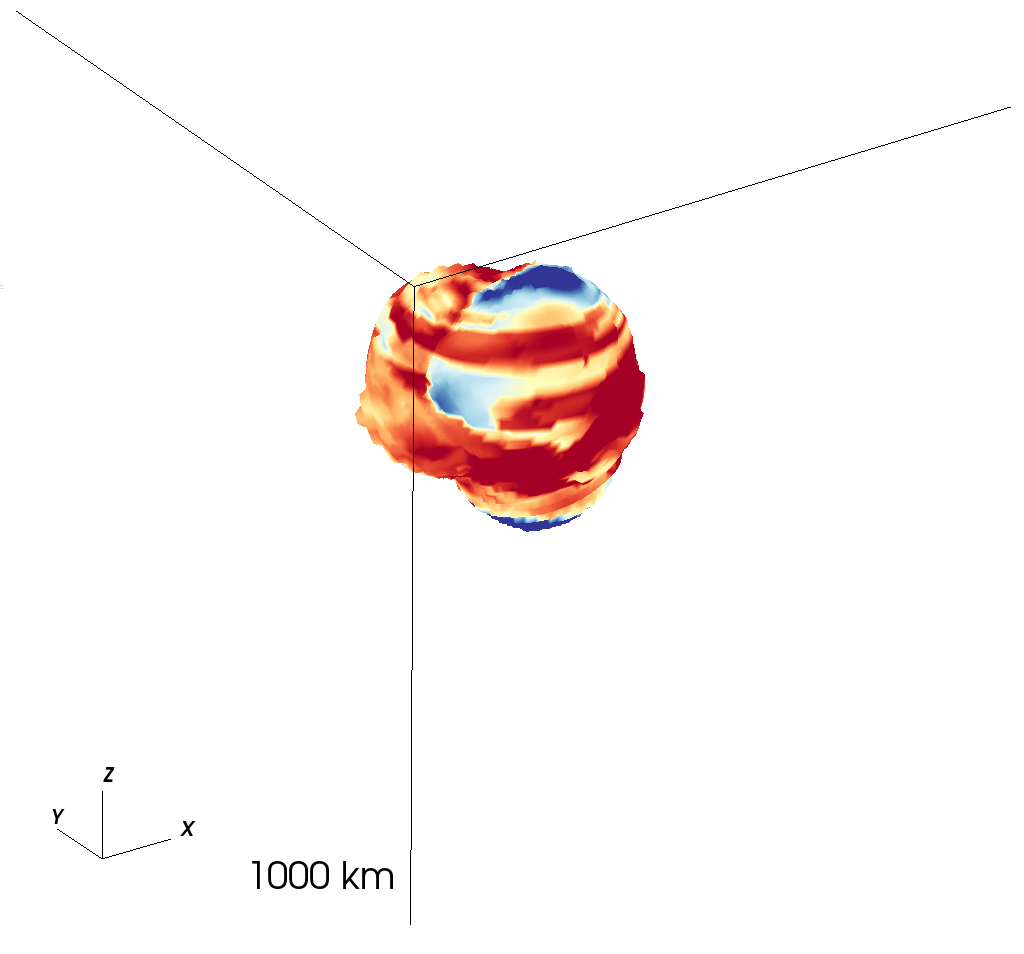}
          \includegraphics[width=0.23\linewidth]{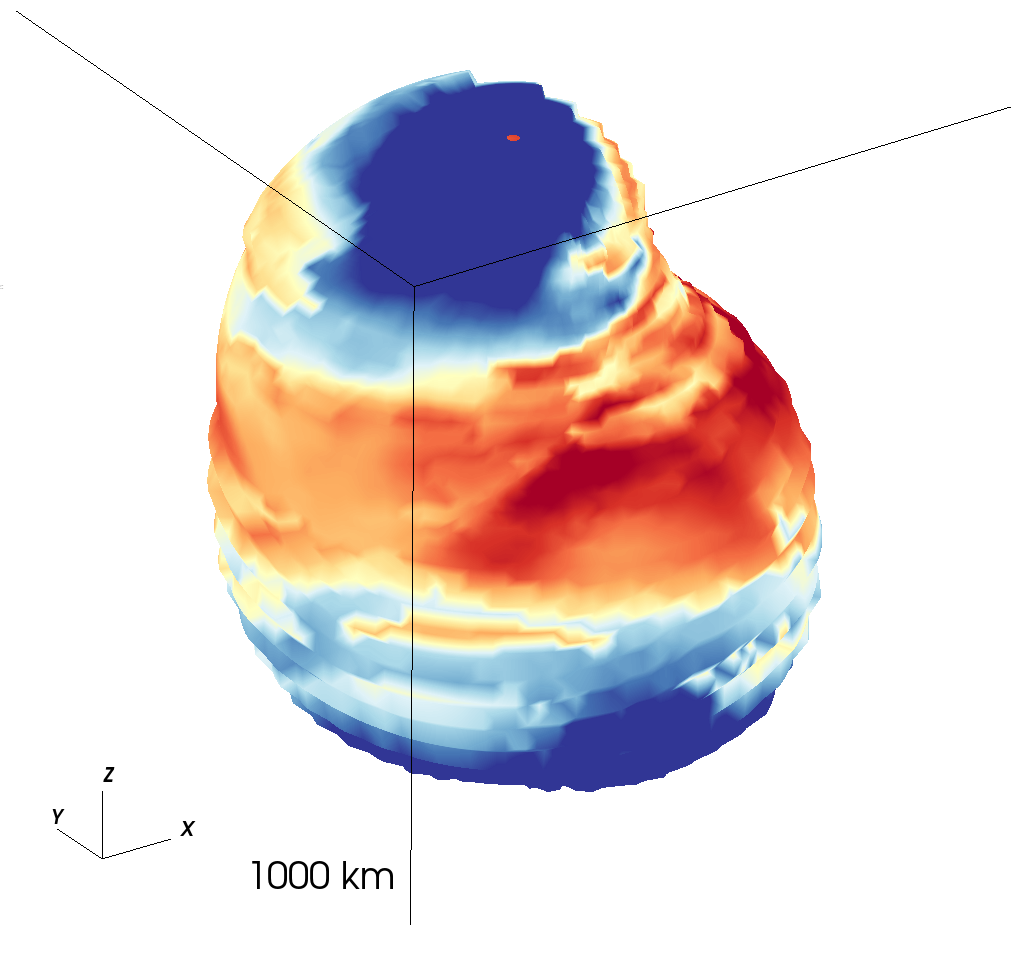}
        }
        \hbox{
          \includegraphics[width=0.05\linewidth]{./W-taurat-colbar.png}
          \includegraphics[width=0.23\linewidth]{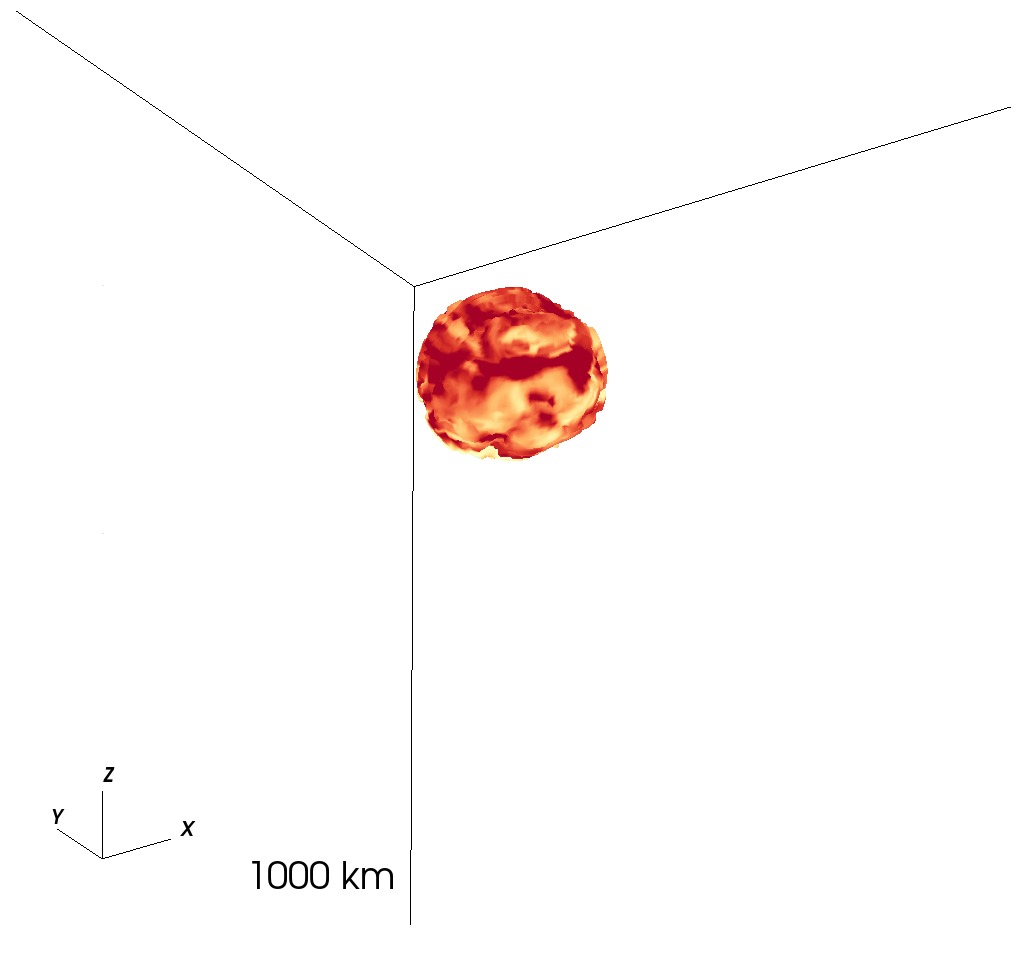}
          \includegraphics[width=0.23\linewidth]{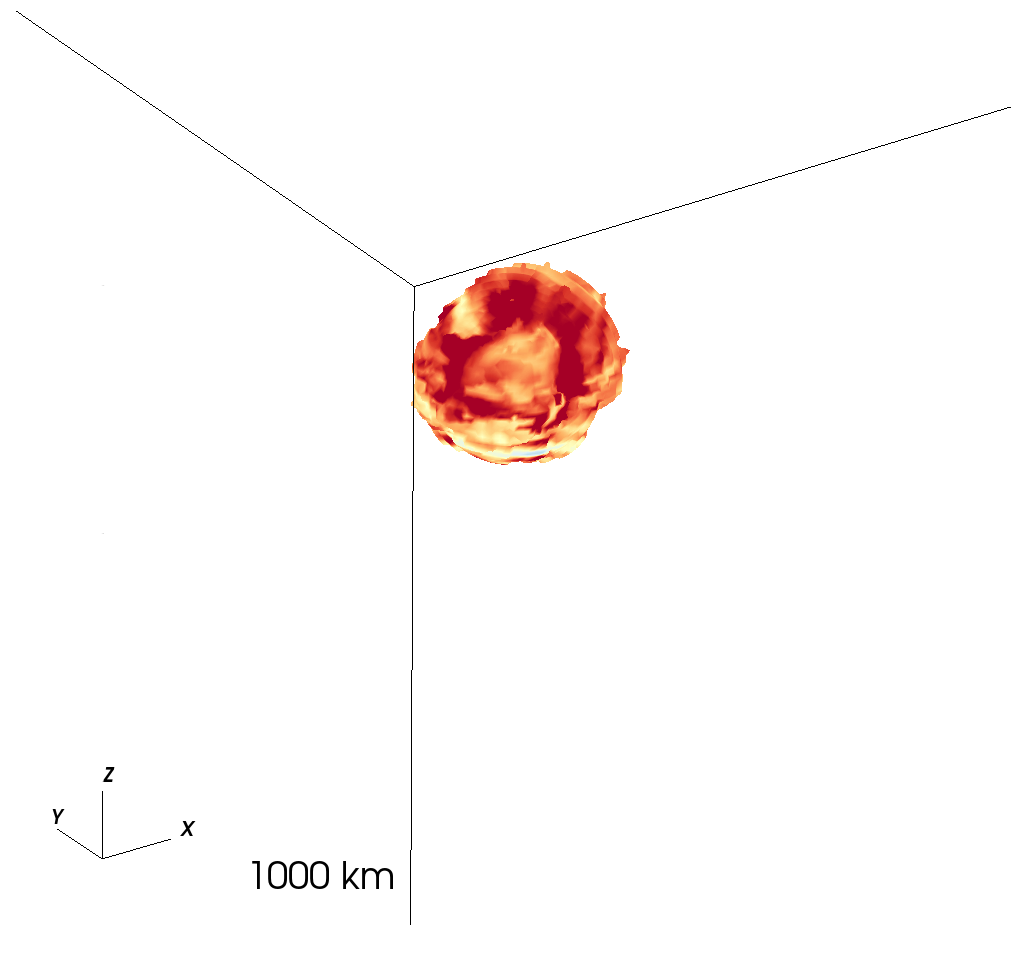}
          \includegraphics[width=0.23\linewidth]{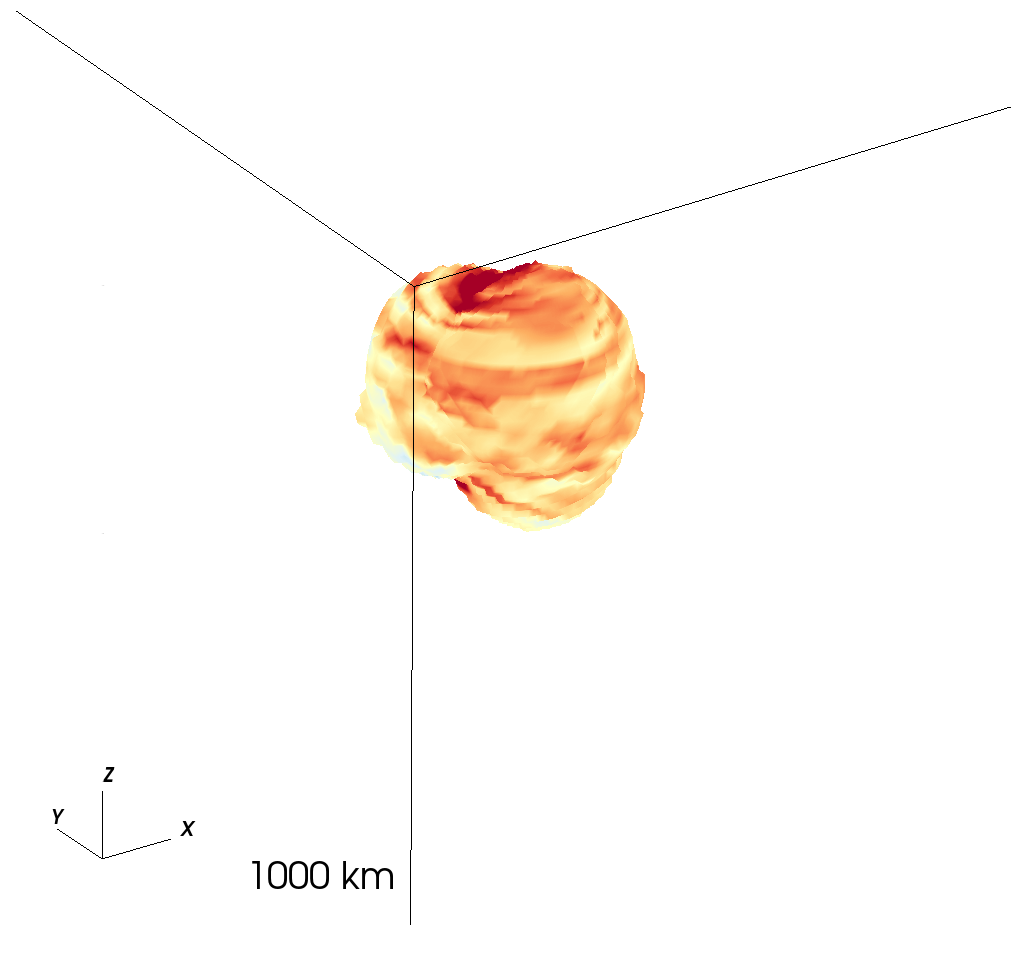}
          \includegraphics[width=0.23\linewidth]{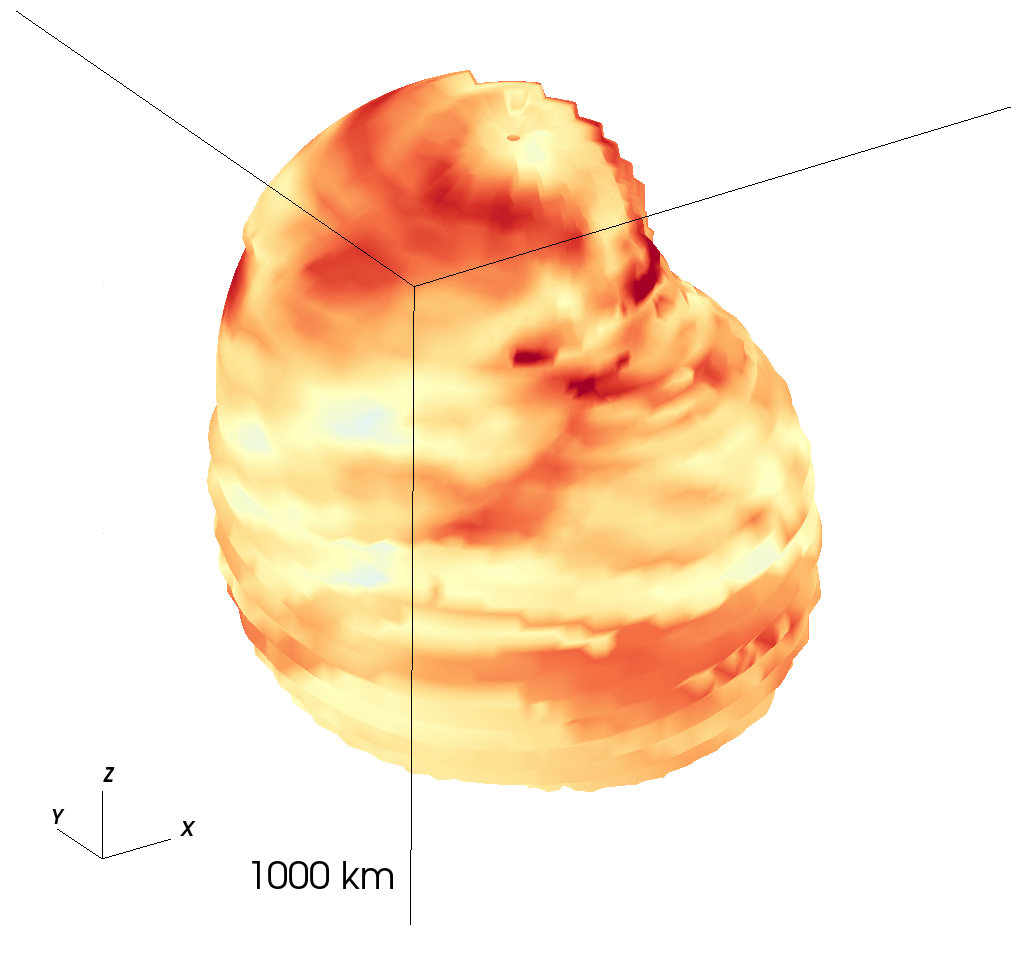}
        }
      }
    }
    \node[fill=white, opacity=0, text opacity=1] at (-8.2,+5.3) {$\tautau$};
    \node[fill=white, opacity=1, text opacity=1] at (-5.9,+2.2) {(a)};
    \node[fill=white, opacity=1, text opacity=1] at (-1.7,+2.2) {(b)};
    \node[fill=white, opacity=1, text opacity=1] at (2.4,+2.2) {(c)};	
    \node[fill=white, opacity=1, text opacity=1] at (6.6,+2.2) {(d)};
    \node[fill=white, opacity=0, text opacity=1] at (-8.2,+1.3) {$e [c^2]$};
    \node[fill=white, opacity=0, text opacity=1] at (-5.9,-1.6) {(e)};
    \node[fill=white, opacity=1, text opacity=1] at (-1.7,-1.6) {(f)};
    \node[fill=white, opacity=0, text opacity=1] at (2.4,-1.6) {(g)};	
    \node[fill=white, opacity=1, text opacity=1] at (6.6,-1.6) {(h)};
    \node[fill=white, opacity=0, text opacity=1] at (-8.2,-2.4) {$\tautaum$};
    \node[fill=white, opacity=0, text opacity=1] at (-5.9,-5.45) {(i)};
    \node[fill=white, opacity=1, text opacity=1] at (-1.7,-5.45) {(j)};
    \node[fill=white, opacity=0, text opacity=1] at (2.4,-5.45) {(k)};	
    \node[fill=white, opacity=1, text opacity=1] at (6.6,-5.45) {(l)};
  \end{tikzpicture}  
  \caption{
    Shock surface of \mRO at, from left to right, $\tpb = 0.24, 0.29,
    0.34, 0.41 \, \sek$.  The \banel{top} and \banel{middle} rows
    show the same quantities as in \figref{Fig:Rw-3dtauratio}, \ie,
    the advection-to-heating timescale ratio and the specific total
    energy, respectively.  The \banel{bottom} row adds the ratio between
    the advection and the \Alfven timescales, $\tautaum$. 
  }
  \label{Fig:RO-3dtauratio}
\end{figure*}

Though happening at a similar time as in \mRw (cf.~the evolution of
the shock radii in \figref{Fig:35OC-RO-rad}\panel{a}), the shock
revival in \mRO is the result of a rather different evolution and
explosion mechanism.  The shape of the shock wave before the explosion
(times $\tpb = 0.24 \, \sek, 0.29 \, \sek$, and $\tpb = 0.34 \, \sek$
in \figref{Fig:RO-3dtauratio}) does not exhibit a similarly strong
$m = 1$ modification as in the case of the weakest initial field.
While not completely suppressed, the spiral mode grow to about half
the amplitude as in \mRw, measured in terms of the normalized dipole
coefficients $\ashlm{1}{\pm 1}$ (\figref{Fig:35OC-RO-rad}\panel{c});
note also that these coefficients show quasi-periodic oscillations
similarly to \mRw, albeit with a lower frequency.  As a consequence,
the shock is less oblate and the quadrupole coefficient grows to
$\ashlm{2}{0} \approx 0.08 \, \ashlm{0}{0}$.

Most importantly, neutrino heating does not, by itself, achieve conditions
favourable for the explosion.  The heating timescale usually is longer
than the advection timescale at low latitudes (\banel{top} panels of
\figref{Fig:RO-3dtauratio}) and these regions do not reach positive
total energies in the gain layer (\banel{middle} panels).  The magnetic
field, on the other hand, is dynamically relevant.  Although 
averaged over the gain layer, its energy amounts to only about $5 \,
\%$ of the kinetic energy, it is strong enough to cause the partial
suppression of the $m = 1$ mode.  Near the polar axis, the field is
sufficiently strong for \Alfven waves to propagate faster through the
gain layer than gas is advected towards the PNS as we show in the
blue regions of the shock wave in the \banel{bottom} panels.

The polar regions are, furthermore, subject to the most intense
neutrino radiation.  The PNS has an even more aspherical shape than
that of \mRw.  From \figref{Fig:35OC-RO-rad}\panel{b}, we can extract
a drop of the pole-to-equator axis to about $1:2$ and a growth of the
quadrupole coefficient to
$\aPNSlm{2}{0} \approx - 0.14 \, \aPNSlm{0}{0}$ until
$\tpb = 0.35 \, \sek$.  The larger equatorial radius reduces the the
part of the neutrino luminosity powered by the accretion rather than
thermal cooling of the PNS emitted into low latitudes as well as the
total luminosity.  Close to the pole, on the other hand, the neutrino
fluxes are as high as in \mRw.  Consequently, a favourable ratio of
both the heating and the \Alfven timescales to the advection
timescales develop there (\figref{Fig:RO-3dtauratio} \banel{top} and
\banel{bottom} panels).  These conditions stabilize the shock at the
pole at radii $\rsh \approx 200 \, \km$ in contrast to the gradual
receding it undergoes in \mRw.  Ultimately, they combine to launch a
polar, rather than equatorial, runaway of the shock wave.  The
explosion starts with a peanut-shaped shock wave that is at first
expanding faster into the southern hemisphere ($\ashlm{1}{0} < 0$),
but soon thereafter evolves into a largely symmetric pair of prolate
outflows with the quadrupole coefficient approaching
$\ashlm{2}{0} \approx 0.1 \, \ashlm{0}{0}$.

\begin{figure*}
  \centering
  \begin{tikzpicture}
    \pgftext{\hbox{ 
        \includegraphics[width=0.245\linewidth]{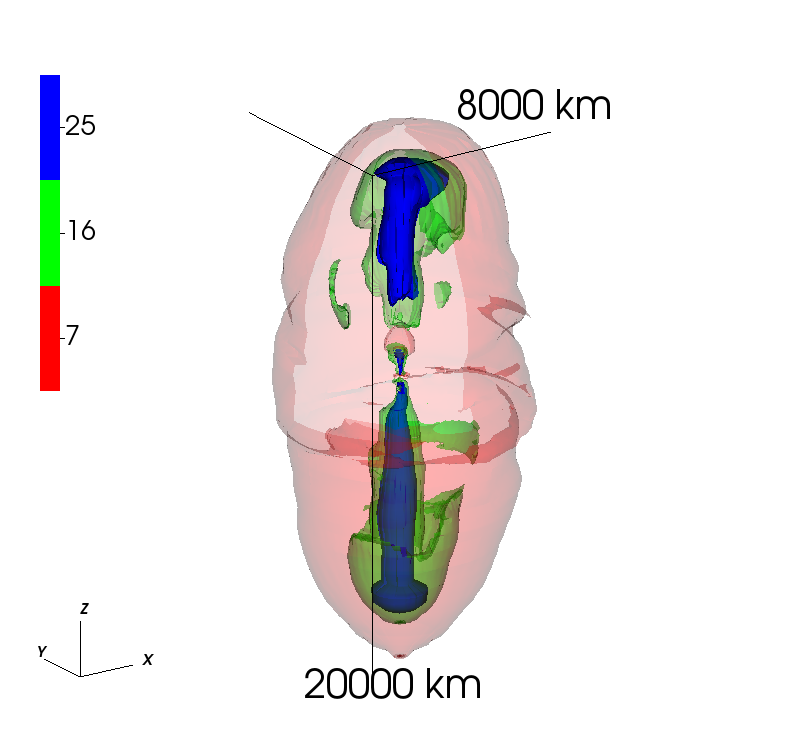}
        \includegraphics[width=0.245\linewidth]{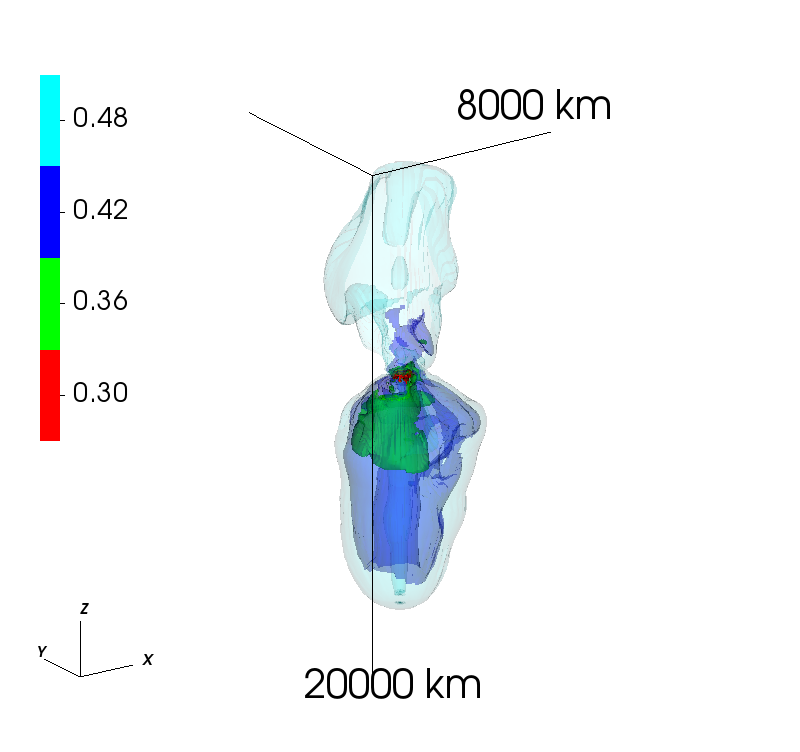}          
        \includegraphics[width=0.245\linewidth]{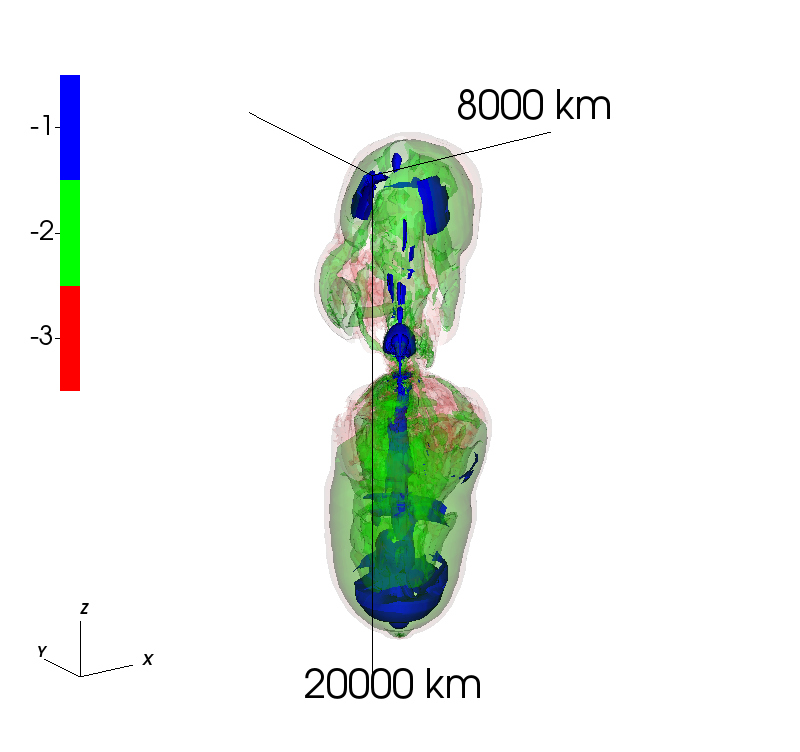}	
        \includegraphics[width=0.245\linewidth]{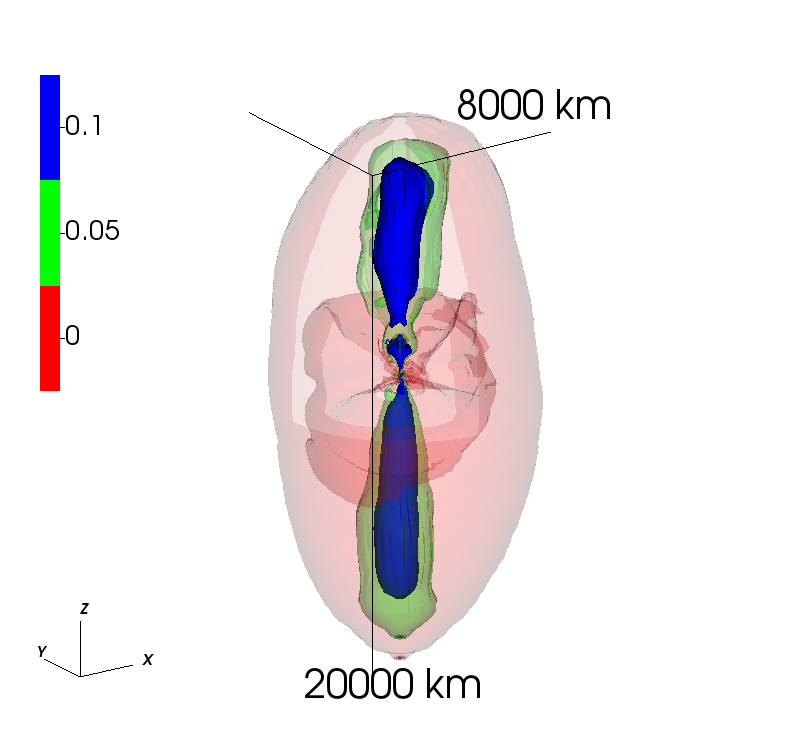}
      }
    }%
    \node[fill=white, opacity=0, text opacity=1] at (-8.75,-1.85) {(a)};
    \node[fill=white, opacity=0, text opacity=1] at (-4.30,-1.85) {(b)};
    \node[fill=white, opacity=0, text opacity=1] at (0.10,-1.85) {(c)};	
    \node[fill=white, opacity=0, text opacity=1] at (4.54,-1.85) {(d)};
    \node[fill=white, opacity=1, text opacity=1] at (-8.45,1.80) {\sffamily\scriptsize $s\,[k_{\textsc{b}}/\text{baryon}]$}; 
    \node[fill=white, opacity=1, text opacity=1] at (-4.07,1.80) {\sffamily\scriptsize $Y_{\rm e}$};	
    \node[fill=white, opacity=1, text opacity=1] at (0.40,1.85) {\sffamily\scriptsize $\log\beta^{-1}$};	
    \node[fill=white, opacity=1, text opacity=1] at (4.6,1.7) {\sffamily\scriptsize energy flux [$c$]};
    \node[fill=white, opacity=0, text opacity=1] at (4.6,1.9) {\sffamily\scriptsize specific};
    %
  \end{tikzpicture}
  \caption{
    Same as \figref{Fig:Rw-3dplots}, but for \mRO at $\tpb = 0.8 \, \sek$.
  }
  \label{Fig:RO-3dplots}
\end{figure*}

We present the structure of the ejecta of \mRO at a late stage in the
simulation ($\tpb \approx 0.8 \, \sek$) in \figref{Fig:RO-3dplots}.
Their bipolar morphology offers a stark contrast to \mRw
(\figref{Fig:Rw-3dplots}).  The shock wave (red surface in
\figref{Fig:RO-3dplots}\panel{a} showing the entropy) possesses a
mostly axisymmetric shape with an equatorial radius of
$R_{\mathrm{sh}}^{\mathrm{eqtl}} \approx \zehnh{4.5}{3} \, \km$ and a
polar elongation of
$R^{\mathrm{pol}}_{\mathrm{sh}} \approx \zehnh{8.9}{3} \, \km$.  In
the largest part of the volume of the post-shock region, the gas has
only a small positive or negative radial velocity.  The PNS is fed by
an oblate accretion flow at low latitudes with peak velocities
exceeding $\zehn{9} \, \cms$.  It extends to a distance of about 2000
km.  In the innermost 1000 km, the downflow deviates strongly from
axisymmetry.  In this region, matter falls onto the PNS in the form of
a spiral mode that coexists with pockets of outwards moving matter.
The downdrafts decelerate considerably at $r \approx 300 \, \km$,
where the local rotational energy reaches half of the local
gravitational energy and, hence, matter is nearly entirely supported
by centrifugal forces.  The gas with positive radial velocity in the
equatorial region does not manage to overcome the ram pressure of the
downflows.  Instead, the core ejects gas in fast
($v^r \gtrsim \zehnh{4}{9} \, \cms$, blue surfaces in
\figref{Fig:RO-3dplots}\panel{a}) bipolar outflows along the
rotational axis.  The magnetisation of the jets is moderate, reaching
$\ateb \gtrsim 0.1$ along its beam and $\ateb \approx 0.01$ in the
cocoon (blue and green surfaces in in
\figref{Fig:RO-3dplots}\panel{c}), while the magnetic field is
insignificant in the downflows.

These jets maintain their stability and coherence over a distance of
several 1000 km.  Within the first 1000 km of their propagation, they
expand laterally to a diameter of 500 km.  Thereafter, the lateral
spreading slows down and the outflows resemble collimated jets.
Though non-axisymmetric modes modify their geometry, they do not
suffice to quench the outflow in a similar manner as observed by
\cite{Mosta_et_al__2014__apjl__MagnetorotationalCore-collapseSupernovaeinThreeDimensions}.
Since this stability is a common feature of all our magneto-rotational
outflows, we defer a discussion to a separate section comparing models
\RO, \Rp, and \Rs.

The outflows are more neutron-rich than in \mRw
(\figref{Fig:RO-3dplots}\panel{b}).  The beam of the jet contains
relatively hot, almost symmetric ($\Ye \approx 0.5$) matter and is
surrounded by cooler gas with an electron fraction as low as
$\Ye \approx 0.32$.  The composition of the outflows shows a higher
north-south asymmetry than the velocity.  The southern outflow
contains a large shroud of material with $\Ye < 0.4$ enclosing the
beam (dark blue surface) as well as a cloud with $\Ye < 0.35$ (green)
expanding non-axisymmetrically at the edge of the ejecta.  In their
high electron fraction, our jets differ from previous models
\citep[e.g.,][]{Winteler_et_al__2012__apjl__MagnetorotationallyDrivenSupernovaeastheOriginofEarlyGalaxyr-processElements,Moesta_et_al__2018__apj__r-processNucleosynthesisfromThree-dimensionalMagnetorotationalCore-collapseSupernovae,Halevi_Moesta__2018__mnras__r-Processnucleosynthesisfromthree-dimensionaljet-drivencore-collapsesupernovaewithmagneticmisalignments}.
This feature, seen in all magnetorotationally driven jets, will be
discussed in the next subsection.

\begin{figure}
  \centering
  \begin{tikzpicture}
    \pgftext{
      \includegraphics[width=0.99\linewidth]{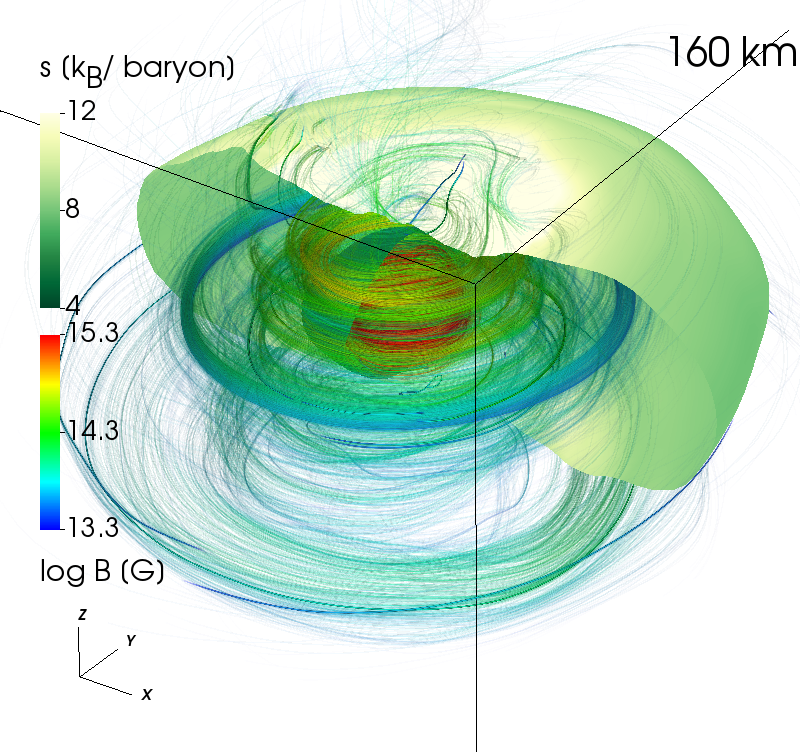}
    }
  \end{tikzpicture}
  \caption{
    Same as \figref{Fig:Rw-PNS3d}, but for the \mRO at $\tpb \approx 0.8
    \, \sek$.
  }
  \label{Fig:RO-3dPNS}
\end{figure}

The PNS evolves in an even more oblate manner than in \mRw
(\figref{Fig:35OC-RO-rad}\panel{b} and \panel{d}).  The polar radii
(black lines, \figref{Fig:35OC-RO-rad}\panel{b}) continuously contract
to $R_{\mathrm{PNS}, \mathrm{pol}} \approx 23 \, \km$ at the end of
the simulation, while the equatorial radii saturates at
$R_{\mathrm{PNS}, \mathrm{eqtr}} \gtrsim 50 \, \km$ (blue lines).  The
high degree of flattening is reflected in the drift of the quadrupole
coefficient to $\aPNSlm{2}{0} \approx -0.12 \, \aPNSlm{0}{0}$.  We do
not observe a gyration of the PNS axis similar to the case of \mRw.
As a consequence, the magnetic field retains a more ordered geometry.
Inside of and around the PNS, it mainly consists of an $m = 1$ mode
that spirals out from its centre at $r = 0$ through the entire
equatorial plane of the PNS into the aforementioned partially
centrifugally supported region up to $r \approx 300 \, \km$
(\figref{Fig:RO-3dPNS} where we represent the inner $\sim 160\,$km).
Inside this flux tube, $\ateb$ can reach values of
$\ateb \gtrsim 0.1$.  This structure can account for a
significant modification of the rotational profile.  Around the PNS
surface, its Maxwell stress component
$\mathcal{M}^{r \phi} = b^r b^{\phi}$ corresponds to a relative local
rate of change of the angular momentum of
$\tau_{J}^{-1} := J^{-1}\partial_t J \sim J^{-1} r^{-2} \partial_{r} (
\varpi - r^2 \mathcal{M}^{r \phi} ) \gtrsim 10 \, \sek^{-1}$.  This
rate decreases by one order of magnitude towards the centre of the
PNS, but remains comparable to the secular timescales of the evolution
of the PNS.  Similarly to many of our axisymmetric models, the outward
transport of angular momentum increases the centrifugal support in the
outer layers of the PNS
\citepalias{Aloy_Obergaulinger_2020__mnras_PaperII}.  Along the
rotational axis, the field lines form a helical structure that
connects the PNS with the polar jets.

Similarly to \mRw, albeit at a smaller amplitude, the strong
variations of $\EmPNS$ after $t \approx 0.4 \, \sek$ occur in
coincidence with deviations of $\aPNSlm{1}{0}$ from zero.  As
$\aPNSlm{1}{0}$ rises relatively quickly, $\EmPNS$ assumes an
intermediate maximum.  Later ($t \gtrsim 0.45 \, \sek$), the magnetic
energy decreases in parallel to the decrease of $\aPNSlm{1}{0}$.
Though this parallelism suggests an explanation of the field growths
in terms of PNS deformations, it is more difficult to ascribe the
field evolution to a single dynamical mode than in the case of the
tilting PNS of \mRw.

\subsection{Models \Rp and \Rs: MHD explosions}
\label{sSek:Rps}

\begin{figure*}
  \centering
  \begin{tikzpicture}
    \pgftext{
      \hbox{ 
        \includegraphics[width=0.49\linewidth]{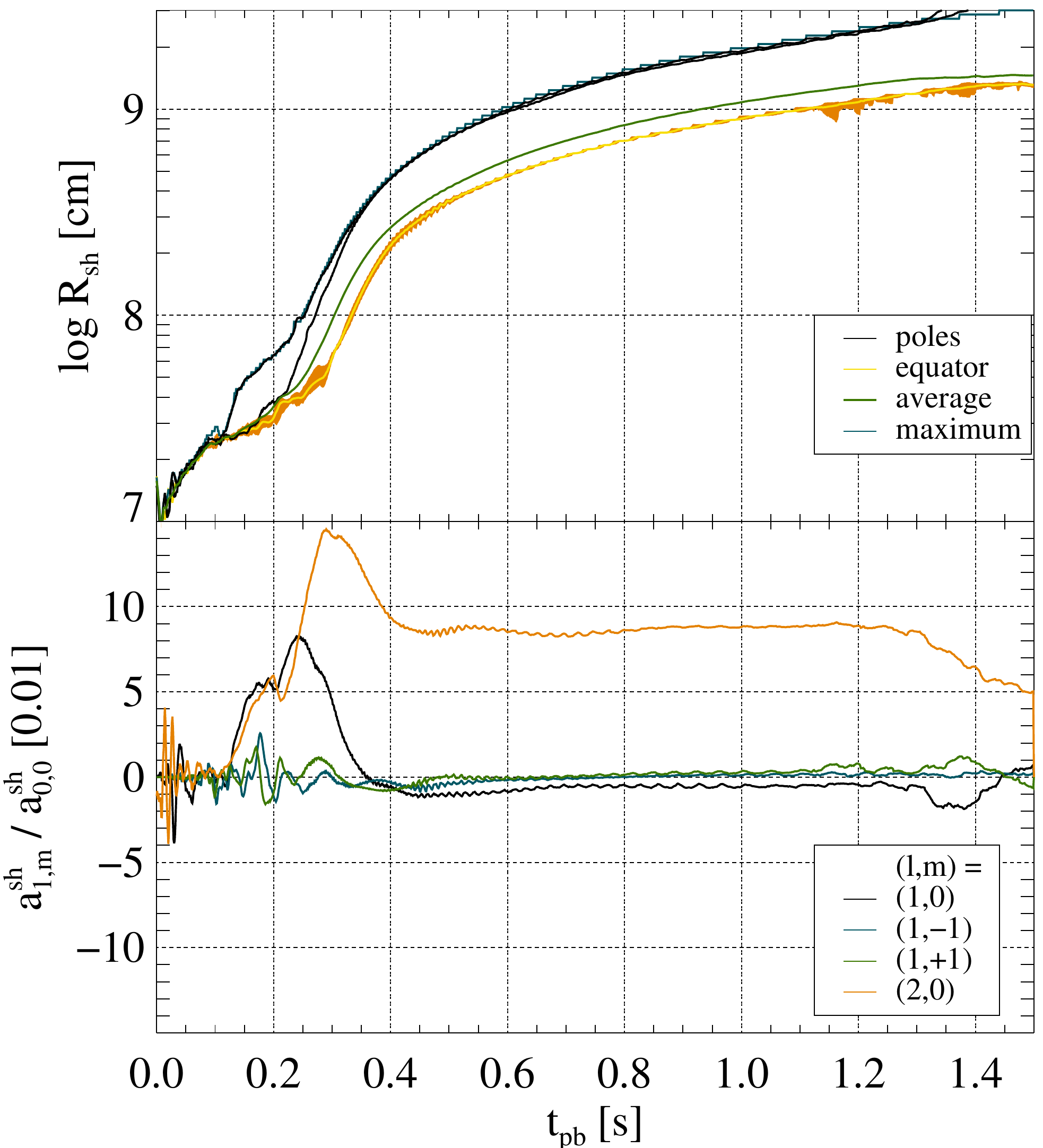}
        \includegraphics[width=0.49\linewidth]{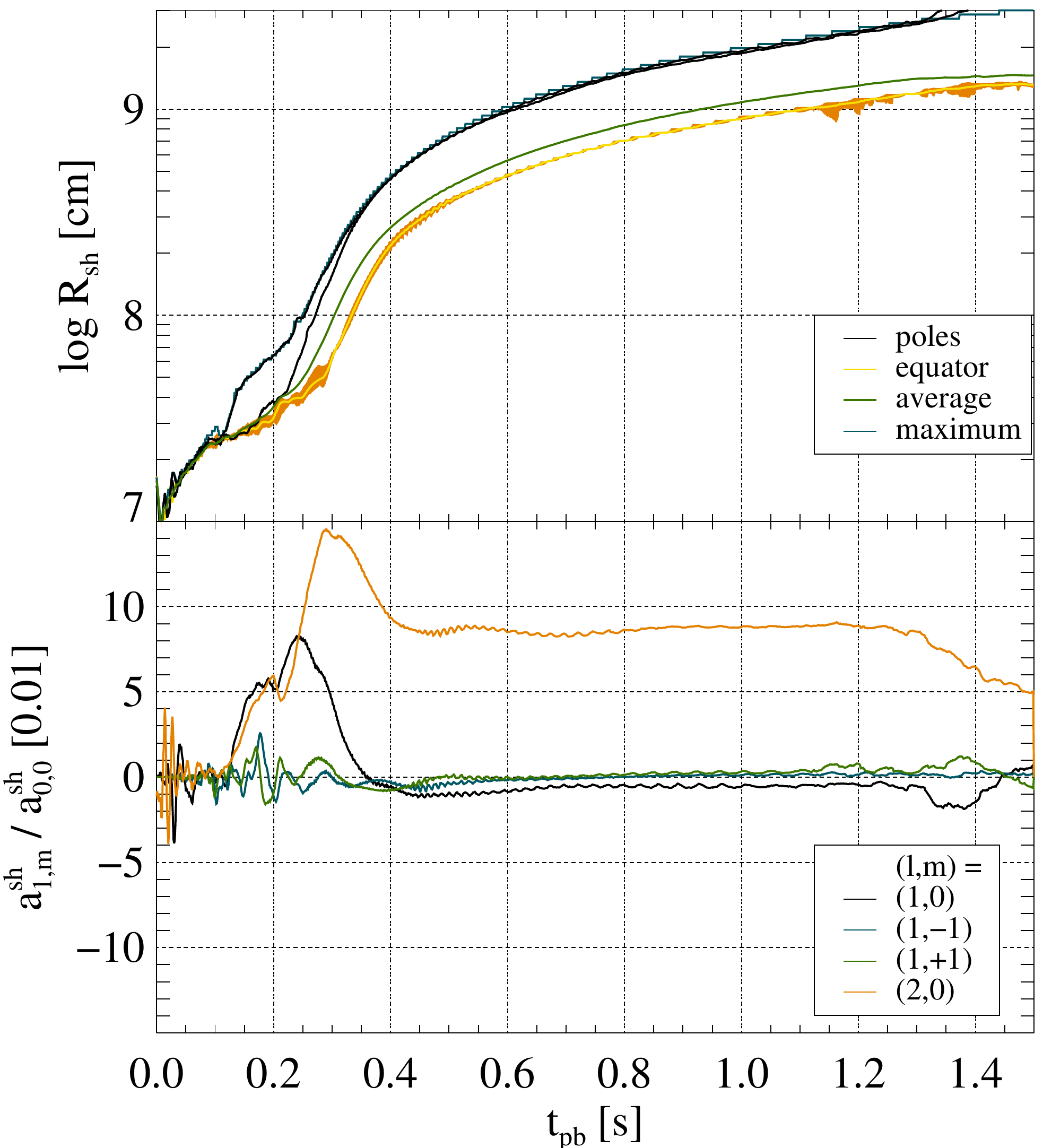}
      }
    }%
    \node[fill=white, opacity=1, text opacity=1] at (-8.25,+4.49) {\large (a)};
    \node[fill=white, opacity=1, text opacity=1] at (0.54,+4.49) {\large (b)};
    \node[fill=white, opacity=0, text opacity=1] at (-8.25,-3.99) {\large (c)};
    \node[fill=white, opacity=0, text opacity=1] at (0.54,-3.99) {\large (d)};	
  \end{tikzpicture}
  \caption{
    Same as \figref{Fig:35OC-Rw-rad}, but for \mRp.
  }
  \label{Fig:35OC-Rp3-rad}
\end{figure*}

\begin{figure*}
  \centering
  \begin{tikzpicture}
    \pgftext{
      \hbox{ 
        \includegraphics[width=0.49\linewidth]{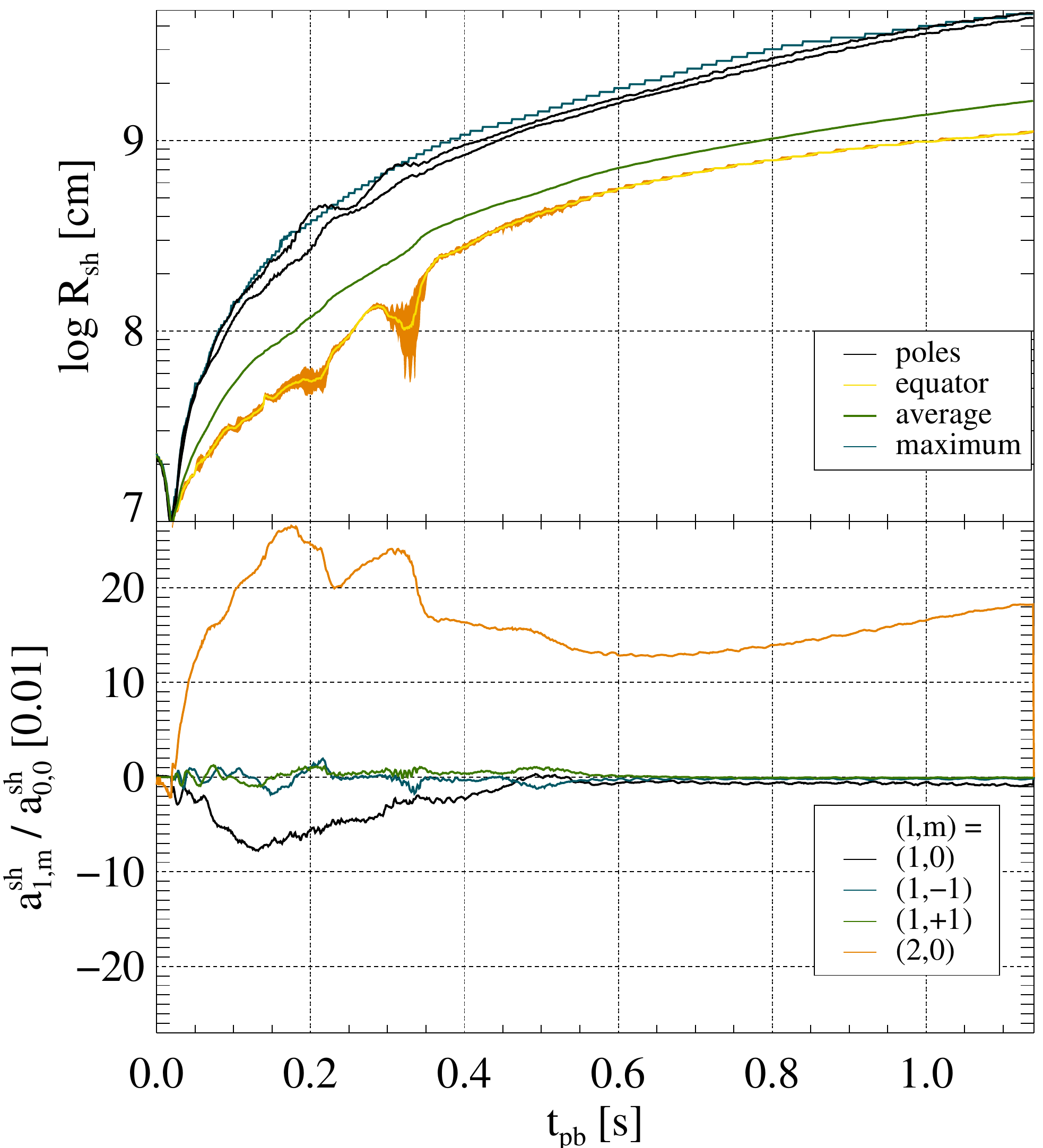}
        \includegraphics[width=0.49\linewidth]{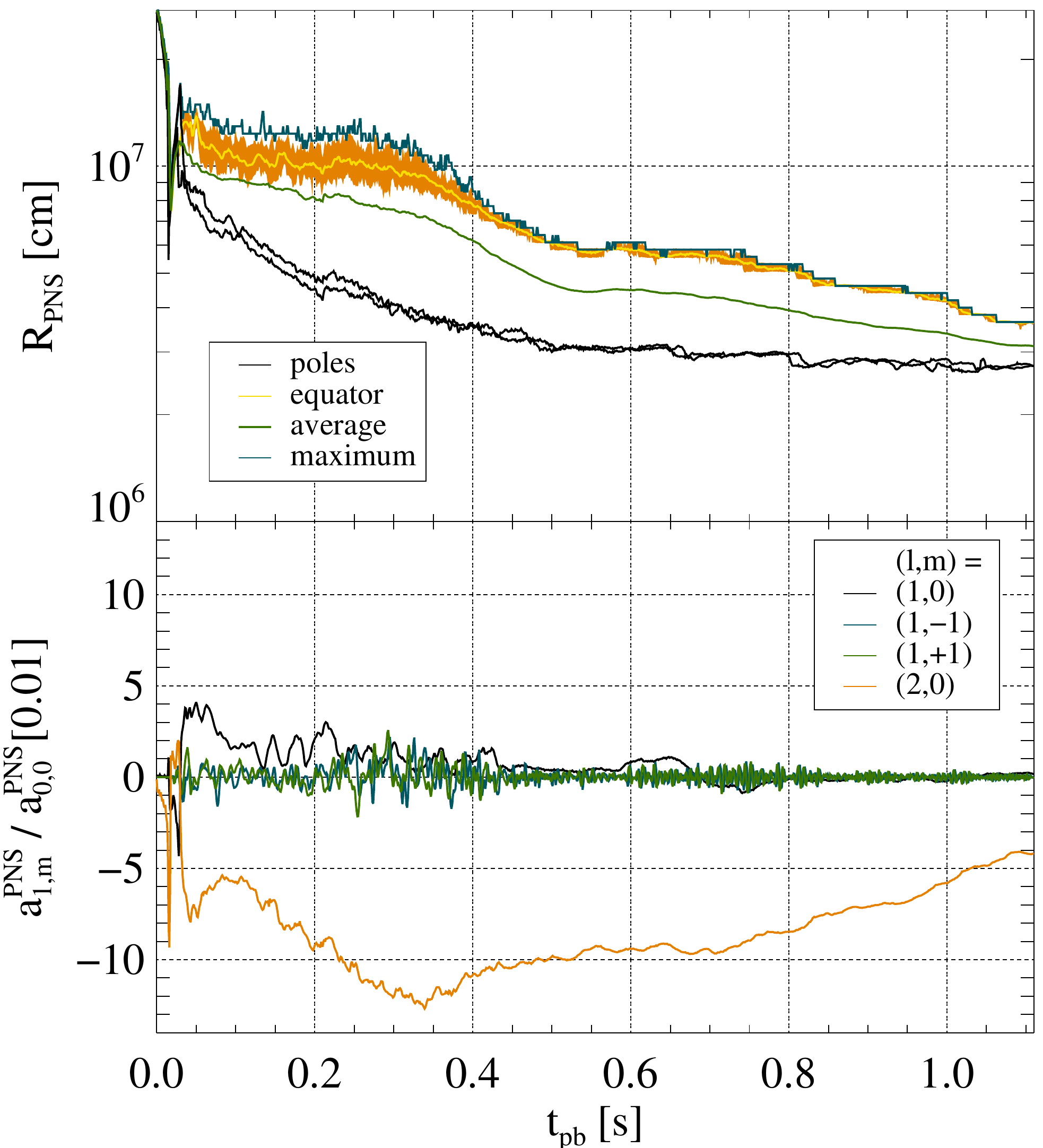}
      }
    }%
    \node[fill=white, opacity=1, text opacity=1] at (-8.25,+4.49) {\large (a)};
    \node[fill=white, opacity=1, text opacity=1] at (0.54,+4.49) {\large (b)};
    \node[fill=white, opacity=0, text opacity=1] at (-8.25,-3.99) {\large (c)};
    \node[fill=white, opacity=0, text opacity=1] at (0.54,-3.99) {\large (d)};	
  \end{tikzpicture}
  \caption{
    Same as \figref{Fig:35OC-Rw-rad}, but for \mRs.
  }
  \label{Fig:35OC-Rs-rad}
\end{figure*}

\begin{figure*}
  \centering
  \begin{tikzpicture}
    \pgftext{\vbox{
        \hbox{ 
          \includegraphics[width=0.05\linewidth]{./W-taurat-colbar.png}
          \includegraphics[width=0.23\linewidth]{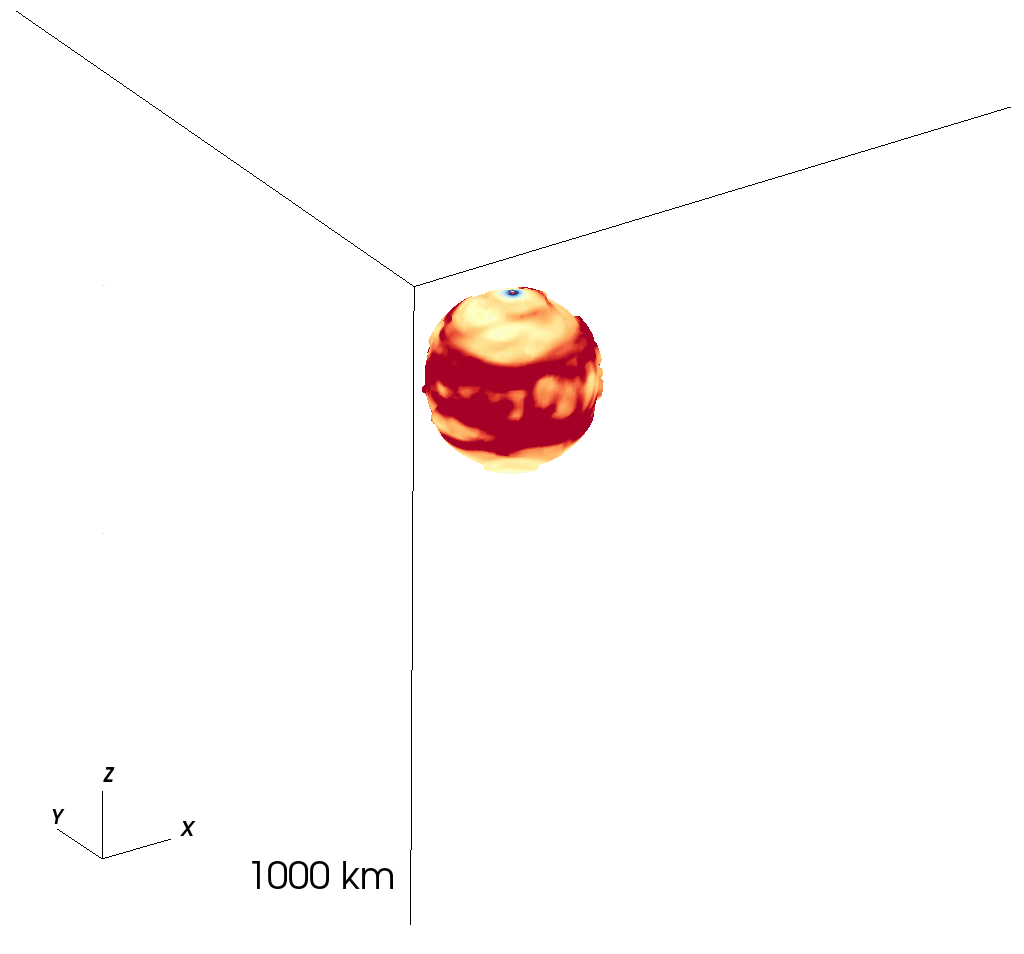}
          \includegraphics[width=0.23\linewidth]{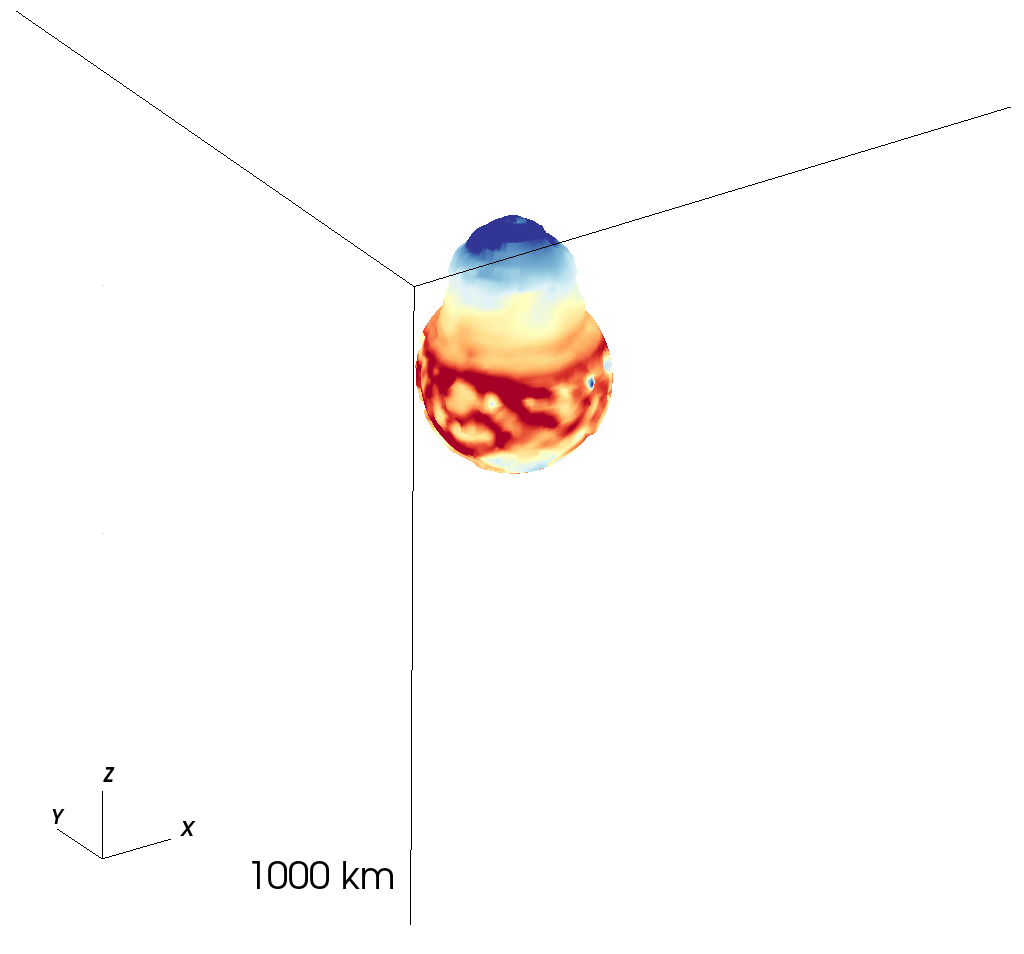}
          \includegraphics[width=0.23\linewidth]{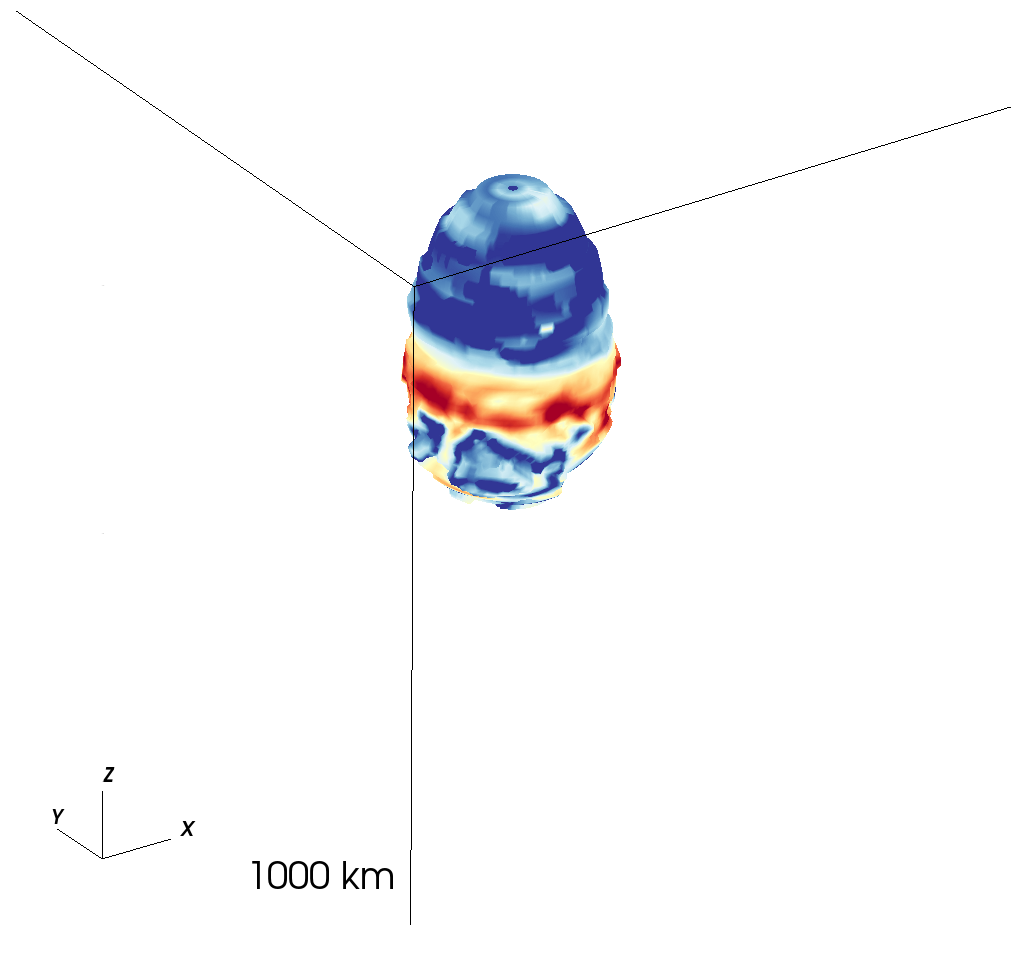}
          \includegraphics[width=0.23\linewidth]{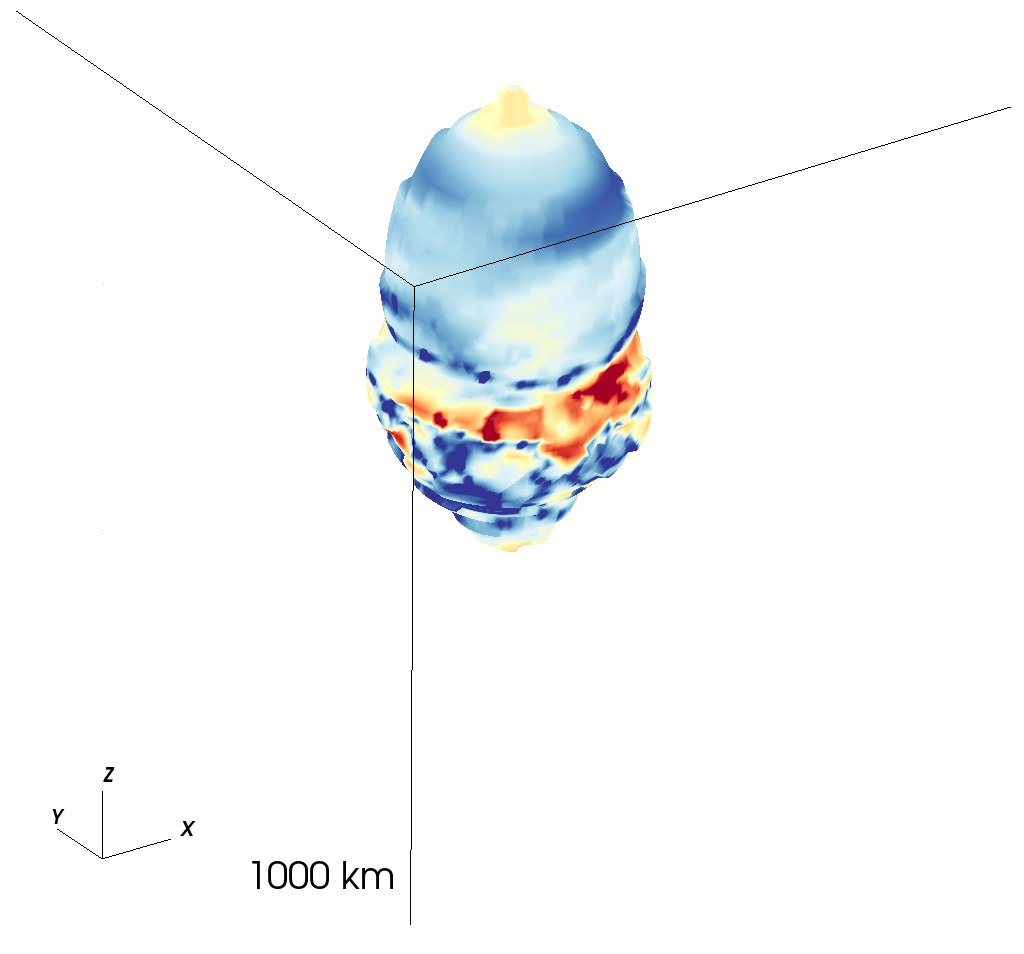}
        }
        \hbox{
          \includegraphics[width=0.05\linewidth]{./W-espec-colbar.png}
          \includegraphics[width=0.23\linewidth]{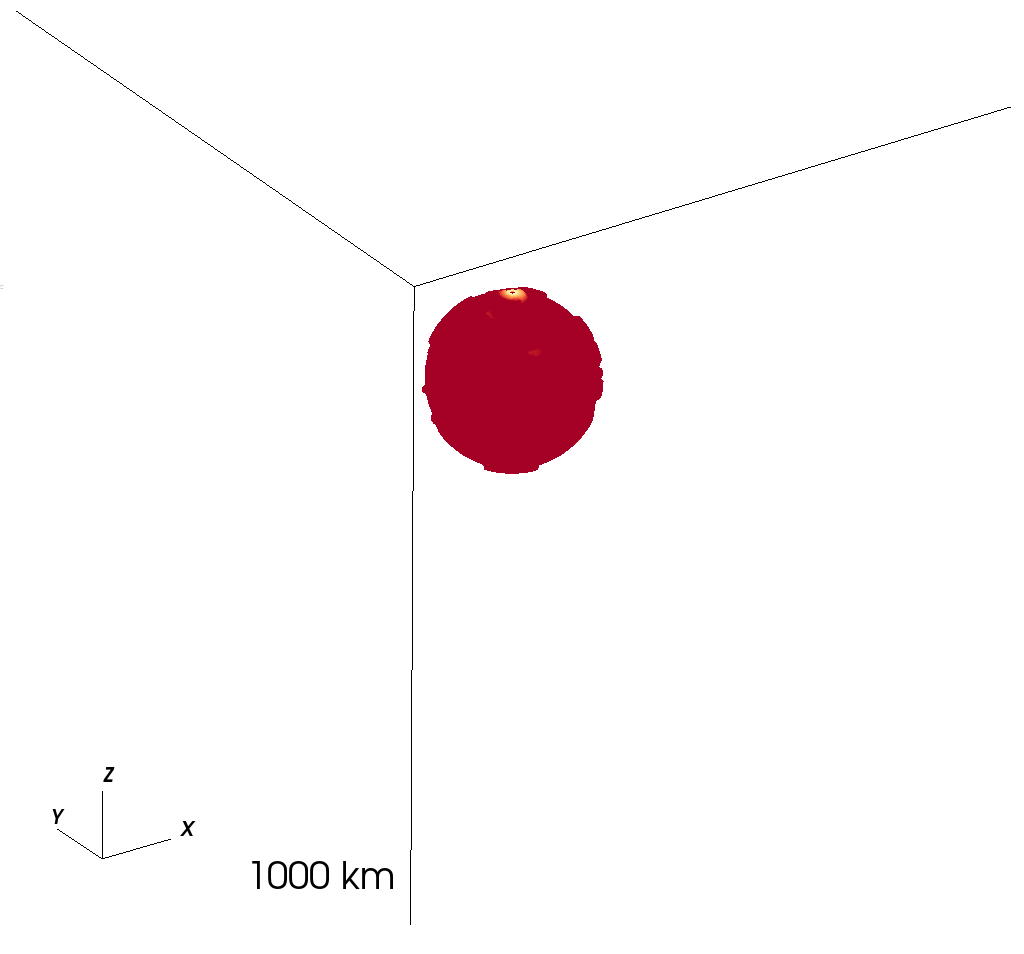}
          \includegraphics[width=0.23\linewidth]{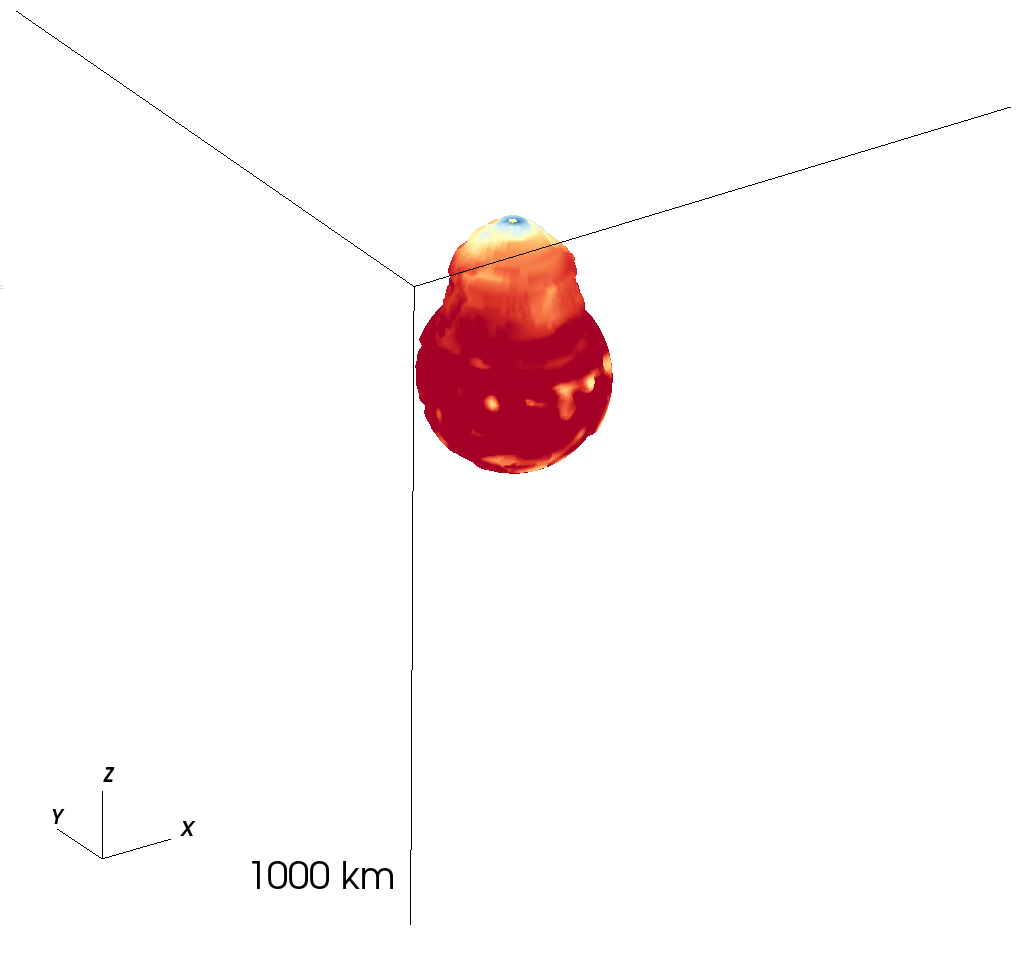}
          \includegraphics[width=0.23\linewidth]{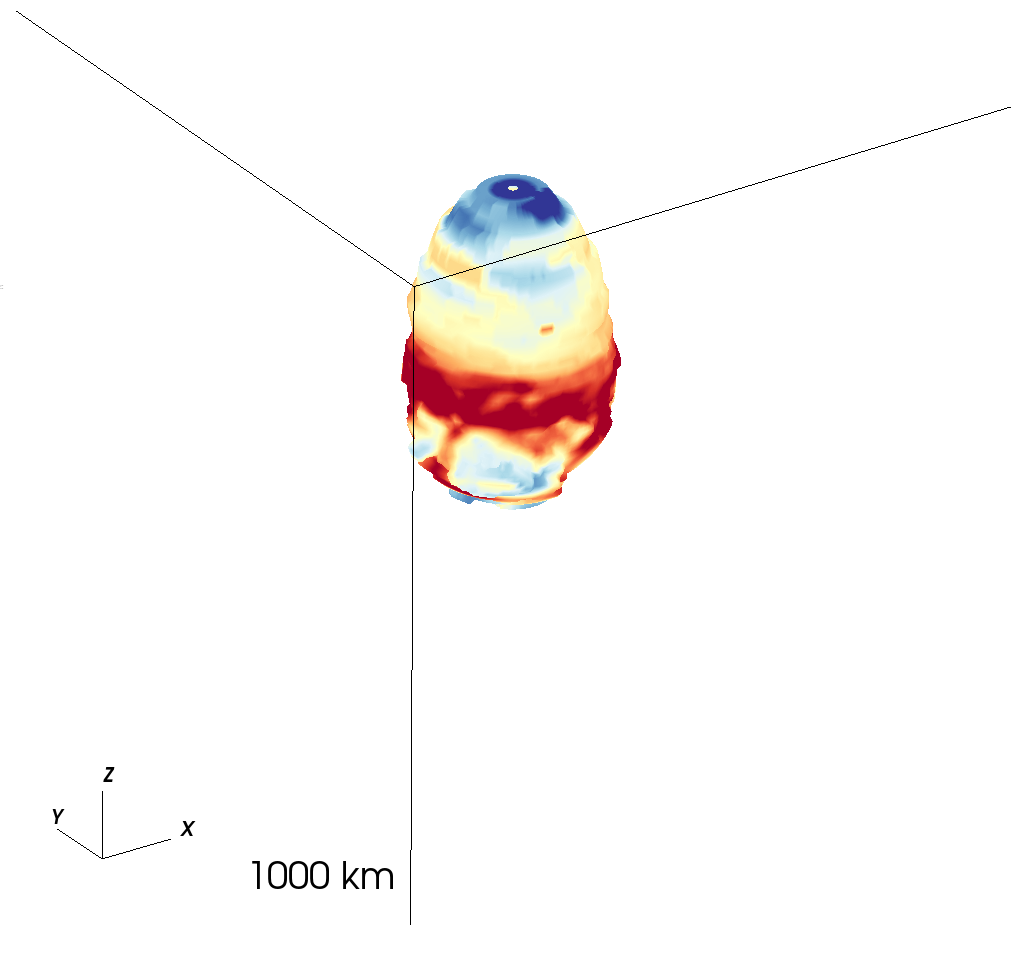}
          \includegraphics[width=0.23\linewidth]{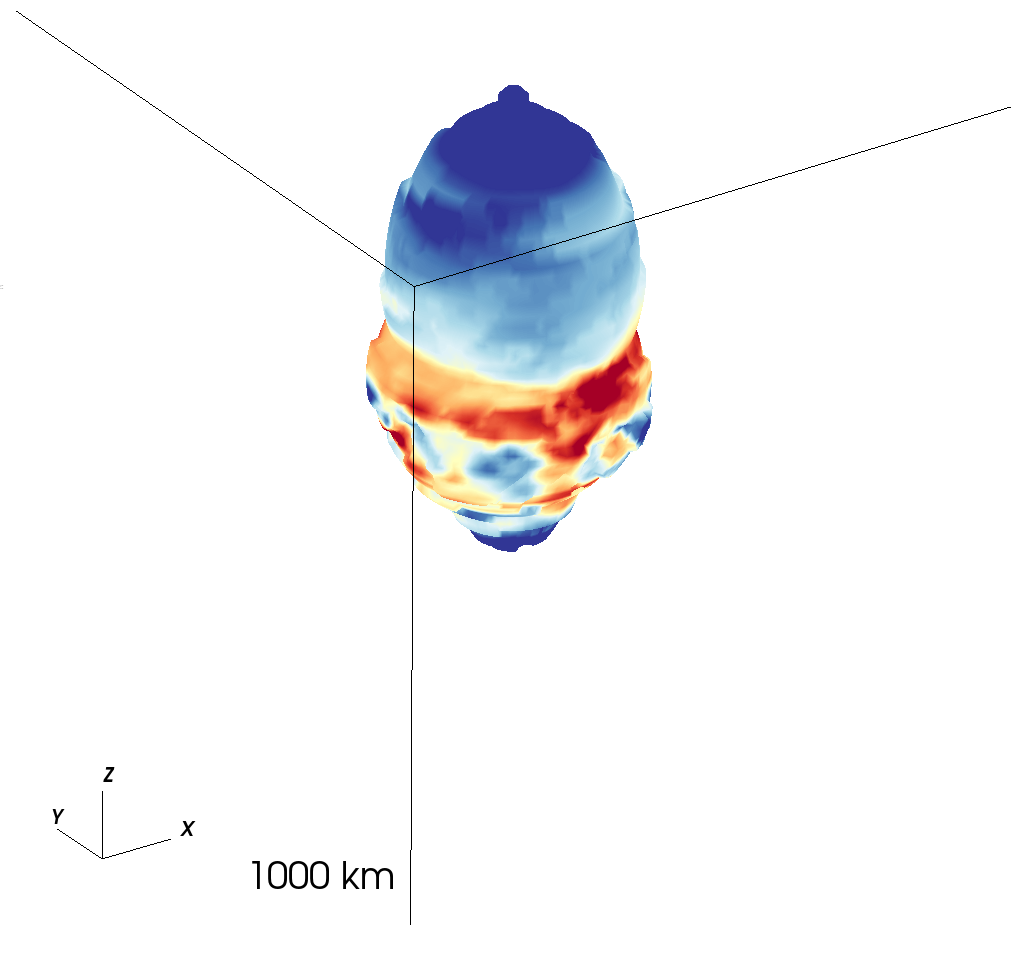}
        }
        \hbox{
          \includegraphics[width=0.05\linewidth]{./W-taurat-colbar.png}
          \includegraphics[width=0.23\linewidth]{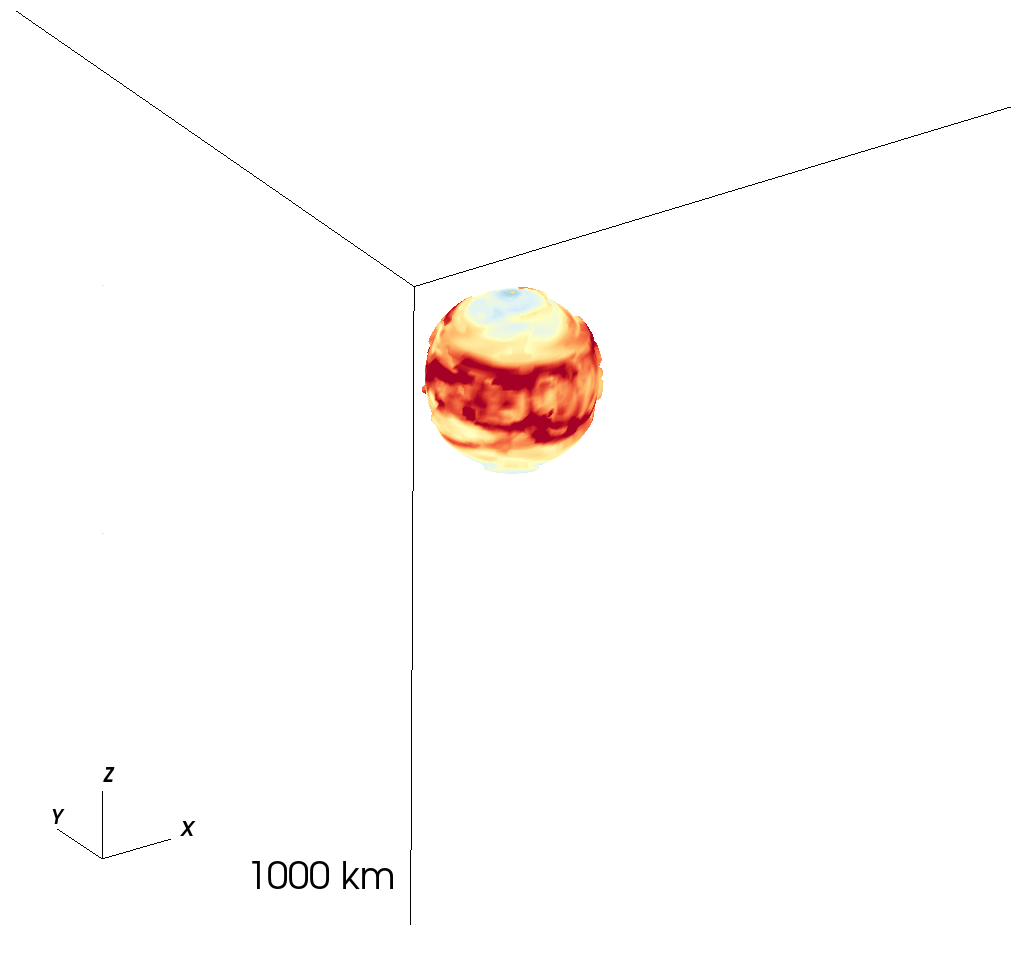}
          \includegraphics[width=0.23\linewidth]{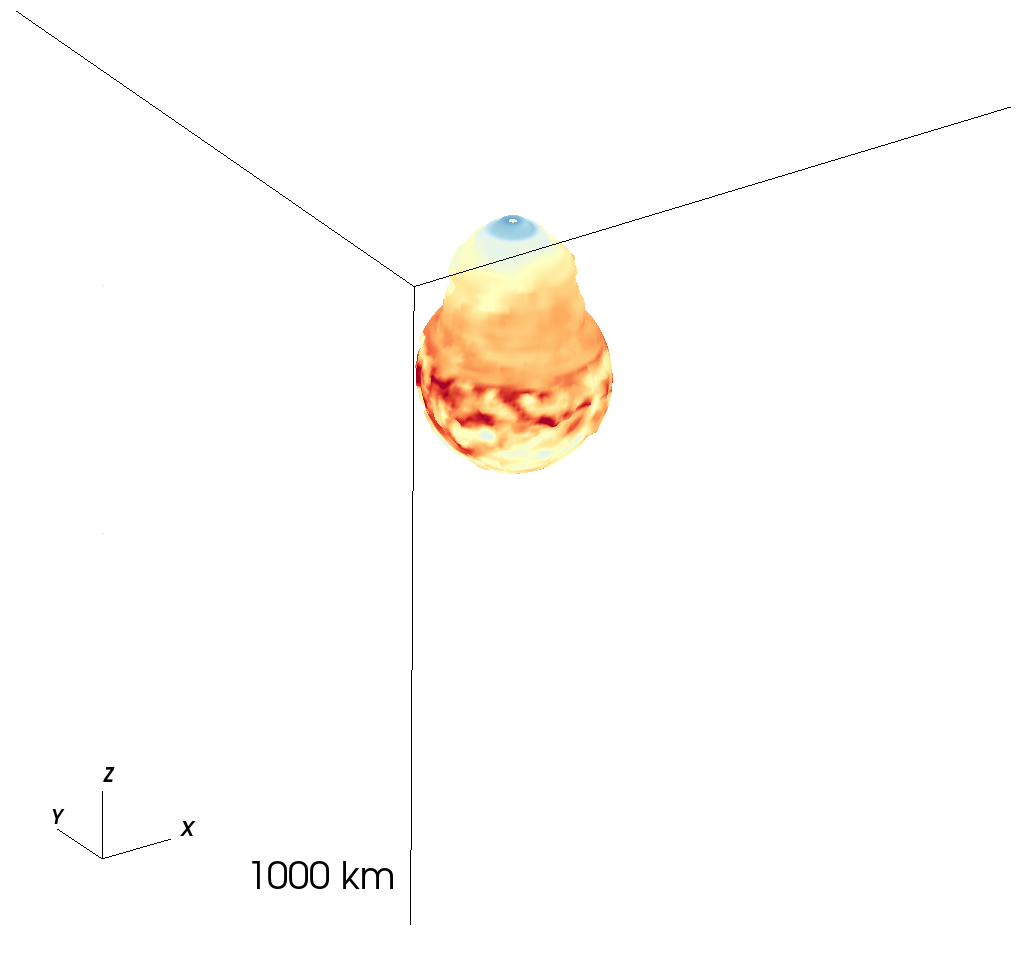}
          \includegraphics[width=0.23\linewidth]{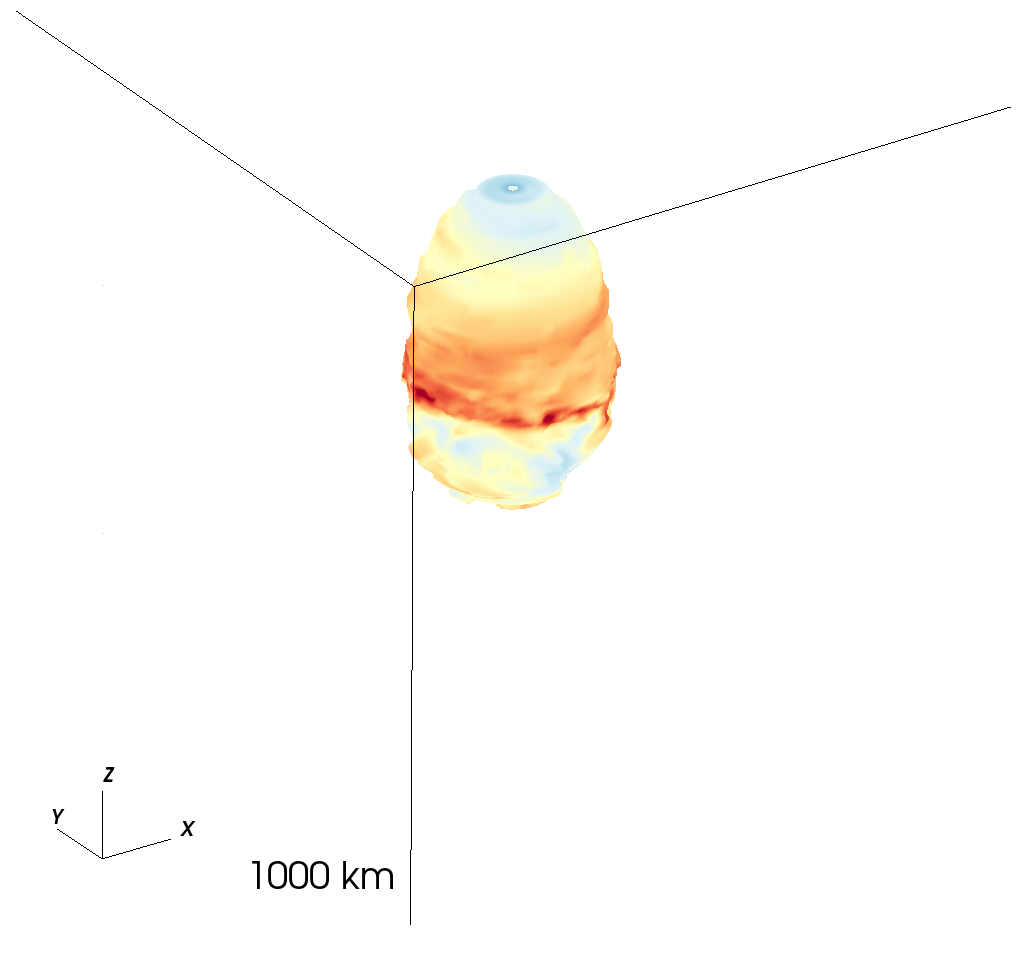}
          \includegraphics[width=0.23\linewidth]{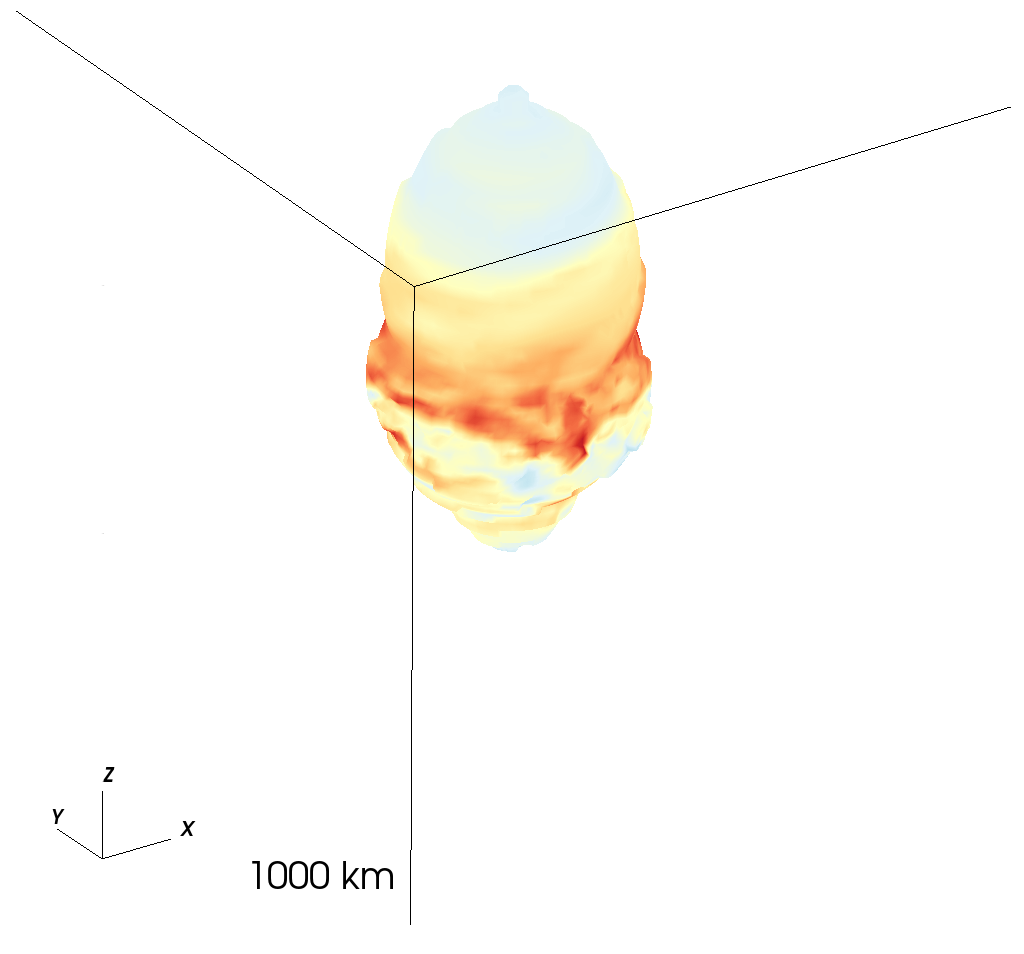}
        }
      }
    }
    \node[fill=white, opacity=0, text opacity=1] at (-8.2,+5.3) {$\tautau$};
    \node[fill=white, opacity=1, text opacity=1] at (-5.9,+2.2) {(a)};
    \node[fill=white, opacity=1, text opacity=1] at (-1.7,+2.2) {(b)};
    \node[fill=white, opacity=1, text opacity=1] at (2.4,+2.2) {(c)};	
    \node[fill=white, opacity=1, text opacity=1] at (6.6,+2.2) {(d)};
    \node[fill=white, opacity=0, text opacity=1] at (-8.2,+1.3) {$e [c^2]$};
    \node[fill=white, opacity=0, text opacity=1] at (-5.9,-1.6) {(e)};
    \node[fill=white, opacity=1, text opacity=1] at (-1.7,-1.6) {(f)};
    \node[fill=white, opacity=0, text opacity=1] at (2.4,-1.6) {(g)};	
    \node[fill=white, opacity=1, text opacity=1] at (6.6,-1.6) {(h)};
    \node[fill=white, opacity=0, text opacity=1] at (-8.2,-2.4) {$\tautaum$};
    \node[fill=white, opacity=0, text opacity=1] at (-5.9,-5.45) {(i)};
    \node[fill=white, opacity=1, text opacity=1] at (-1.7,-5.45) {(j)};
    \node[fill=white, opacity=0, text opacity=1] at (2.4,-5.45) {(k)};	
    \node[fill=white, opacity=1, text opacity=1] at (6.6,-5.45) {(l)};
  \end{tikzpicture} 
  \caption{
    Same as \figref{Fig:RO-3dtauratio}, but for \mRp at, from left to
    right, $\tpb = 0.11, 0.15, 0.19, 0.23 \, \sek$.
  }
  \label{Fig:Rp3-3dtauratio}
\end{figure*}

\begin{figure*}
  \centering
  \begin{tikzpicture}
    \pgftext{\vbox{
        \hbox{ 
          \includegraphics[width=0.05\linewidth]{./W-taurat-colbar.png}
          \includegraphics[width=0.23\linewidth]{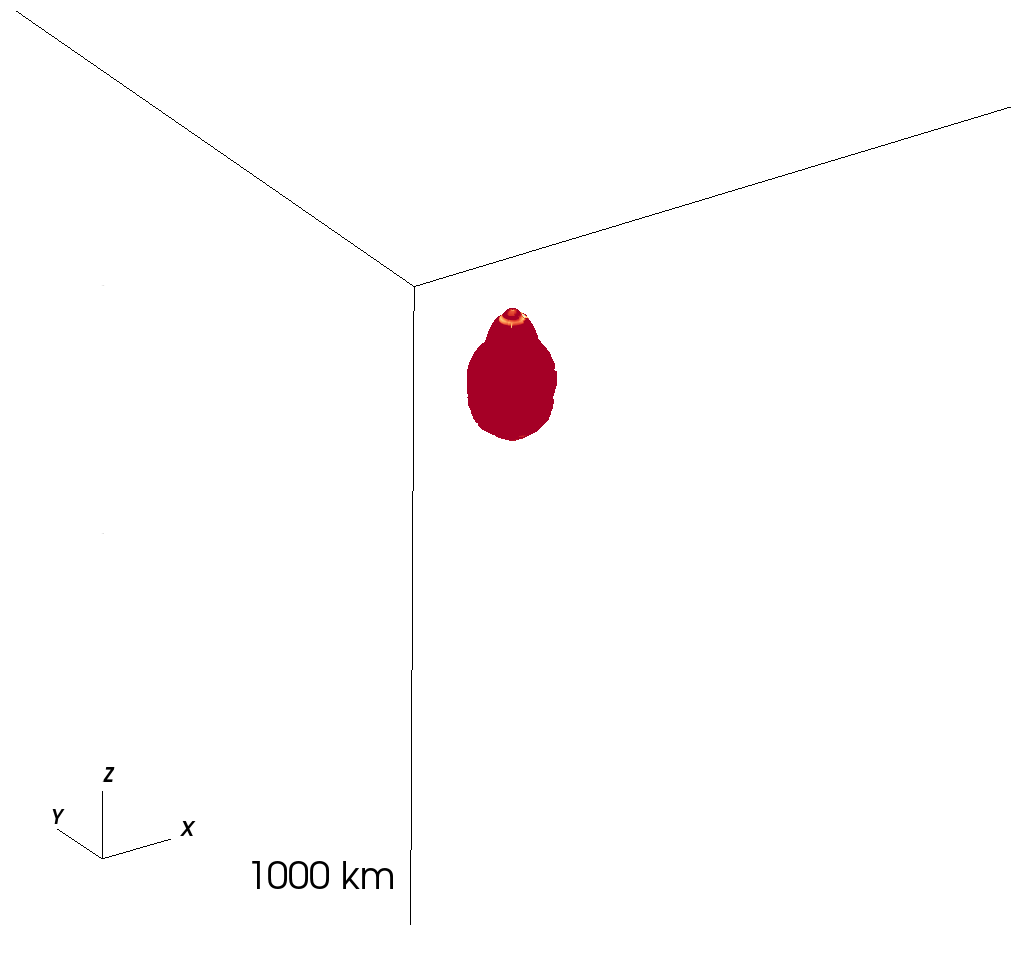}
          \includegraphics[width=0.23\linewidth]{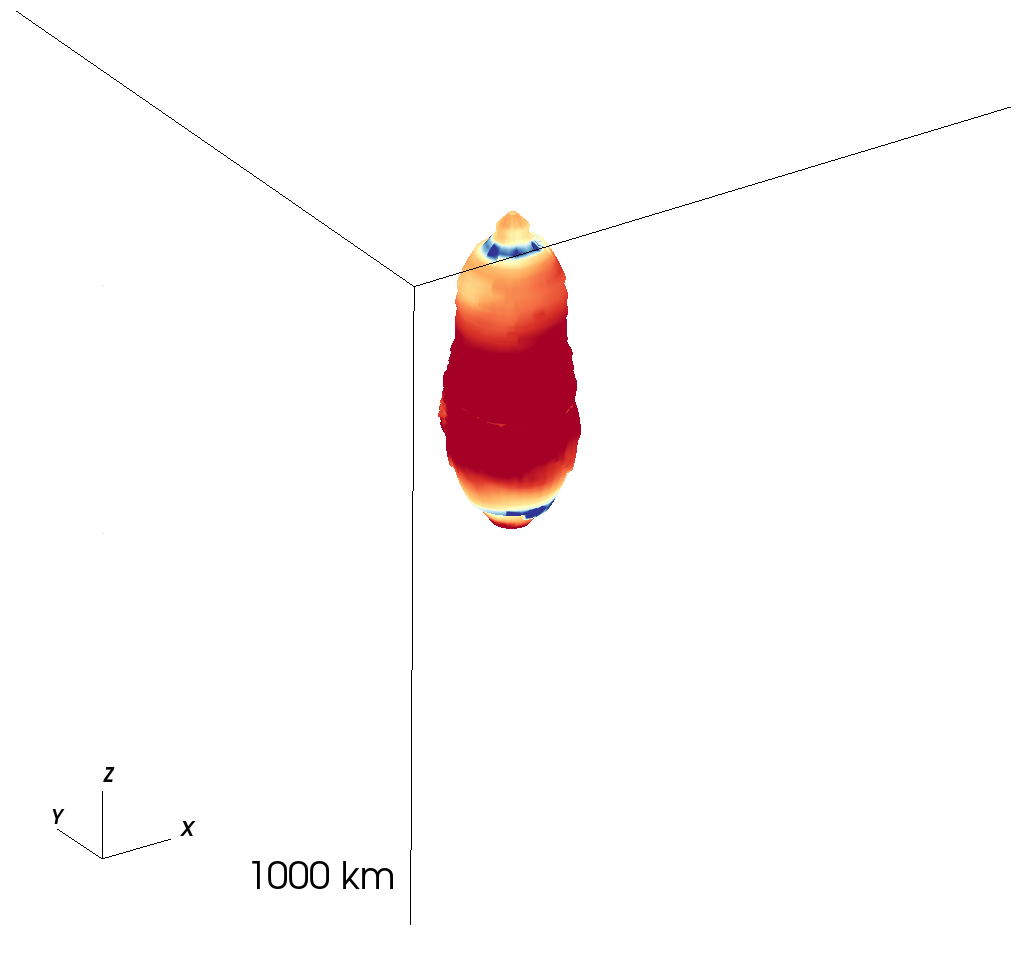}
          \includegraphics[width=0.23\linewidth]{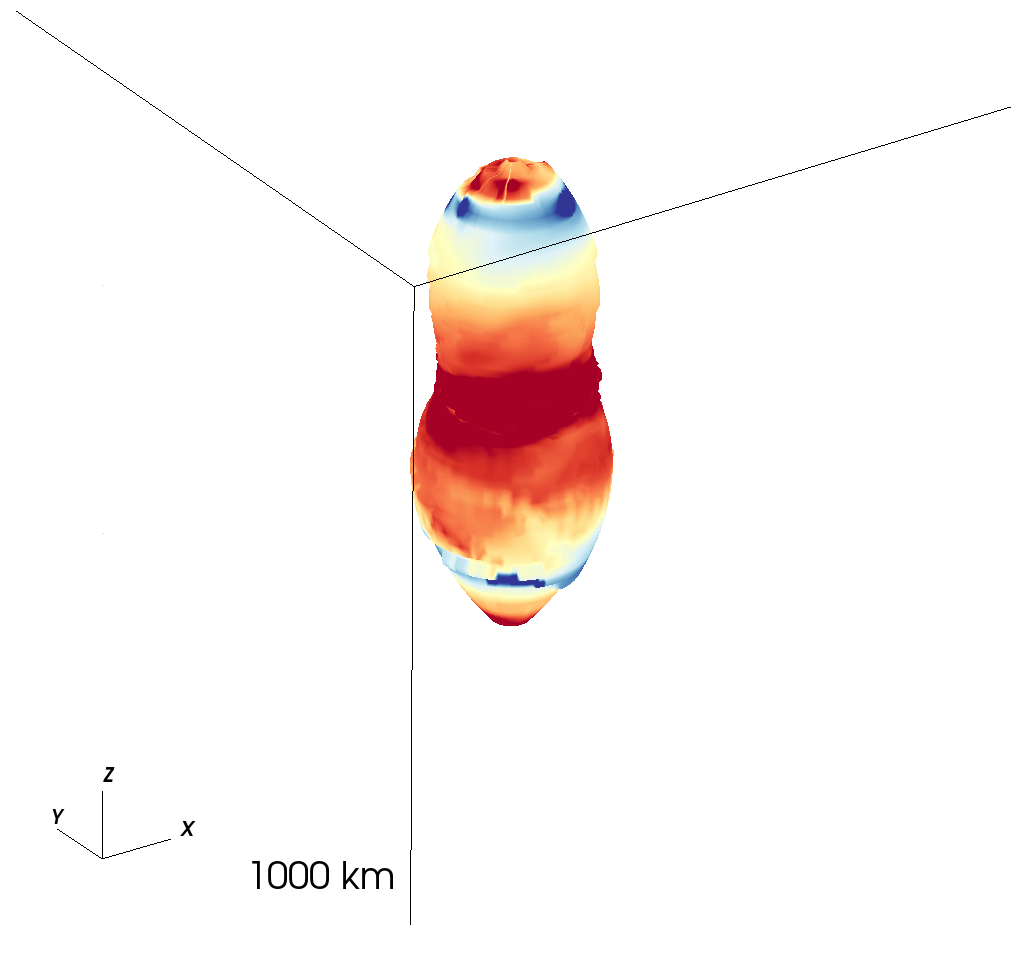}
          \includegraphics[width=0.23\linewidth]{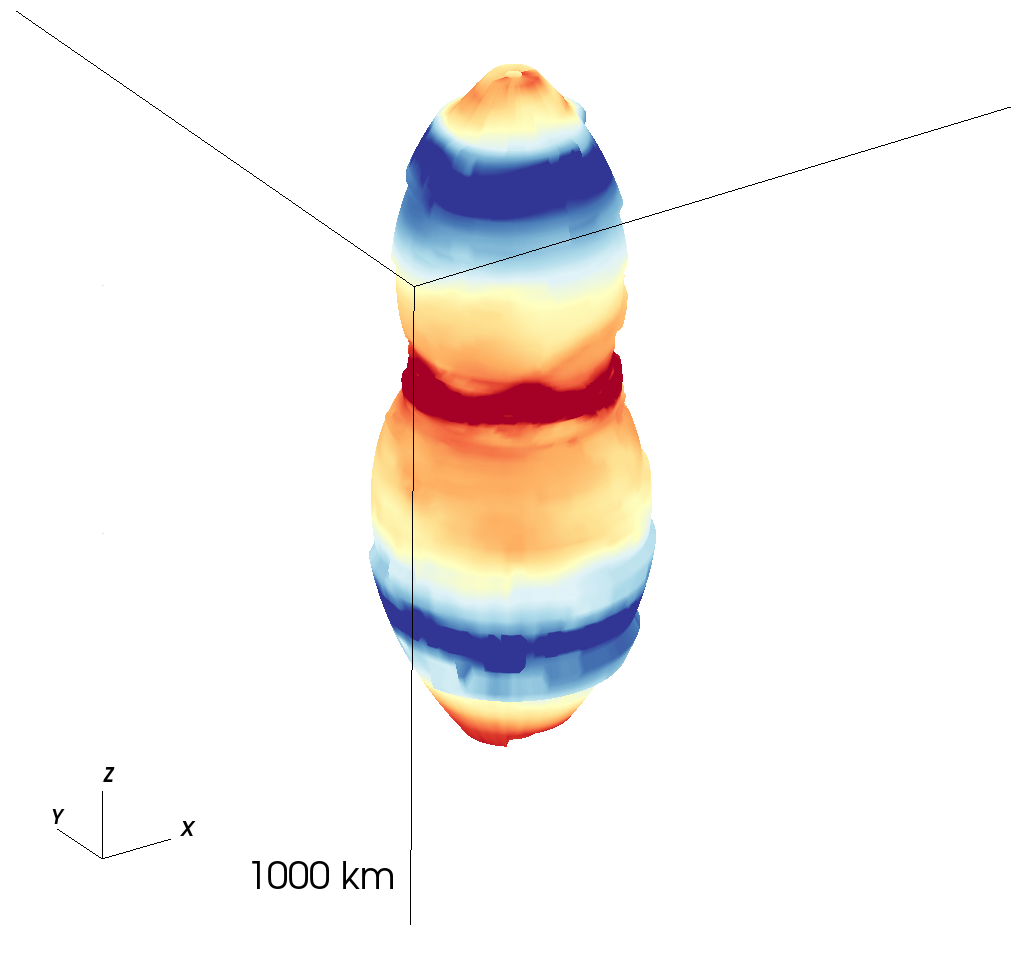}
        }
        \hbox{
          \includegraphics[width=0.05\linewidth]{./W-espec-colbar.png}
          \includegraphics[width=0.23\linewidth]{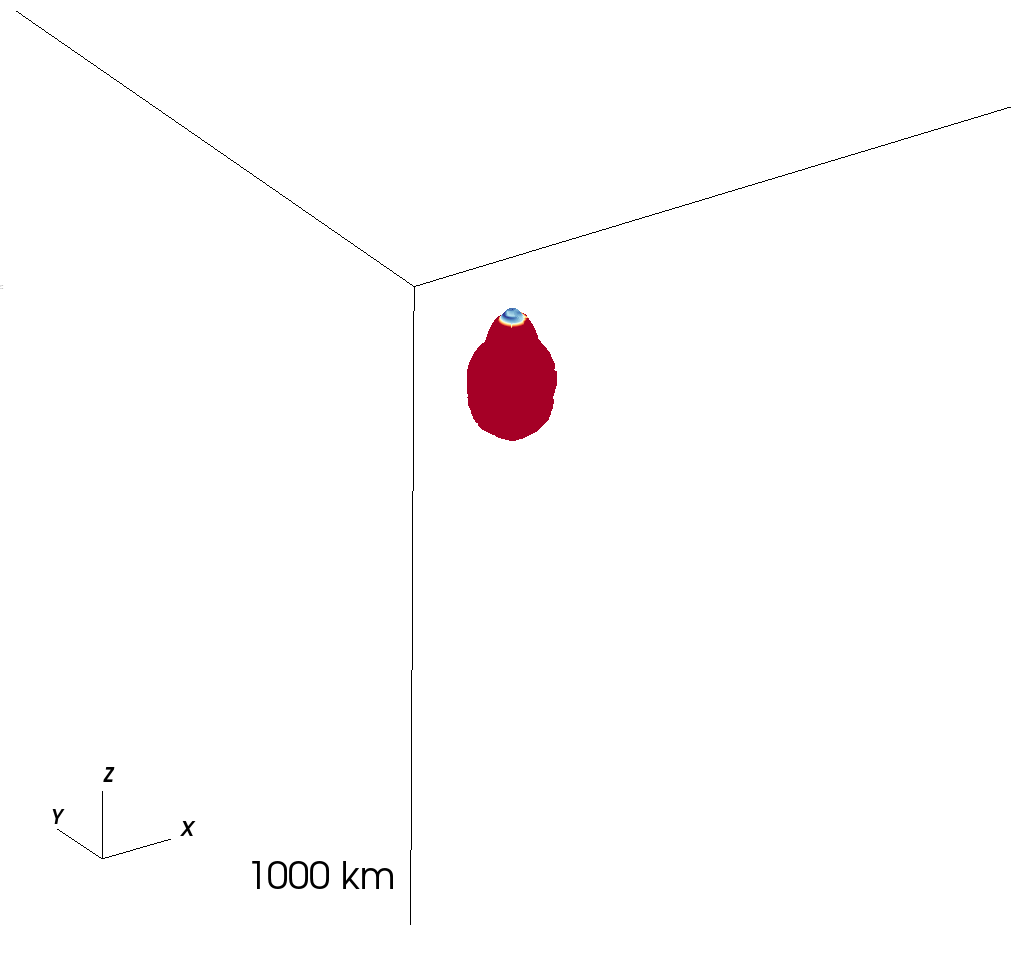}
          \includegraphics[width=0.23\linewidth]{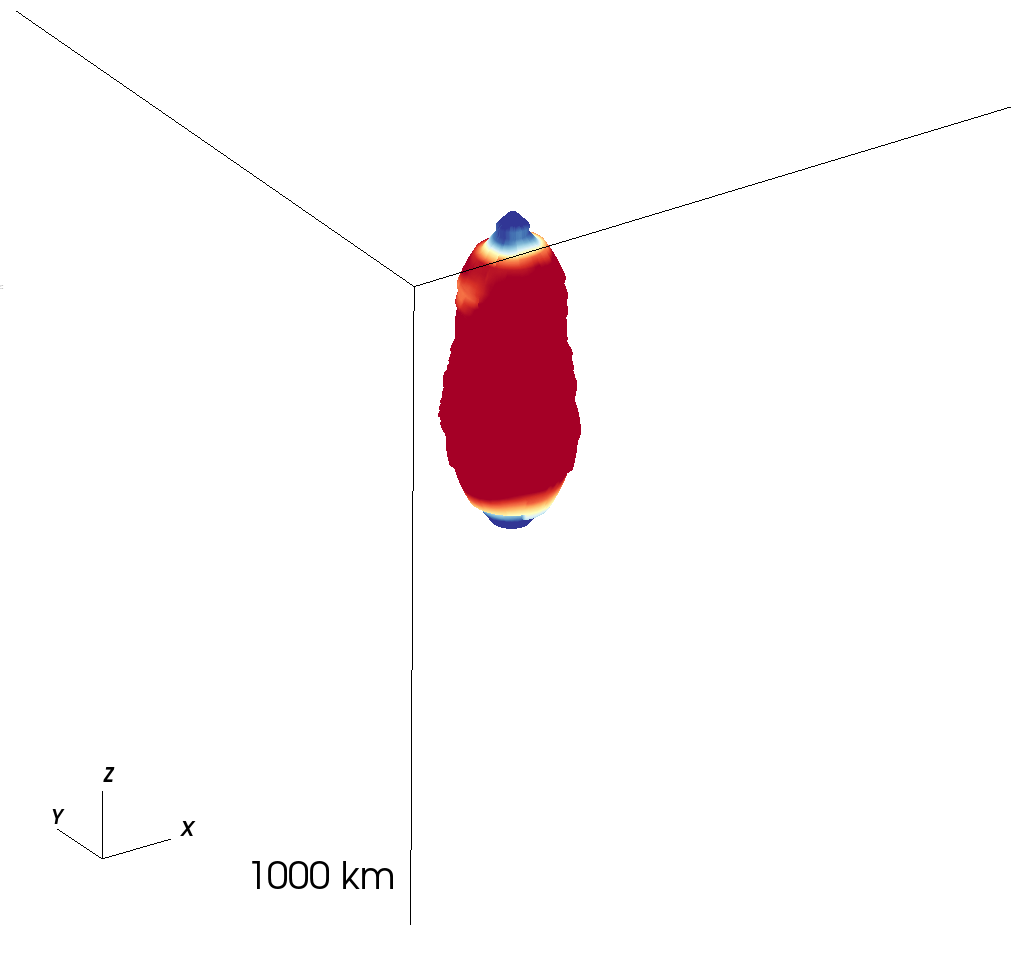}
          \includegraphics[width=0.23\linewidth]{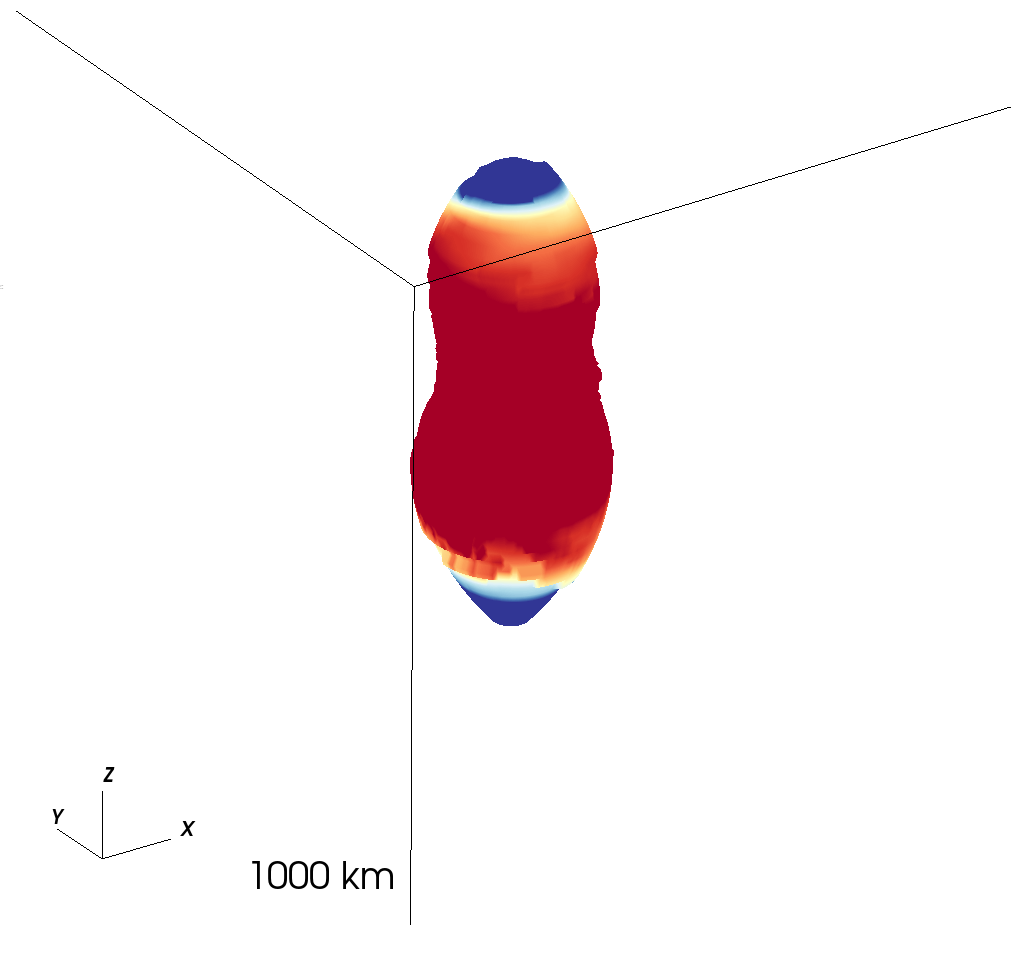}
          \includegraphics[width=0.23\linewidth]{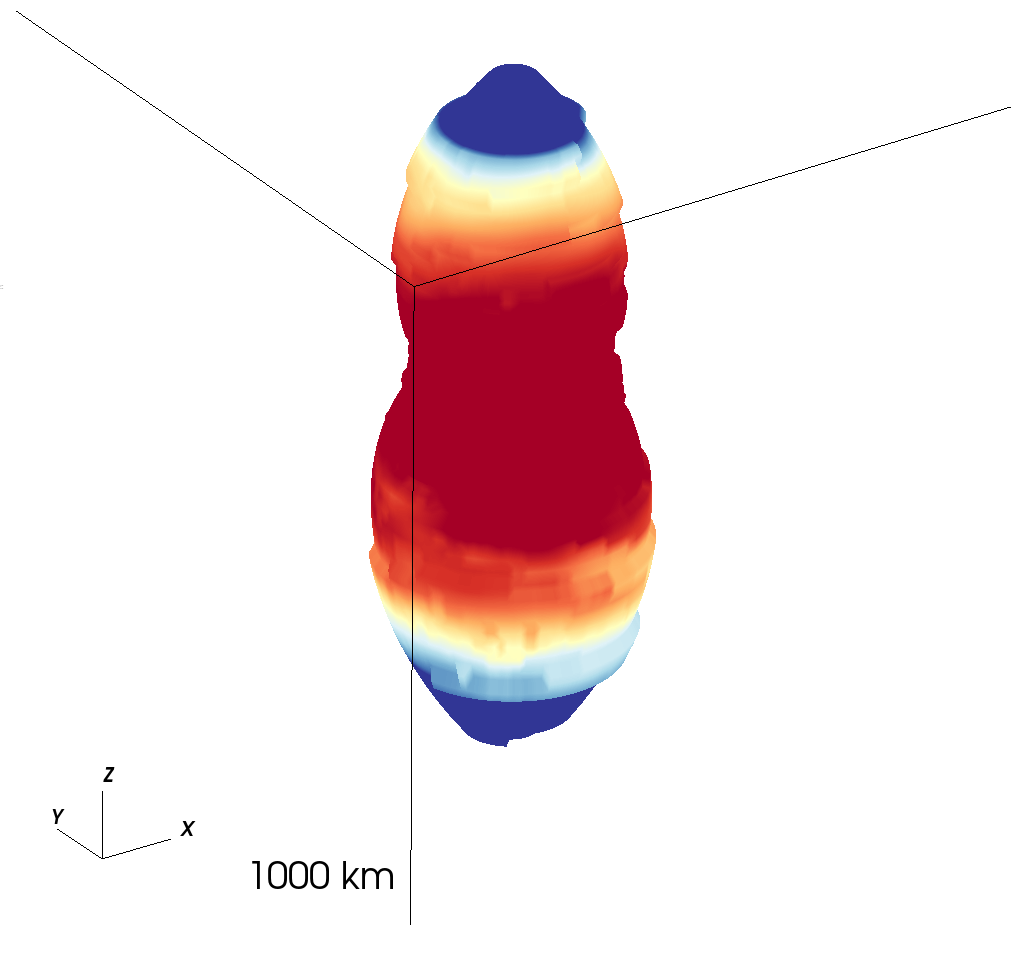}
        }
        \hbox{
          \includegraphics[width=0.05\linewidth]{./W-taurat-colbar.png}
          \includegraphics[width=0.23\linewidth]{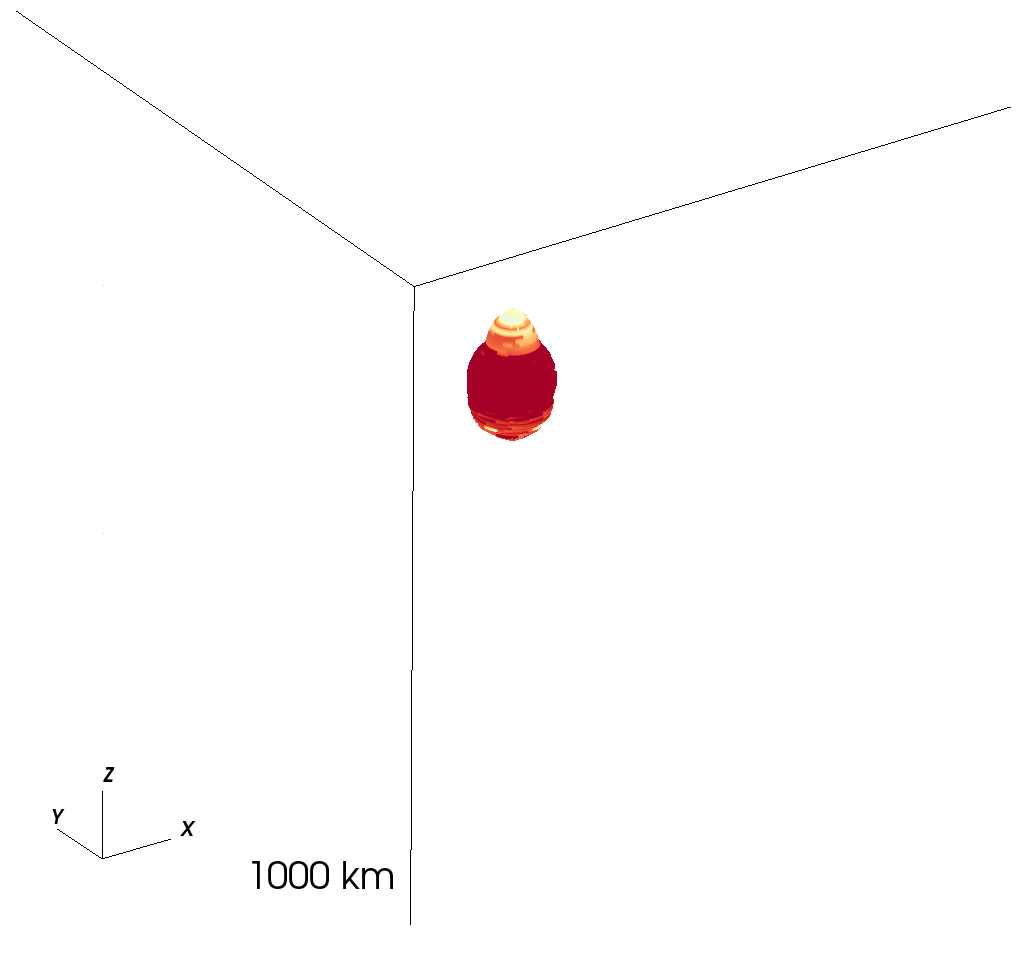}
          \includegraphics[width=0.23\linewidth]{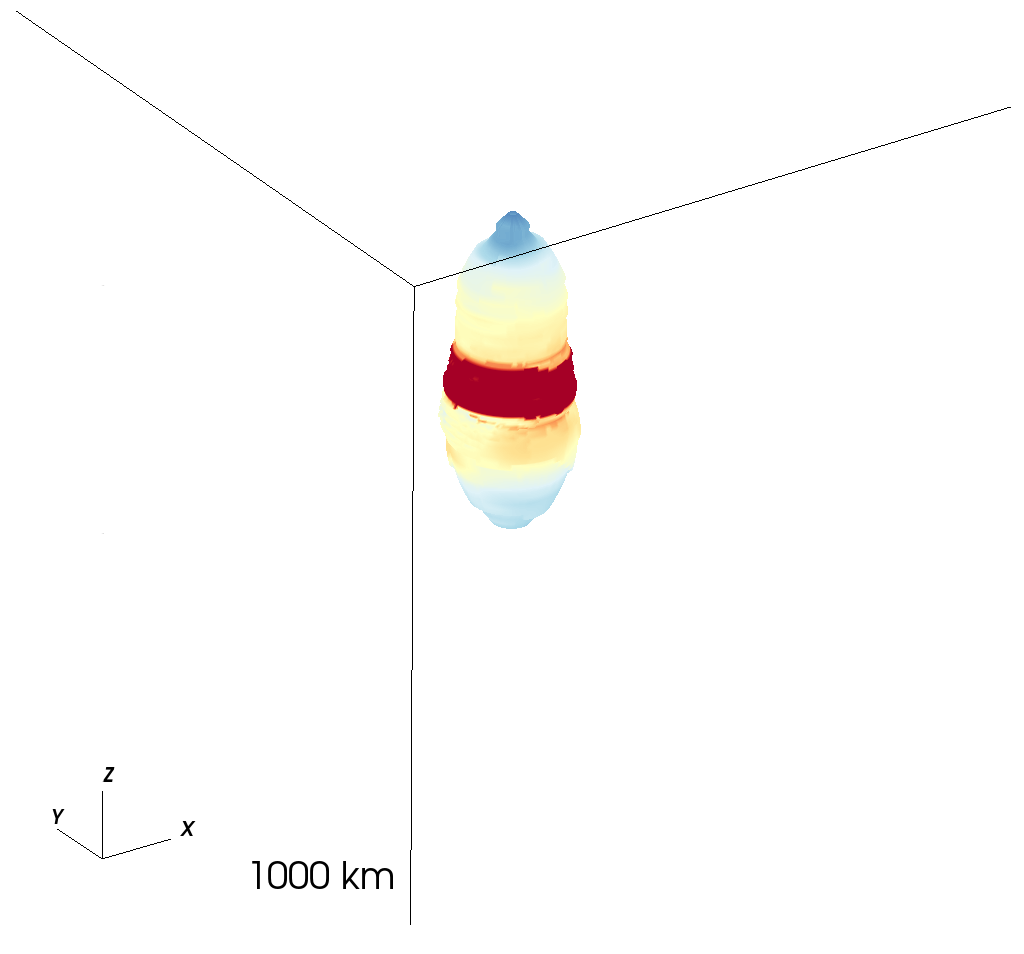}
          \includegraphics[width=0.23\linewidth]{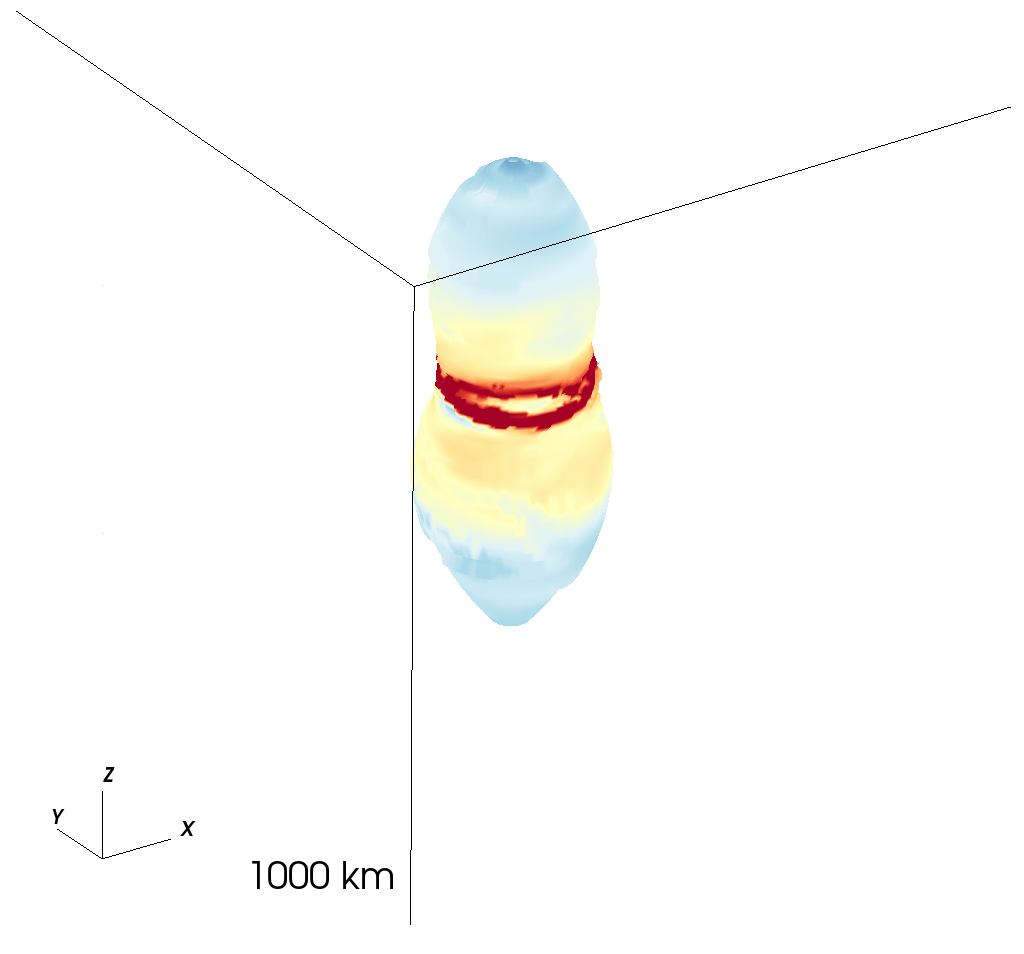}
          \includegraphics[width=0.23\linewidth]{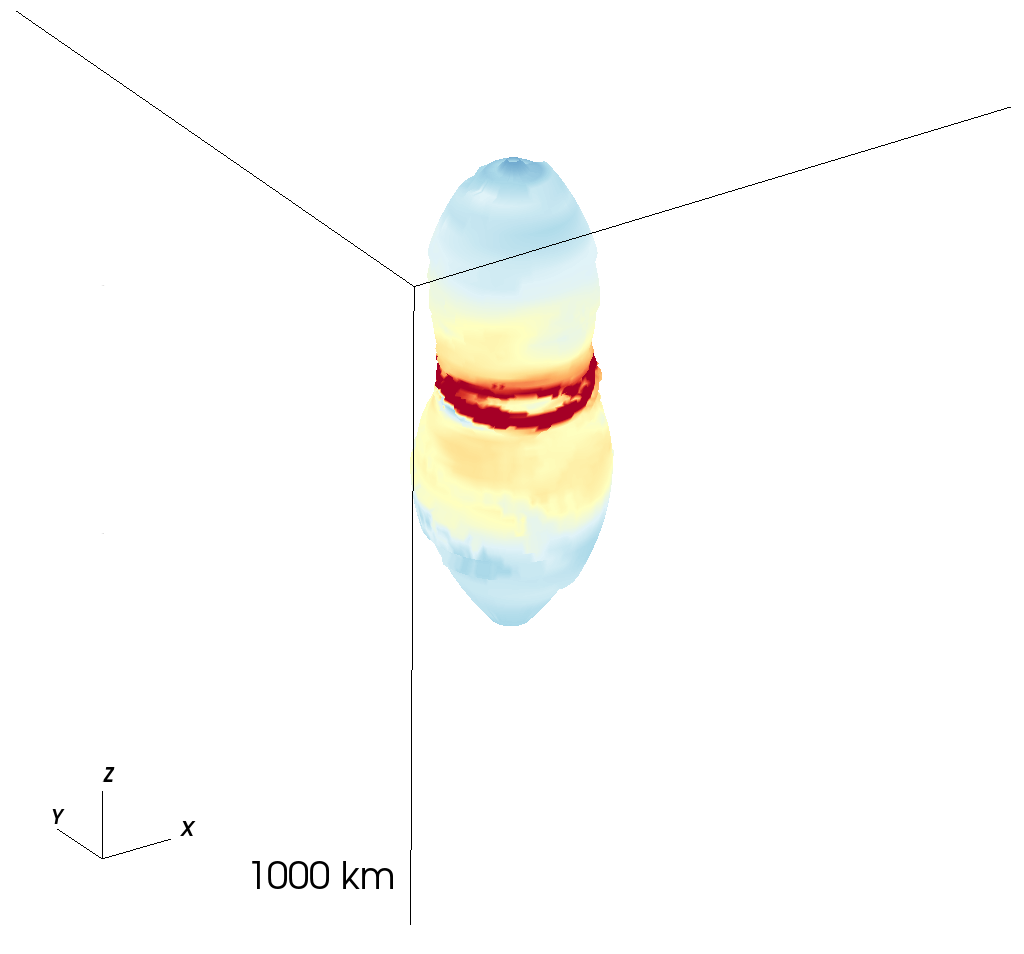}   }
      }
    }
    \node[fill=white, opacity=0, text opacity=1] at (-8.2,+5.3) {$\tautau$};
    \node[fill=white, opacity=1, text opacity=1] at (-5.9,+2.2) {(a)};
    \node[fill=white, opacity=1, text opacity=1] at (-1.7,+2.2) {(b)};
    \node[fill=white, opacity=1, text opacity=1] at (2.4,+2.2) {(c)};	
    \node[fill=white, opacity=1, text opacity=1] at (6.6,+2.2) {(d)};
    \node[fill=white, opacity=0, text opacity=1] at (-8.2,+1.3) {$e [c^2]$};
    \node[fill=white, opacity=0, text opacity=1] at (-5.9,-1.6) {(e)};
    \node[fill=white, opacity=1, text opacity=1] at (-1.7,-1.6) {(f)};
    \node[fill=white, opacity=0, text opacity=1] at (2.4,-1.6) {(g)};	
    \node[fill=white, opacity=1, text opacity=1] at (6.6,-1.6) {(h)};
    \node[fill=white, opacity=0, text opacity=1] at (-8.2,-2.4) {$\tautaum$};
    \node[fill=white, opacity=0, text opacity=1] at (-5.9,-5.45) {(i)};
    \node[fill=white, opacity=1, text opacity=1] at (-1.7,-5.45) {(j)};
    \node[fill=white, opacity=0, text opacity=1] at (2.4,-5.45) {(k)};	
    \node[fill=white, opacity=1, text opacity=1] at (6.6,-5.45) {(l)};
  \end{tikzpicture}  
  \caption{
    Same as \figref{Fig:RO-3dtauratio}, but for \mRs at, from left to right, $\tpb = 0.01,
    0.03, 0.05, 0.07 \,  \sek$.
  }
  \label{Fig:Rs-3dtauratio}
\end{figure*}

\begin{figure*}
  \centering
  \begin{tikzpicture}
    \pgftext{\hbox{ 
        \includegraphics[width=0.245\linewidth]{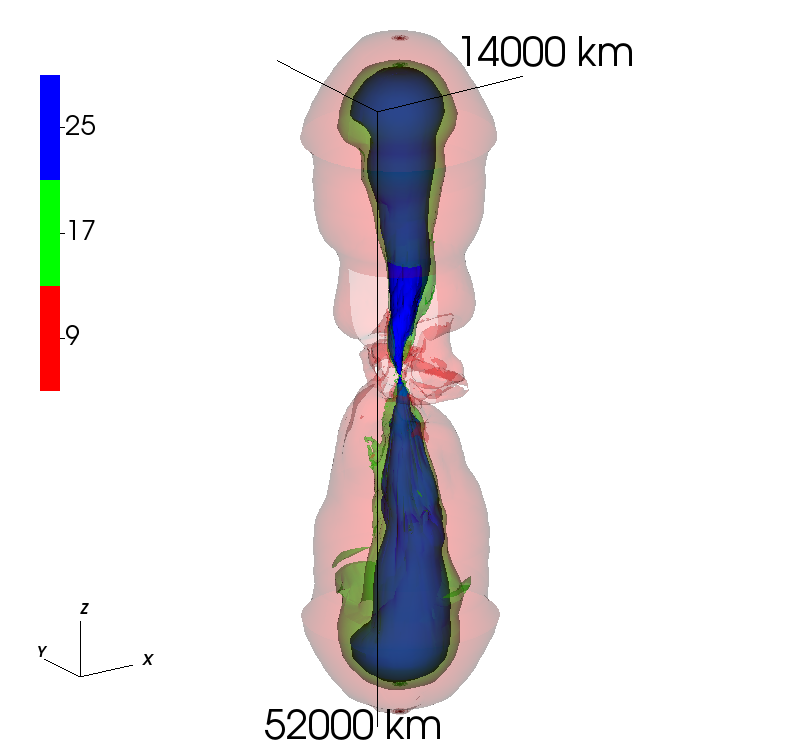}
        \includegraphics[width=0.245\linewidth]{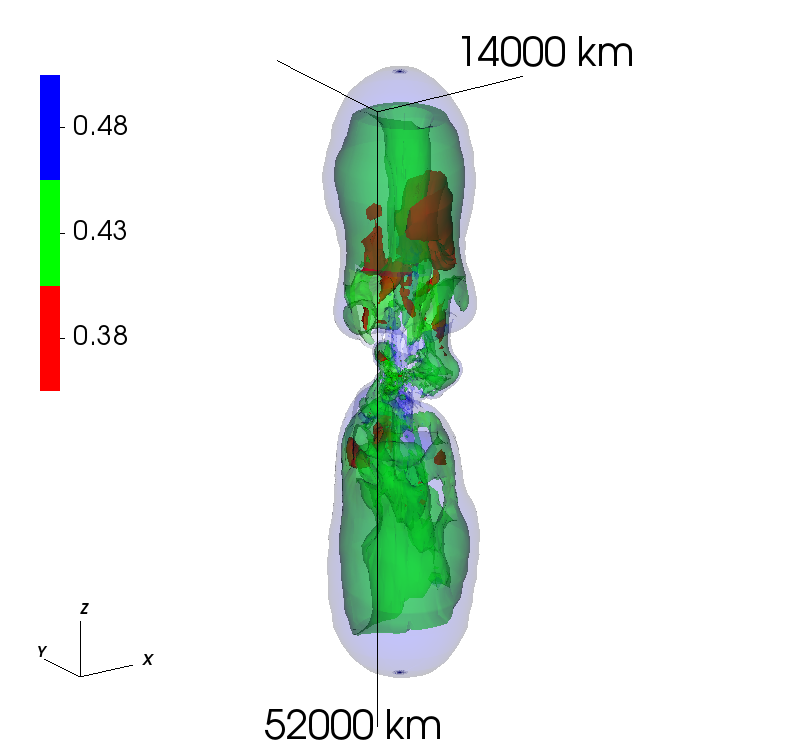}
        \includegraphics[width=0.245\linewidth]{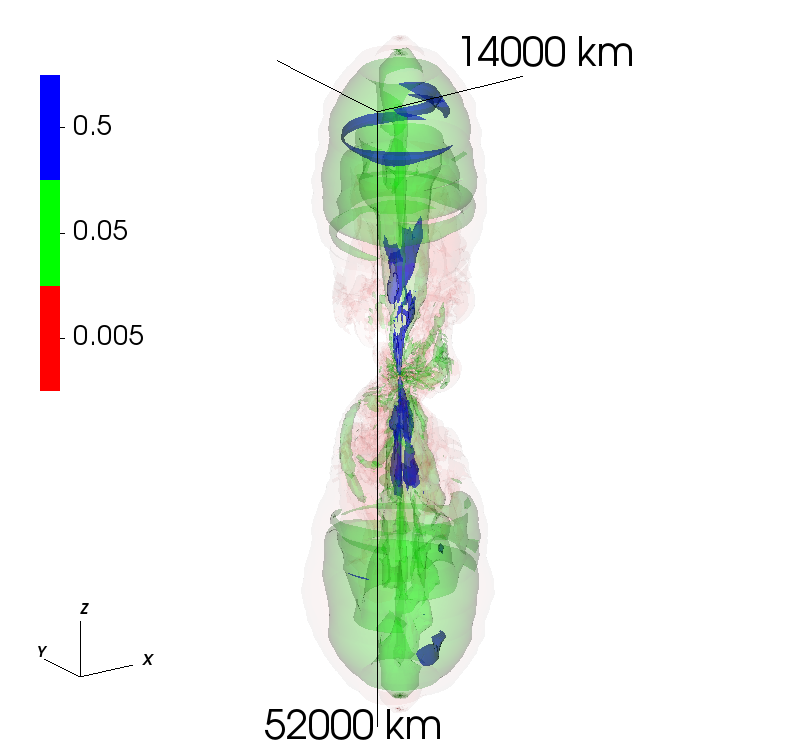}
        \includegraphics[width=0.245\linewidth]{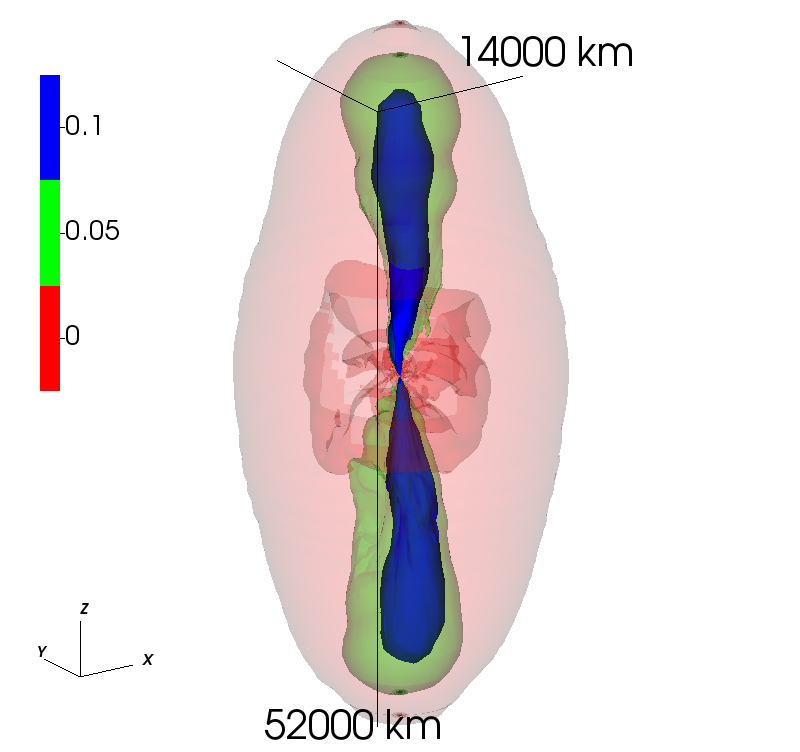}
      }
    }%
    \node[fill=white, opacity=0, text opacity=1] at (-8.75,-1.85) {(a)};
    \node[fill=white, opacity=0, text opacity=1] at (-4.30,-1.85) {(b)};
    \node[fill=white, opacity=0, text opacity=1] at (0.10,-1.85) {(c)};	
    \node[fill=white, opacity=0, text opacity=1] at (4.54,-1.85) {(d)};
    \node[fill=white, opacity=1, text opacity=1] at (-8.45,1.80) {\sffamily\scriptsize $s\,[k_{\textsc{b}}/\text{baryon}]$}; 
    \node[fill=white, opacity=1, text opacity=1] at (-4.07,1.80) {\sffamily\scriptsize $Y_{\rm e}$};	
    \node[fill=white, opacity=1, text opacity=1] at (0.40,1.85) {\sffamily\scriptsize $\beta^{-1}$};	
    \node[fill=white, opacity=1, text opacity=1] at (4.6,1.7) {\sffamily\scriptsize energy flux [$c$]};
    \node[fill=white, opacity=0, text opacity=1] at (4.6,1.9) {\sffamily\scriptsize specific};
    %
  \end{tikzpicture}
  \caption{
    Same as \figref{Fig:Rw-3dplots}, but for \mRp at $\tpb = 1.45 \, \sek$.
  }
  \label{Fig:Rp-3dplots}
\end{figure*}

\begin{figure*}
  \centering
  \begin{tikzpicture}
    \pgftext{\hbox{ 
        \includegraphics[width=0.245\linewidth]{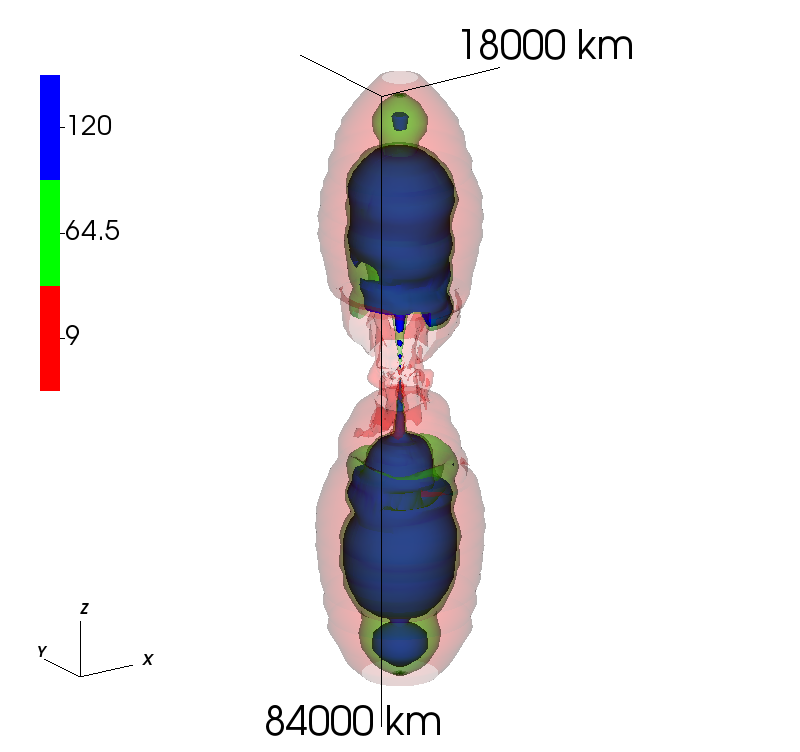}
        \includegraphics[width=0.245\linewidth]{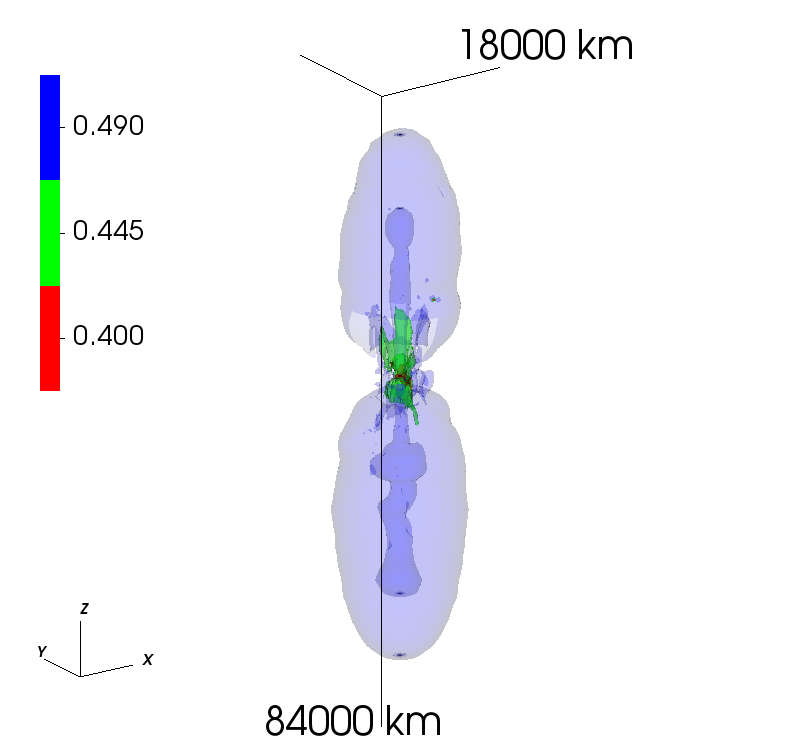}
        \includegraphics[width=0.245\linewidth]{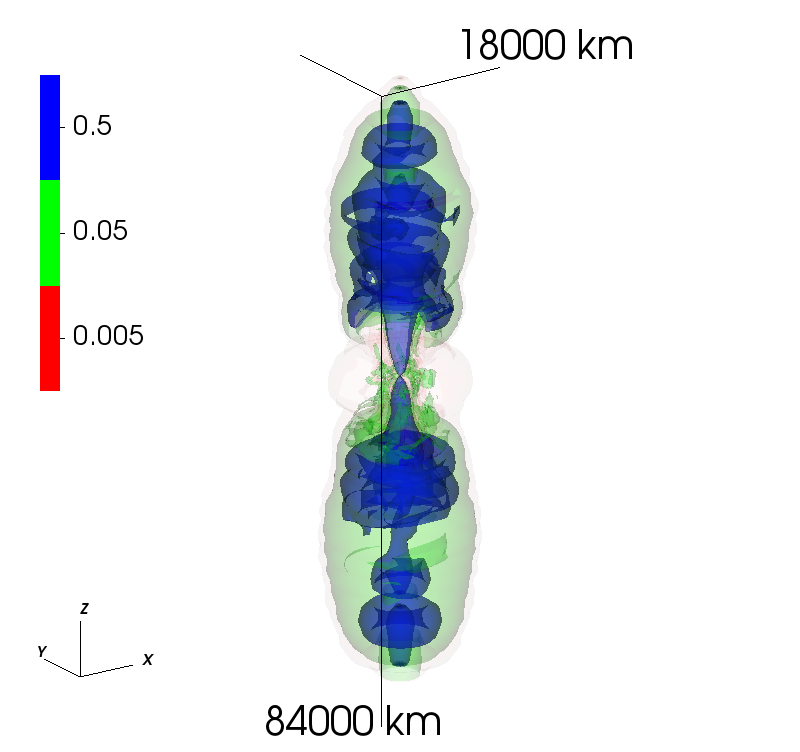}
        \includegraphics[width=0.245\linewidth]{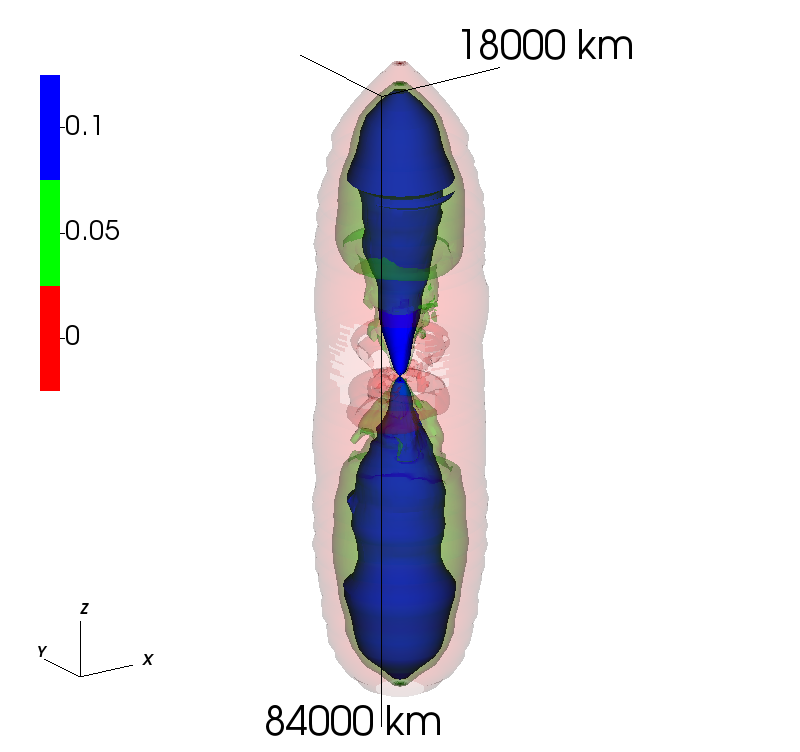}
      }
    }%
    \node[fill=white, opacity=0, text opacity=1] at (-8.75,-1.85) {(a)};
    \node[fill=white, opacity=0, text opacity=1] at (-4.30,-1.85) {(b)};
    \node[fill=white, opacity=0, text opacity=1] at (0.10,-1.85) {(c)};	
    \node[fill=white, opacity=0, text opacity=1] at (4.54,-1.85) {(d)};
    \node[fill=white, opacity=1, text opacity=1] at (-8.45,1.80) {\sffamily\scriptsize $s\,[k_{\textsc{b}}/\text{baryon}]$}; 
    \node[fill=white, opacity=1, text opacity=1] at (-4.07,1.80) {\sffamily\scriptsize $Y_{\rm e}$};	
    \node[fill=white, opacity=1, text opacity=1] at (0.40,1.85) {\sffamily\scriptsize $\beta^{-1}$};	
    \node[fill=white, opacity=1, text opacity=1] at (4.6,1.7) {\sffamily\scriptsize energy flux [$c$]};
    \node[fill=white, opacity=0, text opacity=1] at (4.6,1.9) {\sffamily\scriptsize specific};
    %
  \end{tikzpicture}
  \caption{
    Same as \figref{Fig:Rw-3dplots}, but for \mRs at $\tpb = 1.05 \, \sek$.
  }
  \label{Fig:Rs-3dplots}
\end{figure*}

Models \Rp and \Rs produce explosions setting in with hardly any shock
stagnation (at $t_{\mathrm{exp}} \approx 100 \,$ms, for \mRp) and
promptly after bounce (for \mRs).  As a result, there is no time for
the development of an $m = 1$ mode in the shock before the onset of
the explosion (\figref{Fig:35OC-Rp3-rad}; note the difference with
\modls{W} and \RO in Figs.\,\ref{Fig:35OC-Rw-rad} and
\ref{Fig:35OC-RO-rad}, respectively).  Shock revival starts from a
virtually spherical shape (note that all lines in the
\figref{Fig:35OC-Rp3-rad}\panel{a} cluster around the same values and
the comparably low magnitude of $\ashlm{2}{0}$ between
$\tpb \approx 0.05 \, \sek$ and $\tpb \approx 0.1 \, \sek$ in
\figref{Fig:35OC-Rp3-rad}\panel{c}).  Thereafter, the asymmetry rises
quickly as a prolate explosion sets in, first in the northern, and
later also in the southern hemisphere.  At
$\tpb \approx 0.35 \, \sek$, the northern and southern shock radii
start to agree and expand at the same velocity.  In the equatorial
region, the shock expands slower, leading to a relatively constant
pole-to-equator axis ratio of $2:1$ and a very high quadrupole
coefficient maintaining a value of
$\ashlm{2}{0} \approx 0.09 \, \ashlm{0}{0}$ for a long time.  The
shock expansion proceeds faster in \mRs and without any intermediate
state of a more or less spherical shape.  The shock radii show a minor
north-south asymmetry (\cf black lines in
\figref{Fig:35OC-Rs-rad}\panel{a} for the polar shock radii and the
moderate value of $\ashlm{1}{0}$ in
\figref{Fig:35OC-Rs-rad}\panel{c}).  The ejecta show the most extreme
morphology of all models with an axis ratio around $4:1$ and a
quadrupole coefficient in the range
$0.12 \lesssim \ashlm{2}{0} / \ashlm{0}{0} \lesssim 0.25$.

In both models, neutrino heating contributes very little to the
explosion.  Immediately after bounce, the PNS does not become highly
oblate.  Consequently, the neutrino emission is relatively isotropic
(\figref{Fig:nulums}\panel{c}, \panel{d}) and we do not find an
enhancement of the neutrino heating in the polar regions in the same
way as in \mRO.  The ratios between advection and heating timescales,
presented in the \banel{top} rows of panels of Figs.\xspace
\ref{Fig:Rp3-3dtauratio} and \ref{Fig:Rs-3dtauratio}, do not favour
shock revival at any location.  In \mRp, we only find an increase of
$\tautau$ beyond unity once the advection timescale rises after the
shock expands along the rotational axis. In \mRs, $\tautau$ remains
below unity until well after the explosion has started.  Then the
shock expansion leads to an increase of $\tauadv$ and, consistently,
an increase of $\tautau$, in qualitative agreement with the evolution
of $\tautau$ displayed by the corresponding axisymmetric
\modl{35OC-Rs}
\citepalias{Obergaulinger_Aloy__2020__mnras__MagnetorotationalCoreCollapseofPossibleGRBProgenitorsIExplosionMechanisms}.
The strong magnetic fields lead to short \Alfven timescales and, thus,
$\tautaum > 1$ at high latitudes (in qualitative similarity to the
respective axisymmetric models), indicating a magnetically driven
launch of the shock (\banel{bottom} rows of panels).  They cause the
gas in these regions to become gravitationally unbound as we show in
the blue regions in the \banel{middle} rows of panels.

The outflows of both models, visualized in \figref{Fig:Rp-3dplots} and
\figref{Fig:Rs-3dplots}, present a typical jet-like morphology
\citepalias[as in axial
symmetry;][]{Obergaulinger_Aloy__2020__mnras__MagnetorotationalCoreCollapseofPossibleGRBProgenitorsIExplosionMechanisms}
with a narrow beam with maximum velocities up to
$v^r_{\mathrm{max}} \approx c/2$ and similar values of the normalised
energy fluxes along the rotational axis.  The jet heads reach
propagation speeds of
$v_{\mathrm{jet}} \approx \zehnh{2.3}{9} \, \cms$ and
$v_{\mathrm{jet}} \approx \zehnh{4.8}{9} \, \cms$ for models \Rp and
\Rs, respectively.  The beams of thee jets are surrounded by a cocoon
moving at much smaller radial speeds.  \MRp shows a relatively wide
cocoon morphology similar to \mRO and a slightly curved beam.  \MRs
features a narrower cocoon and a beam of a conical shape, widening as
it propagates outwards.  Similarly to \mRO, \mRp hosts a large, very
anisotropic downflow near the equator transporting matter from a
radius of $r \approx 10000 \, \km$ to under $1000 \, \km$.  Like in
\mRO, the downflow is interspersed with matter with positive radial
velocities.  However, unlike in that model, an $m = 1$ mode is less
prominent and restricted to a smaller region around the PNS.  In \mRs,
we find only small clumps of gas falling towards the PNS rather than a
large-scale downflows.  Both the equatorial expansion of the bow shock
driven by the jets and the injection of part of the downflowing mass
into the polar outflows act as a feedback mechanism on the mass
accretion rate onto the PNS. Due to this feedback, the mass accretion
rate is reduced and, consistently, \modls{P} and \Rs display smaller
PNS masses than \modls{W} and \RO (\figref{Fig:globvars4}).

In \mRp, the early and later phases of high and low energy fluxes
($F^{+}_{\mathrm{E}} \sim \zehn{53} \, \ergs$ and
$F^{+}_{\mathrm{E}} \approx \zehnh{2}{52} \, \ergs$, respectively; \cf
\panel{c} of \figref{Fig:globvars2}) correspond to two different
geometries of the outflow.  During the early phase, the jet has a
large opening angle and deposits energy in a relatively wide region
around the rotational axis.  As the energy flux decreases, the jet
opening angle reduces and the energy injection is more concentrated
towards the axis.  Hence, some of the outer parts of the outflows
receive less energy from the interior regions.  Consequently, the
total energy of these regions ceases to increase and even undergoes a
slight decrease, which is reflected in the maximum of $\eej$.
However, energy injection goes on, albeit at a lower rate, and hence
the explosion energy will continue to rise.

Both outflows are highly magnetised at their base.  In the case of
\mRp, $\ateb \approx 20$ over the polar caps of the PNS where the gas
is accelerated.  It drops, however, with radius.  The magnetisation is
strongest at the outer edge of the beam, where a sheath-like region
with $\ateb$ larger than unity extends up to $z \gtrsim 2000 \, \km$.
Beyond that point, this condition can be fulfilled in clumps
propagating outwards in the jet.  This geometry is similar in \mRs,
though in a modified manner owing to its more coherent magnetic field.
Near the jet base, the magnetisation exceeds $\ateb \gtrsim 200$, and
the continuous sheath of super-equipartition magnetic fields
($\ateb > 1$) dissolves into smaller clumps at radii
$r > 10000 \, \km$.

The jets contain very hot ejecta with entropies around
$s \sim 200 \, \kbb$ in the beam of \mRp.  The jet of \mRs is very
inhomogeneous with $s \gtrsim 100 \, \kbb$ in large parts of it, even
reaching values $s \gtrsim 500 \, \kbb$ occasionally.  \MRs
ejects predominantly matter with $\Ye \approx 0.5$ (although close to
the flanks of the jets matter with $0.4\lesssim \Ye \lesssim 0.45$ is
also launched; blue and green shades in
\figref{Fig:Rs-3dplots}\panel{b}), whereas \mRp produces a significant
amount of neutron-rich ejecta with an electron fraction down to
$\Ye \approx 0.33$.  Like in \mRO, this component is distributed in
several large bubbles at mid latitudes around the beam of the jets
where, typically, $\Ye \approx 0.5$.

\begin{figure}
  \centering
  \begin{tikzpicture}
    \pgftext{\vbox{
        \includegraphics[width=\linewidth]{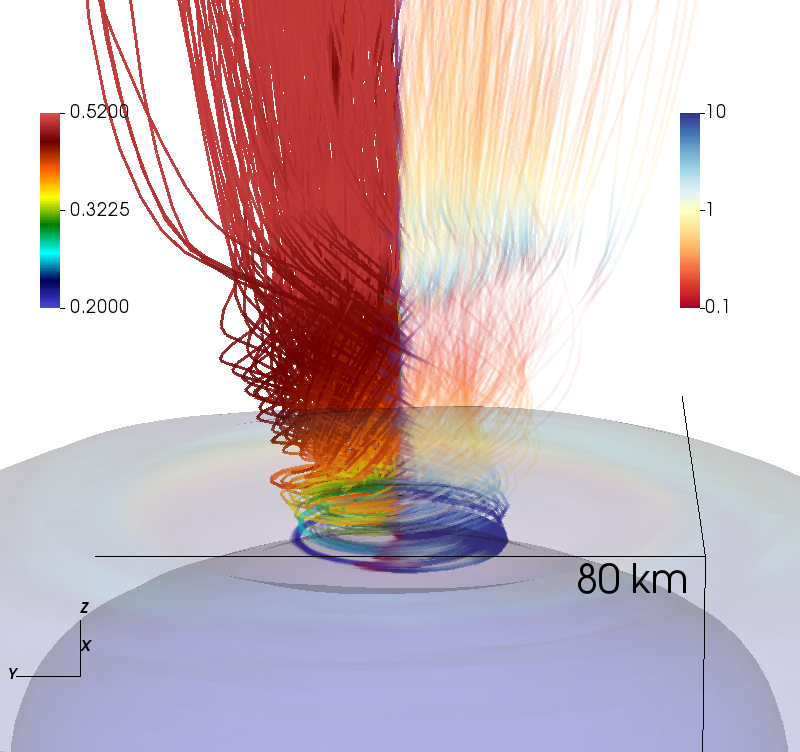}
      }
    }
    \node[fill=white, opacity=1, text opacity=1,font=\fontsize{15}{7.2}] at (-3.8,3.4) {$\Ye$};
    \node[fill=white, opacity=1, text
    opacity=1,font=\fontsize{15}{7.2}] at (+3.6,3.4) {$\frac{\tau_{\mathrm{rad}}}{\tau_{\Ye}}$};
  \end{tikzpicture}
  \caption{
    Releptonization of a jet in \mRs.  We show the base of the
    northern outflow at $\tpb = 0.8 \, \sek$.  The surface of the PNS
    is shown by two blueish iso-density surfaces for $\rho =
    \zehn{10,11} \, \gccm$.  We integrate streamlines of the velocity
    field starting near the polar caps of the PNS.  Their colours
    display the electron fraction of the gas (left, rainbow colour
    table) and the ratio of releptonization and expansion timescales
    (right, red/blue colours correspond to regions where
    releptonization is faster/slower than expansion).
  }
  \label{Fig:Rs-jet-ye}
\end{figure}

For an explanation of the high electron fraction, we refer to
\figref{Fig:Rs-jet-ye} showing streamlines of the velocity field at
the base of one of the two jets of \mRs at $\tpb = 0.8 \, \sek$.
Matter that will for the beam of the jet is ejected from a region very
close to the polar cap of the PNS and, thus, has a very low
$Y_e \approx 0.2$ (dark blue in the left half of the figure) at the
beginning of its trajectory.  Within a few km of upward propagation,
however, $Y_e$ increases to $Y_e \approx 0.5$ (red).  Neutrino fluxes
through the surface of the PNS are dominated by $\bar{\nu}_e$, causing
a rapid releptonization of the gas.  Their effect is enhanced by the
rotational flattening of the PNS focusing of the neutrino fluxes into
the polar direction.  At the PNS surface, a strong velocity shear,
$\partial_{r} v^{\phi}$, winds up the magnetic field into a very
strong toroidal component.  Hence, the helical magnetic field lines
threading the jet have a very low pitch angle.  Thus, the gas forced
by the strong field to follow the field lines, has a predominantly
toroidal velocity and orbits the rotational axis several times while
it propagates upward, therefore spending a long time exposed to the
highest neutrino fluxes.  We quantify the relative importance of the
radial propagation and the releptonization by comparing two time
scales, $\tau_{\mathrm{rad}} = r / v^r$ for the radial expansion, and
$\tau_{\Ye} = \Ye / q_{\Ye}$ for the releptonization ($q_{\Ye}$ is the
source term for the electron fraction due to neutrino reactions).  The
right half of the figure shows the ratio
$\tau_{\mathrm{rad}} / \tau_{\Ye}$ on the streamlines.  In the jet
base, the rapid rise of $Y_e$ corresponds to
$\tau_{\mathrm{rad}} / \tau_{\Ye} > 1$.  Outside the radial shear
layer in the vicinity of the shear layer, the pitch angle is larger,
leading to faster expansion and diminishing the role of neutrino
reactions \wrt the radial motion.  Thus, the electron fraction
maintains the high value attained during the first few km of its
propagation.

\begin{figure}
  \centering
  \begin{tikzpicture}
    \pgftext{\vbox{
        \includegraphics[width=\linewidth]{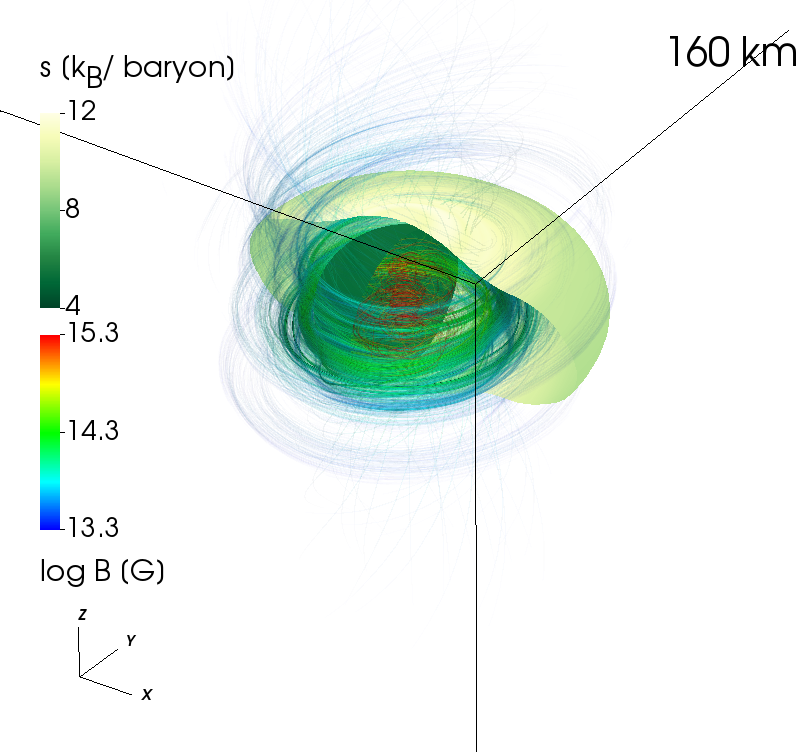}
        \includegraphics[width=\linewidth]{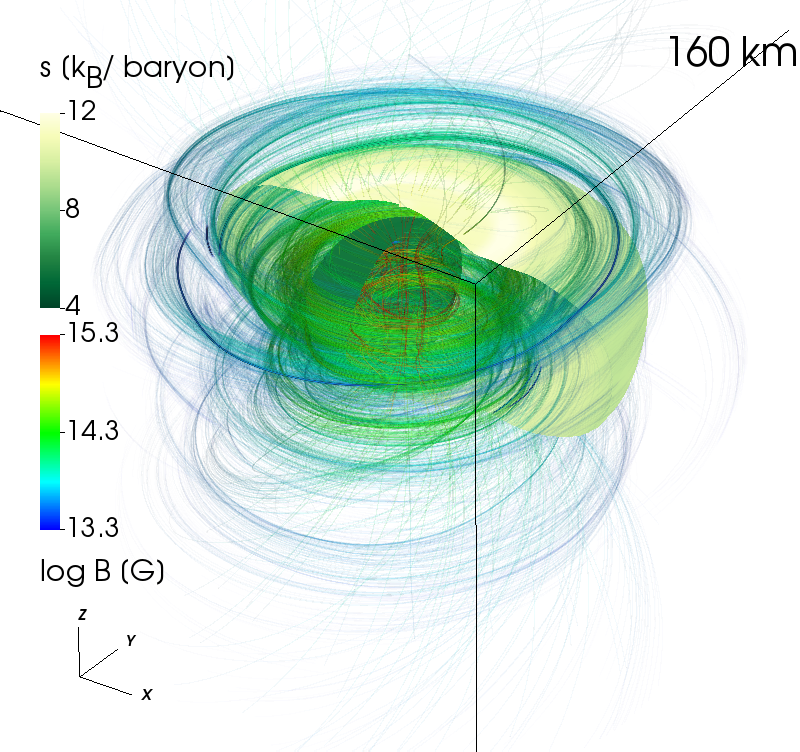}
      }
    }

  \end{tikzpicture}
  \caption{
    Same as \figref{Fig:Rw-PNS3d}, but for  \mRp ($\tpb \approx 1.45 \,
    \sek$, top) and \mRs ($\tpb \approx 1.04 \, \sek$, bottom).
  }
  \label{Fig:PS-PNS}
\end{figure}

The PNSs of the two models is prolate, albeit less than in \mRO at
late times due to their lower rotational energies
(\figref{Fig:35OC-Rp3-rad}\panel{b}, \panel{d} and
\figref{Fig:35OC-Rs-rad}\panel{b}, \panel{d}).  The polar radii
continuously contract to $R_{\mathrm{PNS, pol}} \approx 25 \, \km$
over the course of the simulation.  This process is accompanied by
shrinking equatorial radii.  The PNSs possess a moderate
pole-to-equator axis ratio of up to $1:1.6$ (\Rp) and $1:2.5$ (\Rs),
but decreasing towards the end of the simulation.  The quadrupole
coefficient achieves peak values around
$\ashlm{2}{0} \approx -0.06 \ashlm{0}{0}$ (\Rp) and
$\ashlm{2}{0} \approx -0.13 \ashlm{0}{0}$ (\Rs).  Both measures of the
asymmetry decrease during the last few hundred ms of the simulation
(more pronouncedly in \mRs) as the contraction slows down along the
rotational axis while continuing at a higher rate in the equatorial
region.  We note that the rotational axes of the PNSs remain aligned
with the original axis and do not change in a similar way as in \mRw.

Both PNSs are strongly magnetized with maximum field strengths of up
to $b^{\mathrm{max}} \lesssim \zehn{16} \, \Gauss$.  The PNS is
threaded by field lines wound up around the rotation axis with a much
stronger toroidal than poloidal component (see \figref{Fig:PS-PNS}).
The poloidal field is most notable close to the axis and in the
regions above the polar caps of the PNS from which the jets are
launched.  Both the dominance of the toroidal over the poloidal
magnetic field and the reinforcement of the poloidal component close
to the axis are in qualitative agreement with the axisymmetric results
of \citetalias{Aloy_Obergaulinger_2020__mnras_PaperII}.  From the
polar caps, a helical field extends in the surrounding gas.  In both
models, the magnetic field makes a strong contribution to the
transport of angular momentum, in particular via the helical
components in the polar regions.

\subsection{Jet stability}
\label{sSek:jetstab}

As described above, three of our models develop magnetically driven
jets.  While they differ in important properties such as the time of
explosion, the propagation speed, or the magnetization, a common
feature is their stability.  Unlike the MHD jets found in the models
of
\cite{Mosta_et_al__2014__apjl__MagnetorotationalCore-collapseSupernovaeinThreeDimensions}
and \cite{Kuroda_et_al__2020__apj__MagnetorotationalExplosionofaMassiveStarSupportedbyNeutrinoHeatinginGeneralRelativisticThreeDimensionalSimulations}, once generated our jets propagate outwards at
high speeds without being disrupted by non-axisymmetric
instabilities.  While we could not follow our models until the jets
have entered the outer layers of the stars or until break-out
from the stellar surface, it stands to reason that our models would
produce much more asymmetric and polar explosions than the ones found
by these authors, for which the jets are quenched after a comparably
short distance and a more roundish explosion ensues.  The following
subsection is dedicated to an inquiry into this difference.

We first point out several of the many physical and numerical
differences between our simulations and the others.  Among them, the
most important may be the following:
\begin{itemize}
\item numerical grid: our simulations were performed on spherical
  grids, whereas the other authors used Cartesian coordinates and an
  adaptive mesh refinement;
\item though as a consequence, it is difficult to compare the grid
  resolutions, we note that our simulations employ a grid that is
  finer ($\Delta r = 500 \, \mtr$) at the centre, but has a coarser
  resolution of $\Delta r \approx 3.5 \, \km$ near the outer edge of
  the region where the analysis of kin modes was performed by
  \cite{Mosta_et_al__2014__apjl__MagnetorotationalCore-collapseSupernovaeinThreeDimensions,Kuroda_et_al__2020__apj__MagnetorotationalExplosionofaMassiveStarSupportedbyNeutrinoHeatinginGeneralRelativisticThreeDimensionalSimulations};
\item the simulations of both
  \cite{Mosta_et_al__2014__apjl__MagnetorotationalCore-collapseSupernovaeinThreeDimensions}
  and
  \cite{Kuroda_et_al__2020__apj__MagnetorotationalExplosionofaMassiveStarSupportedbyNeutrinoHeatinginGeneralRelativisticThreeDimensionalSimulations}
  were run in full general relativity rather than using an approximate
  GR potential in special relativistic simulations as in our case
  (however the instabilities found in the jets by other authors happen
  in a range of radii where the GR effects are small); 
\item the three works employ three different approaches to neutrino
  transport:
  \cite{Mosta_et_al__2014__apjl__MagnetorotationalCore-collapseSupernovaeinThreeDimensions}
  used a leakage scheme, whereas
  \cite{Kuroda_et_al__2020__apj__MagnetorotationalExplosionofaMassiveStarSupportedbyNeutrinoHeatinginGeneralRelativisticThreeDimensionalSimulations}
  and we performed the simulations with a two-moment scheme, though
  with differences to ours at the level of both neutrino-matter
  interactions and energy-coupling terms depending on velocity and
  gravity;
\item all studies started from different pre-collapse models, in terms
  of the progenitor masses ($25 \, \Msol$, $20 \, \Msol$, and
  $35 \, \Msol$ for
  \cite{Mosta_et_al__2014__apjl__MagnetorotationalCore-collapseSupernovaeinThreeDimensions},
  \cite{Kuroda_et_al__2020__apj__MagnetorotationalExplosionofaMassiveStarSupportedbyNeutrinoHeatinginGeneralRelativisticThreeDimensionalSimulations},
  and this work, respectively) and of the initial rotational profiles
  (artificially imposed, parametrized profiles for the other two works
  and rotational profiles taken from the stellar evolution model here)
  and magnetic fields (artificially added fields or ones based on the
  progenitor models).
\end{itemize}

While strictly speaking the shorter simulation times of the other
studies limit a comparison to the first phases after bounce, we will
extend our analysis to the later phases during which the jets
propagate to many 1000 km.  Following
\cite{Mosta_et_al__2014__apjl__MagnetorotationalCore-collapseSupernovaeinThreeDimensions,Kuroda_et_al__2020__apj__MagnetorotationalExplosionofaMassiveStarSupportedbyNeutrinoHeatinginGeneralRelativisticThreeDimensionalSimulations},
we compute, after mapping to Cartesian coordinates $x,y,z$, the
position of the barycentre of the magnetic pressure as a function of
vertical coordinate, $z$, and time, $t$:
\begin{equation}
  \label{Gl:bbary}
  \xi_{\mathrm{b}}  =
  \left(
    \int_{\mathcal{A}(z)} \mathrm{d} A \, \xi b^2
  \right)
  \left(
    \int_{\mathcal{A}(z)} \mathrm{d} A \, b^2
  \right)^{-1},
\end{equation}
where $\xi$ stands for $x$ and $y$ and the integration surface
$\mathcal{A} (z)$ limits the analysis to a region around the axis
defined by the relation $\varpi < \max ( |z|/5, 50 \, \km)$.  We
further set $\vpb = \sqrt{x_{{\mathrm{b}}}^2 + y_{{\mathrm{b}}}^2}$.

\cite{Mosta_et_al__2014__apjl__MagnetorotationalCore-collapseSupernovaeinThreeDimensions}
found an exponential growth of the displacement of the barycentre from
its initial position on the rotational axis at heights below 100 km
already during the first 20 ms.  Besides cases in which these modes
prevented the jets from propagating out, the model sets of
\cite{Moesta_et_al__2018__apj__r-processNucleosynthesisfromThree-dimensionalMagnetorotationalCore-collapseSupernovae,Halevi_Moesta__2018__mnras__r-Processnucleosynthesisfromthree-dimensionaljet-drivencore-collapsesupernovaewithmagneticmisalignments}
include a case with a very strong initial magnetic field (their model
B13) in which kink modes grow to a sufficient amplitude such as to
deform the jet, yet do not become strong enough to quench it.
\cite{Kuroda_et_al__2020__apj__MagnetorotationalExplosionofaMassiveStarSupportedbyNeutrinoHeatinginGeneralRelativisticThreeDimensionalSimulations}
found an increase of $\vpb$ in a similar range of times and positions.
This amplification was attributed to kink-mode instabilities growing
in the outflow on timescales of the order of 1 ms and vertical
lengthscales of few km.

\begin{figure*}
  \centering
  \begin{tikzpicture}
    \pgftext{\hbox{
        \includegraphics[width=0.33\linewidth]{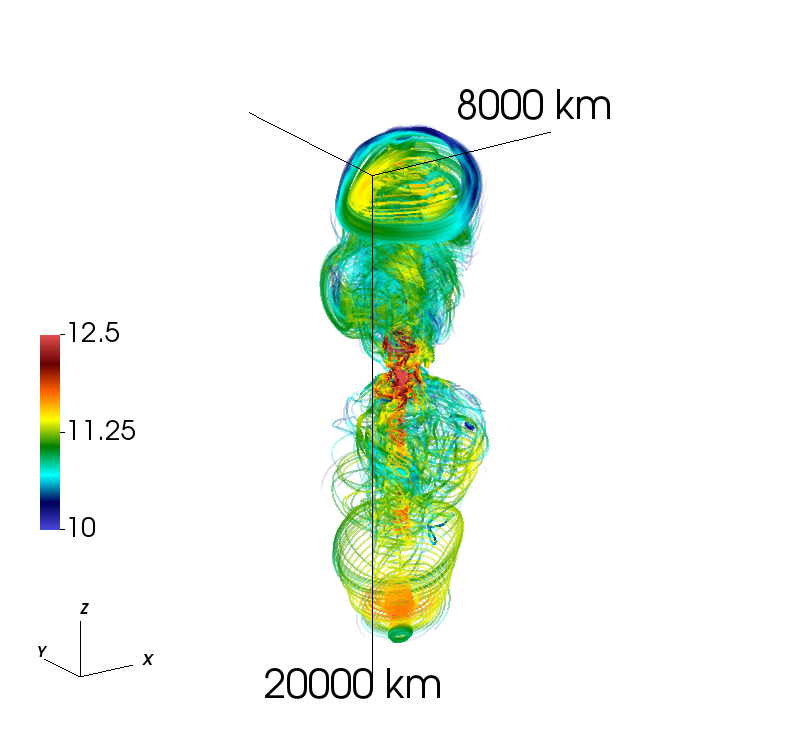}
        \includegraphics[width=0.33\linewidth]{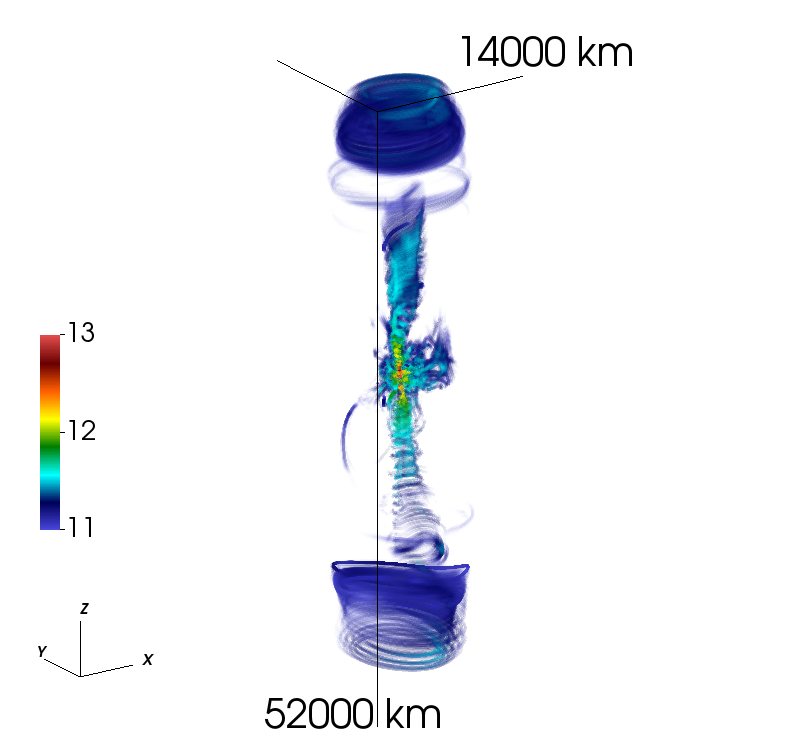}
        \includegraphics[width=0.33\linewidth]{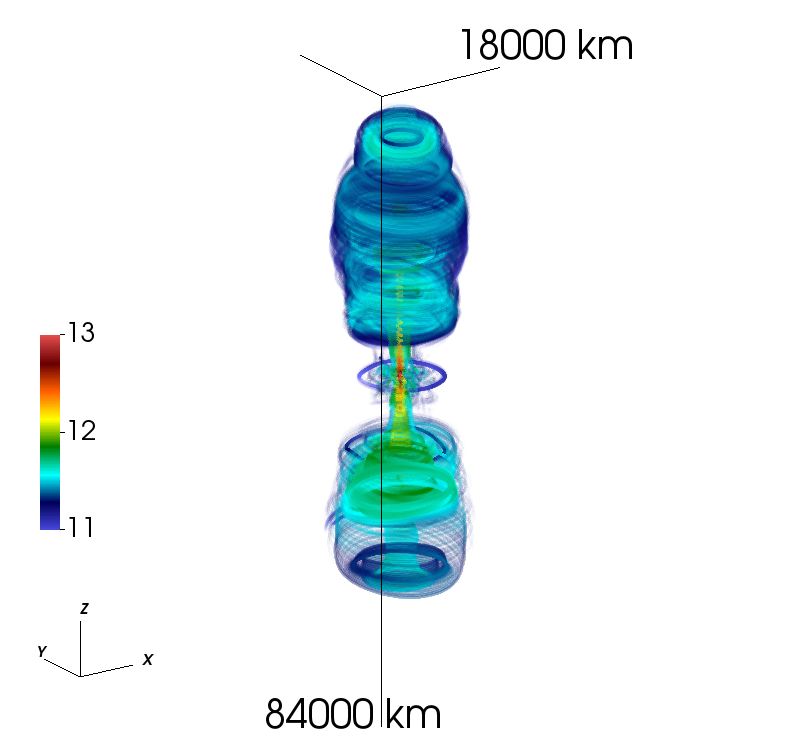}
      }
    }
    \node[fill=white, opacity=1, text opacity=1] at (-8.38,2.18) {(a)};  
    \node[fill=white, opacity=1, text opacity=1] at (-2.38,2.18) {(b)};  
    \node[fill=white, opacity=1, text opacity=1] at (3.58,2.18) {(c)};  
  



    \node[fill=white, opacity=1, text opacity=1] at (-8.15,0.68) {\sffamily\scriptsize $\log\:B\,[10^{10}\text{G}]\qquad\text{   }$};
    \node[fill=white, opacity=1, text opacity=1] at (-2.25,0.68) {\sffamily\scriptsize $\log\:B\,[\text{G}]\qquad\text{   }$};
    \node[fill=white, opacity=1, text opacity=1] at (3.70,0.56)  {\sffamily\scriptsize $\log \:B\,[\text{G}]\qquad\text{   }$};

  \end{tikzpicture}  
  \caption{
    Field lines in the jets of models \RO, \Rp, and \Rs (left to right) at
    the end of the simulations with the colour indicating the field strength.
  }
  \label{Fig:Jetflines}
\end{figure*}

\begin{figure}
  \centering
  \includegraphics[width=\linewidth]{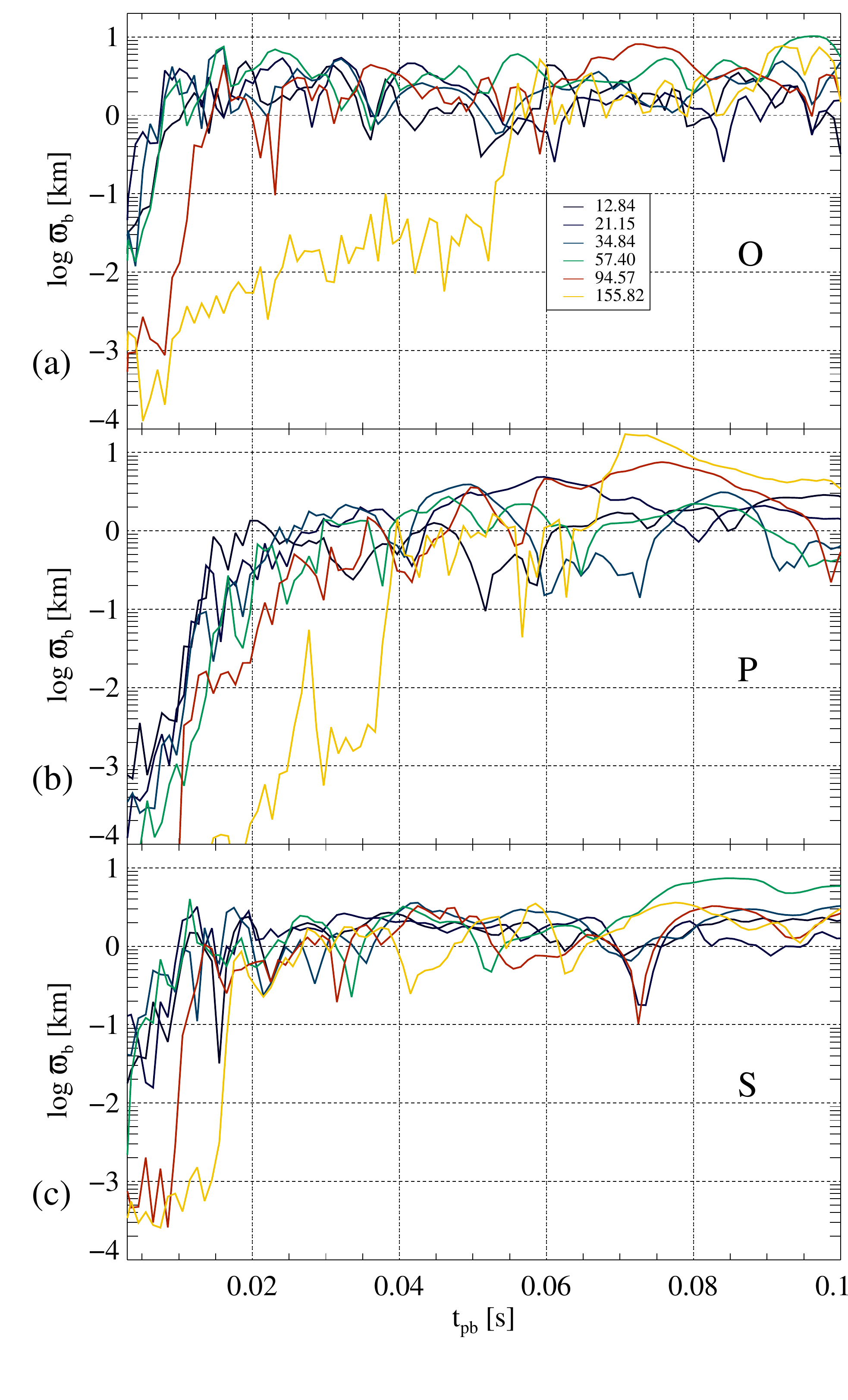}
  \caption{
    Time evolution of $\vpb$ for (from top to bottom)
    models \RO, \Rp, \Rs at various heights in the northern hemisphere
    as indicated in the legend.
  }
  \label{Fig:jet-magbary-t}
\end{figure}

\begin{figure*}
  \centering
  \includegraphics[width=0.33\linewidth]{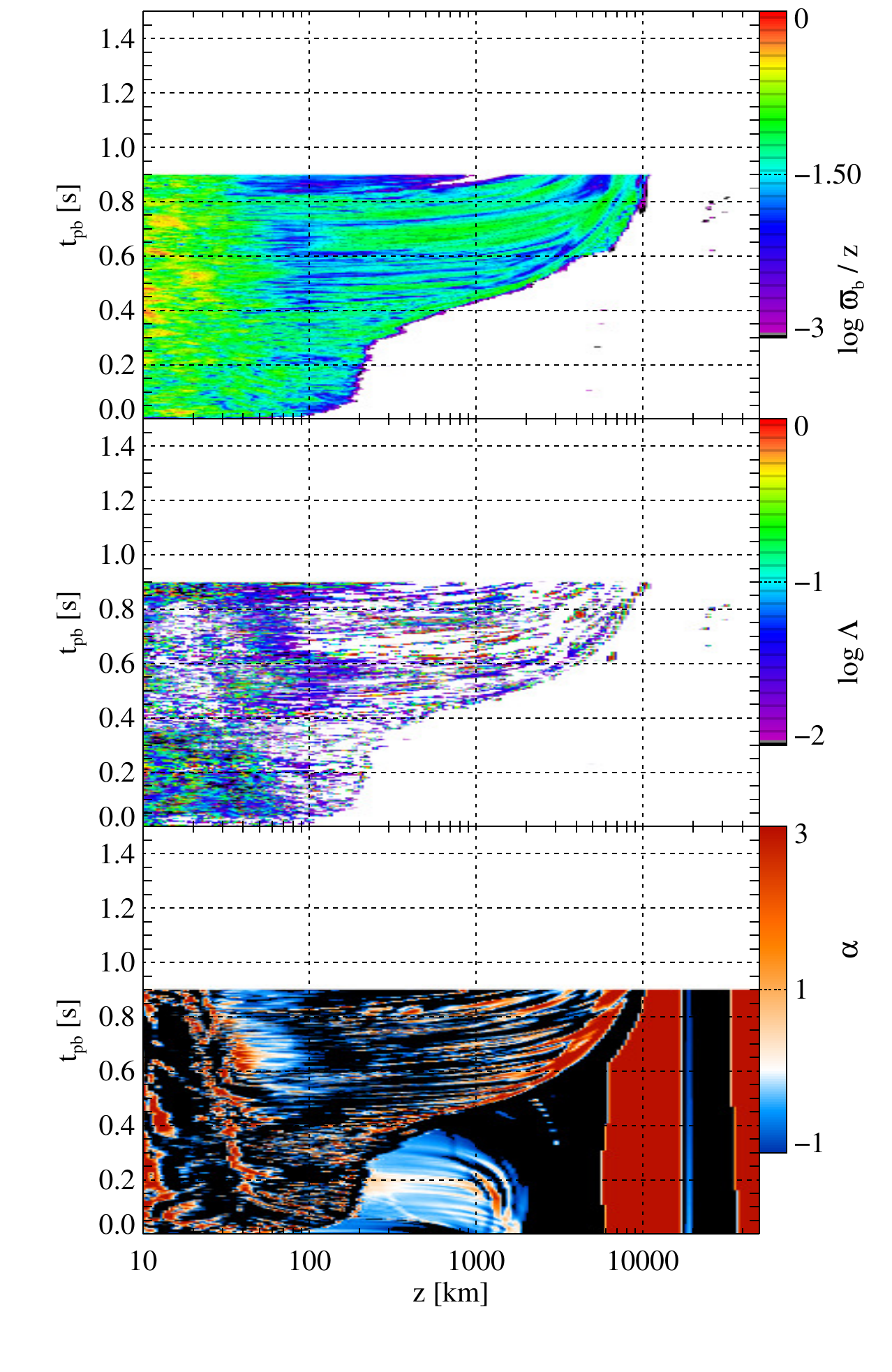}
  \includegraphics[width=0.33\linewidth]{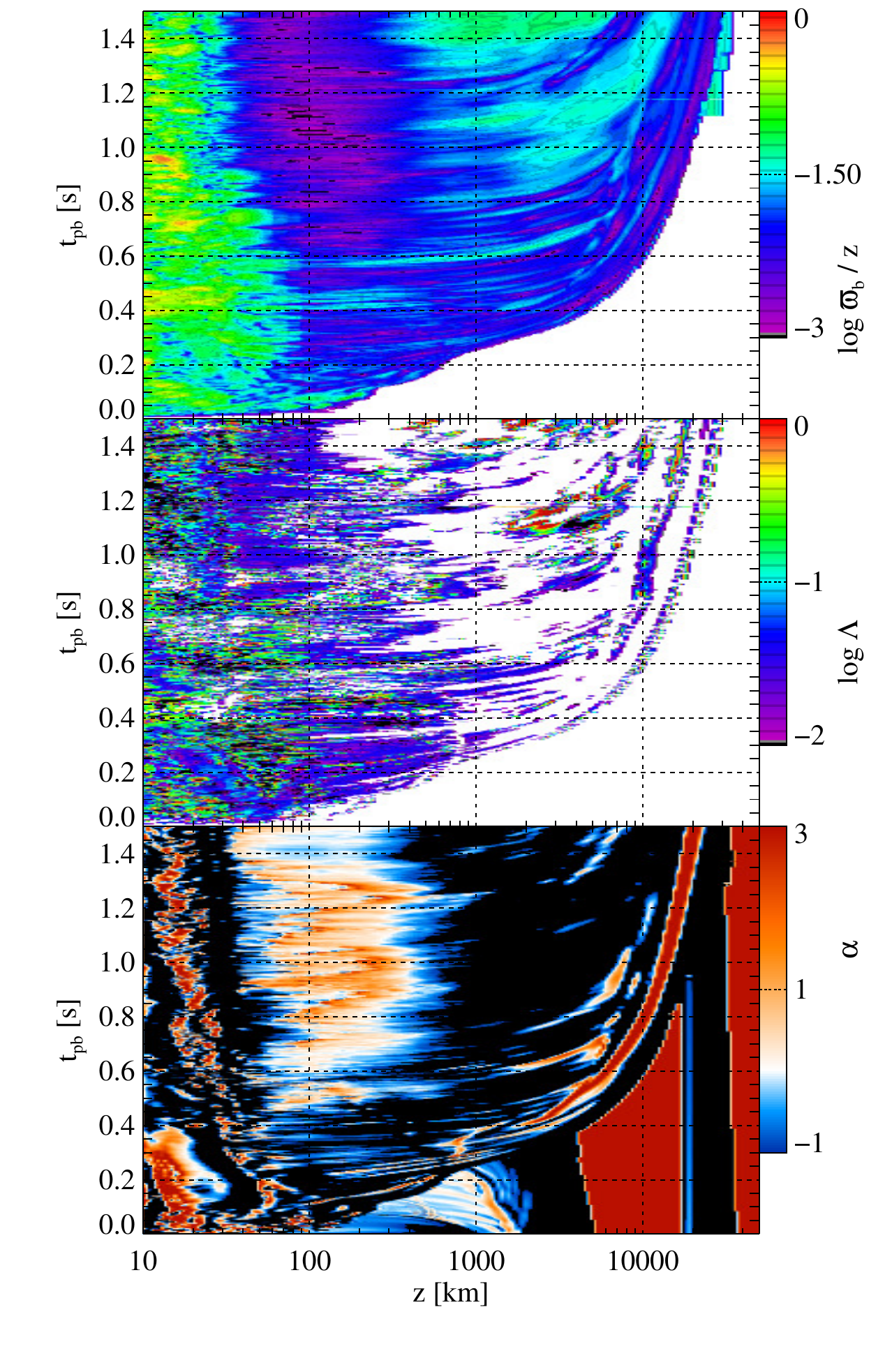}
  \includegraphics[width=0.33\linewidth]{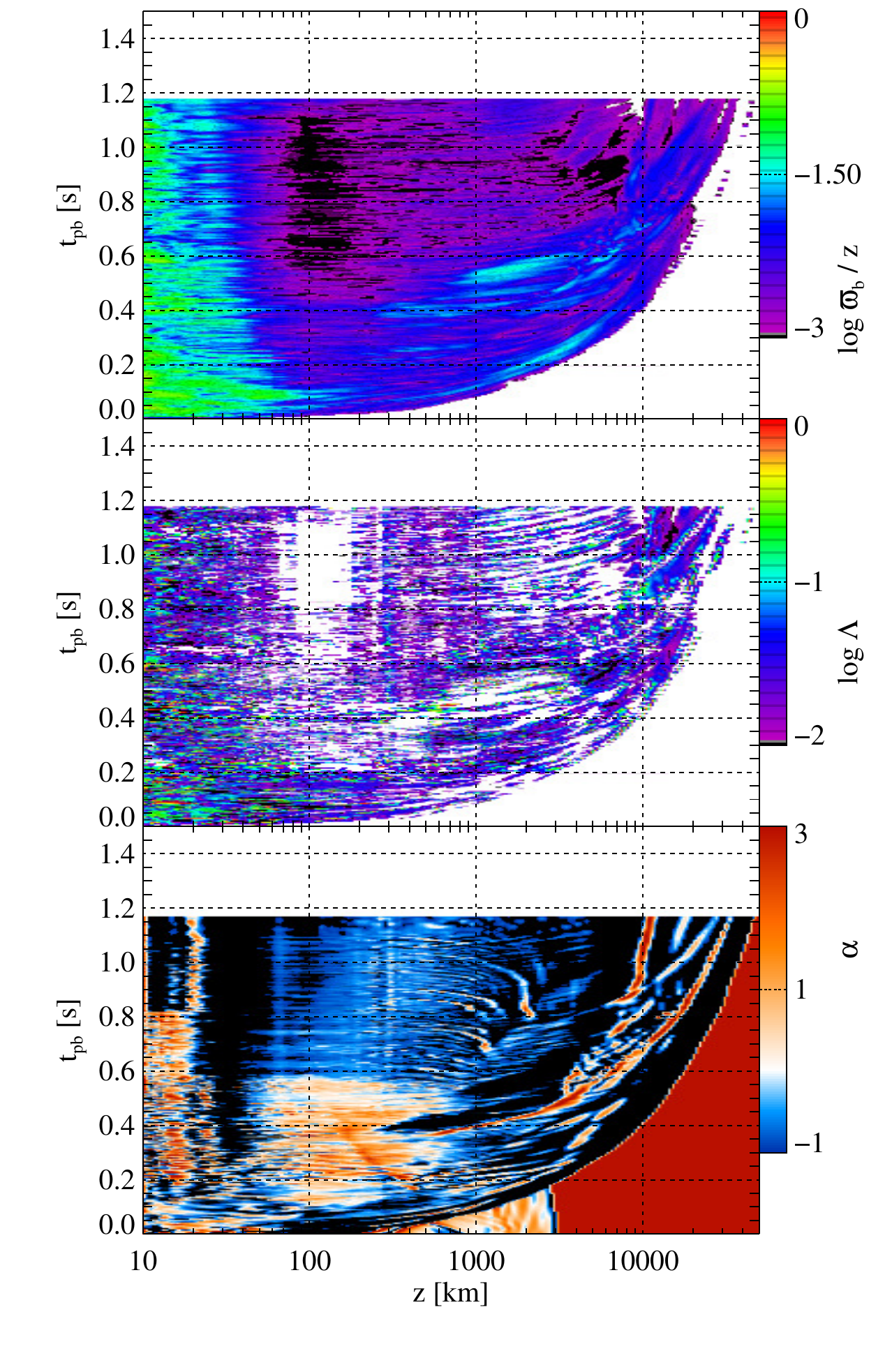}
  \caption{
    Space-time plots of the northern jets of models \RO, \Rp, and \Rs
    (left to right).  The panels show the relative displacement of
    the magnetic barycentre, $\vpb / |z|$ (top), the ratio of kink
    timescale and dynamic time, $\Lambda$, (middle), and the magnetic
    shear parameter, $\alpha$, (bottom) as functions of time and
    height.
  }
  \label{Fig:jet-magbary-2d}
\end{figure*}

We first summarize important results regarding non-axisymmetric
instabilities of the outflows in our simulations.  We show the
structure of the magnetic field of these three models at late times in
\figref{Fig:Jetflines}.  In a striking difference to most of the
aforementioned results, the jets launched by models \mnRO, \mnRp, and
\mnRs close to the PNS propagate over large distances without being
disrupted by strong instabilities.  We note that the propagation speed
of the jet of \mRO, $v_{\mathrm{jet}} \approx \zehnh{2}{9} \, \cms$,
as well as the overall geometry of the jets are comparable to the
results of
\cite{Mosta_et_al__2014__apjl__MagnetorotationalCore-collapseSupernovaeinThreeDimensions}.
For the strongest fields of \mnRp and \mnRs, the
jet beam coincides with a column of helical field roughly aligned with
the rotational axis.  The magnitude of deviations of these columns
from the axis is anti-correlated with the field strength with \mRO and
\mnRs showing the largest and smallest displacements, respectively.
This tendency agrees with the results of the most magnetized model of
\cite{Halevi_Moesta__2018__mnras__r-Processnucleosynthesisfromthree-dimensionaljet-drivencore-collapsesupernovaewithmagneticmisalignments,Halevi_Moesta__2018__mnras__r-Processnucleosynthesisfromthree-dimensionaljet-drivencore-collapsesupernovaewithmagneticmisalignments}
with an initial field strength in a similar range as our models \Rp
and \Rs showing an evolution similar to our models.

As mentioned above, the analyses of
\cite{Mosta_et_al__2014__apjl__MagnetorotationalCore-collapseSupernovaeinThreeDimensions}
and
\cite{Kuroda_et_al__2020__apj__MagnetorotationalExplosionofaMassiveStarSupportedbyNeutrinoHeatinginGeneralRelativisticThreeDimensionalSimulations}
concentrate on the immediate post-bounce phase ($t \le 30 \, \ms$) and
low radii ($r \le 74 \, \km$), i.e. a regime in which the outflow has
not yet fully developed.  Hence, they are dealing with the
instabilities of the magnetic field near the centre of the core
affecting the evolution in the first several tens of kilometres of
their propagation.  We will start our look at the dynamics in a
similar regime.  For a quantitative evaluation, we point to
\figref{Fig:jet-magbary-t} displaying the time evolution of $\vpb$ for
all three models with jets during the first 0.1\,s and at various
heights up to $z = 156 \, \km$.\footnote{Here and in the following,
  results are presented for the northern hemisphere.  We note that
  they equally apply to the southern hemisphere.}  During this period,
the displacement of the barycentre grows on timescales of few ms to
values of $\vpb \gtrsim 1 \, \km$.  The rapid increase sets in after
the shock has passed a given location, \ie almost immediately after
bounce for all lines shown in \figref{Fig:jet-magbary-t} except for
the yellow ones ($z \approx 156 \, \km$) for which the growth is
delayed as the shock takes several tens of ms to reach this height.
We note that the red and yellow lines, respectively corresponding to
$z \approx 95\,\km$, and $156 \, \km$, are already beyond the analysis
heights of
\cite{Mosta_et_al__2014__apjl__MagnetorotationalCore-collapseSupernovaeinThreeDimensions}
and
\cite{Kuroda_et_al__2020__apj__MagnetorotationalExplosionofaMassiveStarSupportedbyNeutrinoHeatinginGeneralRelativisticThreeDimensionalSimulations}
and, most importantly, far outside the PNS.  After the shock wave has
passed, the growth of the barycentre deviation occurs on scales of
$\sim 1\,$ms, similar to the results of
\citeauthor{Mosta_et_al__2014__apjl__MagnetorotationalCore-collapseSupernovaeinThreeDimensions}.

\MRO shows the non-magnetically driven $m = 1$ shock deformations also
found in the least magnetised \mRw, albeit at a lower amplitude.
Disentangling the effects of this non-magnetic instability from
similar $m=1$ modes originated by magnetic effects on the evolution of
$\vpb$ is difficult during this phase.  The increase is similar,
though slightly slower, for \mRp.  \MRs, which launches a prompt
jet-like explosion, presents a very rapid increase of $\vpb$ behind
the shock front, which nevertheless levels off at values
$\vpb\sim$\,few kilometres (similarly to \Modls{O} and \Rp).  The
amplitude as well as the growth times and the location at which the
non-axisymmetric modes grow are similar to the simulations of
\cite{Mosta_et_al__2014__apjl__MagnetorotationalCore-collapseSupernovaeinThreeDimensions}
and
\cite{Kuroda_et_al__2020__apj__MagnetorotationalExplosionofaMassiveStarSupportedbyNeutrinoHeatinginGeneralRelativisticThreeDimensionalSimulations}.

In contrast to the results of
\cite{Mosta_et_al__2014__apjl__MagnetorotationalCore-collapseSupernovaeinThreeDimensions}
and
\cite{Kuroda_et_al__2020__apj__MagnetorotationalExplosionofaMassiveStarSupportedbyNeutrinoHeatinginGeneralRelativisticThreeDimensionalSimulations},
the early development of non-axisymmetric modes does not quench the
jets.  Indeed, our results qualitatively coincide with the ones of
\cite{Bromberg_Tchekhobskoy_2016MNRAS.456.1739}, who find that
relativistic magnetized jets propagating in collapsar progenitors are
relatively immune to \emph{global} kink modes and, hence, able to
maintain their stability well beyond the breakout through the stellar
surface.  Nevertheless, all jets show moderate deviations from
axisymmetry as they propagate through the star \citep[as predicted by
the analytic work of][]{Appl_2000AaA...355..818}.  For the evolution
over a wider range of times and heights, we refer to
\figref{Fig:jet-magbary-2d} (\banel{top} panels).  The importance of
non-axisymmetric modes, expressed in terms of the relative
displacement $\vpb / |z|$, correlates inversely with the power of the
jets in a manner similar to the results of
\cite{Moesta_et_al__2018__apj__r-processNucleosynthesisfromThree-dimensionalMagnetorotationalCore-collapseSupernovae,Halevi_Moesta__2018__mnras__r-Processnucleosynthesisfromthree-dimensionaljet-drivencore-collapsesupernovaewithmagneticmisalignments}:
\begin{itemize}
\item In the case of \mRO, the jets experience strong corrugation of
  their shape with $\vpb \sim \mathcal{O} (10^{-1}) z$ along their
  entire extension.
\item While $\vpb / |z|$ peaks at similar magnitudes, \mRp is
  characterized by lower values at the base of the jet at heights
  between the region where the jet is accelerated outside the PNS at
  several tens of km and about 200\,km as well as near the jet head.
\item This tendency of weaker deformations continues to the most
  energetic jets in \mRs, for which the deviation only occasionally
  reaches the levels of the other models.
\end{itemize}

The estimates for the time and wavelength of the fastest growing modes
in
\cite{Mosta_et_al__2014__apjl__MagnetorotationalCore-collapseSupernovaeinThreeDimensions}
are based on the approximations obtained for cylindrical jets
\citep[e.g.][]{Begelman__1998__apj__InstabilityofToroidalMagneticFieldinJetsandPlerions}. However,
the jets in our models develop a narrow angle conical shape.  The
growth of kink instabilities in conically expanding jets was
investigated by, e.g.,
\cite{Moll_et_al__2008__aap__Kink_instabilities_in_jets_from_rotating_magnetic_fields}.
Despite several differences, their setup and ours are sufficiently
similar for their results to guide our analysis.  Following them, we
define the magnetic pitch as \citep[note the extra factor $2\upi$ with
respect to commonly used definitions in the literature,
e.g.][]{Bodin_1980NucFu..20.1255,Appl_2000AaA...355..818,Bodo_2019MNRAS.485.2909}
\begin{equation}
  \label{Gl:pitch}
  H = 2 \upi \varpi
  \left| \frac{b^r}{b^{\phi}}\right|
\end{equation}
(note that $b^r$ is the radial component of the magnetic field in
spherical coordinates; in our simulations $b^r \approx b^z$)
and the \Alfven crossing time as the time taken by an \Alfven wave to
orbit the jet axis,
\begin{equation}
  \label{Gl:jet-Alfvencrossing}
  \tau_{\mathrm{c}} = \frac{2 \upi \varpi}{c_{\mathrm{A}}^{\phi} - 2 \upi v^{\varpi}}.
\end{equation}
The second term in the denominator of \Eqref{Gl:jet-Alfvencrossing}
accounts for the potential widening of a conical jet, which increases
the travel time of an \Alfven wave along the field lines towards
higher radii.  We note that the estimation of
\Eqref{Gl:jet-Alfvencrossing} is non-relativistic, but sufficient for
our jets, which are only mildly relativistic and, hence, the Lorentz
factor that should multiply the former expression
\citep[e.g.][]{Lyubarskij_1992SvAL...18..356} is approximately one and
has been dropped for the sake of simplicity.  The \Alfven crossing
time is
finite only for
\begin{equation}
  \label{Gl:jet-Alfven-v-crit}
 \frac{ c_{\mathrm{A}}^{\phi}} { v^{\varpi} } > 2 \upi.
\end{equation}
It roughly sets the timescale on which kink modes grow and, thus,
$\tau_{\rm kink}\sim \tau_{\rm c}$.  As noted by
\cite{Moll_et_al__2008__aap__Kink_instabilities_in_jets_from_rotating_magnetic_fields},
the condition in \eqref{Gl:jet-Alfven-v-crit} can be regarded as a
criterion for instability, which may be modified by additional effects
such as the differential rotation of the jet, and the magnetic shear
created by it, may further suppress the growth of kink modes.  This
shear can be quantified in terms of
$\alpha = \frac{d\ln b^{\phi}}{d\ln \varpi}$.  According to
\cite{Begelman__1998__apj__InstabilityofToroidalMagneticFieldinJetsandPlerions}
(eq. 4.2), instability of a mode with vertical wave number $k$ may set
in if $\alpha > \frac{1}{2}( m + k \varpi \frac{b^z}{b^\phi})^2-1$,
with $m=\pm 1$.  Irrespective of $k$, $\alpha > -1$ is an absolute
minimum for the instability.
  
We note, furthermore, that heuristic arguments and numerical evidence
shows that the full development of kink modes, such that the jet is
significantly deformed requires $(5-10)\tau_{\rm kink}$
\citep[e.g.][]{Mizuno_2012ApJ...757...16,Mignone_2013MNRAS.436.1102}.
If \Eqref{Gl:jet-Alfven-v-crit} is fulfilled, kink modes grow on
length scales $\lambda_{\mathrm{kink}} > H$.  Otherwise, the jet
expands too fast for an \Alfven wave to orbit its centre and, thus,
for kink modes to grow.  Taking into account the vertical propagation
of matter in the jet, one may derive the following additional
criterion of instability
\citep{Bromberg_Tchekhobskoy_2016MNRAS.456.1739}
\begin{equation}
  \label{Gl:BTinst}
  \Lambda := \frac{\Xi \tau_{\mathrm{kink}}}{t_{\mathrm{dyn}}} \le 1,
\end{equation}
where $\Xi = 5...10$ is a numerical parameter and
$t_{\mathrm{dyn}} = z/v^z $ is the dynamical timescale for the
expansion of the jet.  In general, this criterion will be more
restrictive than the one stated by Eq.\,\eqref{Gl:jet-Alfven-v-crit}.

In our collimated outflows neither the velocity inside of the jet is
ultrarelativistic, nor the magnetic field dominates the dynamics so
that they would be force-free. Hence, the whole jet is causally
connected in the direction perpendicular to its axis and the causal
restrictions for the growth of kink modes
\citep[e.g.][]{Porth_2015MNRAS.452.1089} are not relevant in our
analysis.

The middle and bottom subpanels of \figref{Fig:jet-magbary-2d} display
the value of $\Lambda$ in regions where it is below unity, \ie, where
an instability is possible, and the magnetic shear parameter $\alpha$,
respectively.  During the first few tens of ms after bounce and at
radii inside the PNS, \Alfven crossing times are around
$\tau_{\mathrm{c}} \sim 1 \, \ms$ and magnetic pitches are in the
range of $H \gtrsim \mathcal{O} (1) \, \km$.  Furthermore,
$\Lambda < 1$ and $\alpha \gtrsim 0$ are indicative of an instability.
These values are consistent with the observed growth we find in this
phase of the three models.  However, the fact that the PNS is
potentially unstable to kink modes does not necessarily mean that the
outflow generated above the PNS surface may be destroyed by magnetic
kinks.

We turn our attention towards the propagation of the outflows outside
of the PNS. The relatively strong kink amplitudes which \mRO
experiences during later phases ($t \gtrsim 0.3 \, \sek$ and
$|z| \gtrsim 100 \, \km$) grow from already rather large seeds at
their base as the jets pass through a layer in which the \Alfven
crossing times are small, allowing for a fast growth of the
instabilities.  We find values of
$\tau_{\mathrm{c}}^{-1} \sim \mathcal{O} (10^3 \, \sek^{-1})$ around
$z \sim 100 \, \km$.  The magnetic pitch varies strongly across the
unstable region between $H \sim \, 100 \km$ and
$H \gtrsim 1000 \, \km$.  Both instability criteria are fulfilled at
the bottom of this region with $\Lambda < 1$ (blue regions below 100
km) and $\alpha \sim 0$ (red-blue region). Further out at
$|z| \gtrsim 200 \,\km$, the growth of kink modes seems to be
unimportant as the normalised barycentre displacement $\vpb / z$ does
not grow further. In this region, the \Alfven crossing times are
longer and $\Lambda$ and $\alpha$ no longer fulfil the criteria for
instability.

In \mRp, the jets develop non-axisymmetric modes in situ at
$|z| \gtrsim 200 \, \km$ from small perturbations of only
$\vpb \sim 1...10 \, \km$ at their bottom.  From these values, $\vpb$
grows over the first $\sim 1000 \, \km$ of propagation, but not to a
point where the perturbations would disrupt the jets.  In the
transition region where the barycentre displacement increases, we find
short \Alfven crossing times corresponding to
$\tau_{\mathrm{c}} ^ {-1} > 100 \,\sek^{-1}$ and long pitches,
$H \sim 300...1000 \, \km$, indicating the possibility of range of
growth rates and unstable modes.  The data for $\Lambda$ and $\alpha$
are rather noisy, but show a tendency towards a growing stability
against kink modes as the jet progresses through the star in \mRp.
The magnetic shear confines the unstable region to the immediate
vicinity of the $z$-axis offering a possible explanation for the
limited growth of kink modes in this model.

\MRs shows a similar picture, though more extreme than in \Rp.  The
jets are subject to very minor deviations from axisymmetry.
Potential regions of fast amplification of kink modes can be found at
their base with \Alfven crossing times in the range of milliseconds
and a wide range of magnetic pitches.  The jets are faster than in any
of the other models, which, together with magnetic shear expressed in
the low values of $\alpha < 0$, with may suppress the
instabilities.

To summarize, we find similarities to the works by
\cite{Mosta_et_al__2014__apjl__MagnetorotationalCore-collapseSupernovaeinThreeDimensions,Moesta_et_al__2018__apj__r-processNucleosynthesisfromThree-dimensionalMagnetorotationalCore-collapseSupernovae,Halevi_Moesta__2018__mnras__r-Processnucleosynthesisfromthree-dimensionaljet-drivencore-collapsesupernovaewithmagneticmisalignments}
and
\cite{Kuroda_et_al__2020__apj__MagnetorotationalExplosionofaMassiveStarSupportedbyNeutrinoHeatinginGeneralRelativisticThreeDimensionalSimulations}
in the growth of $\vpb$ during the early phases of the explosion, but
also a very different evolution thereafter.  Though this phase leads
to similar amplitudes of $\vpb$, the jets are not quenched, but, in
cases of delayed as well as prompt explosions, manage to break out off
the inner core to then propagate over a long distance with only a
minor to moderate influence of non-axisymmetric modes.  We note that
our results are at least qualitatively consistent with an analysis of
the growth of kink modes following
\cite{Moll_et_al__2008__aap__Kink_instabilities_in_jets_from_rotating_magnetic_fields}
and \cite{Bromberg_Tchekhobskoy_2016MNRAS.456.1739}.  The fact that
the nascent jets are able to survive relatively strong deformations is
most apparent for \mRO, in which they are generated at a relatively
late time and in the interior of a stalled shock wave which itself is,
even in the absence of any kink modes, dominated by strong deviations
from axisymmetry.  Although the exact conditions for such modes,
created externally or growing in the jet itself, to destroy the highly
collimated outflow have to be explored further, we can put forward a
tentative explanation.  We attribute the strong stability of the jets
outside the PNSs mostly to a stabilising profile of the toroidal
magnetic field with cylindrical radius with low and, in large regions,
negative values of $\alpha$. Further stabilisation may be provided by
the fast propagation of the jets, leading to a high ratio between
potential kink timescales and dynamic times.

Understanding potential reasons for the aforementioned differences to
previous work on the topic requires a more detailed comparison of
numerical and physical characteristics of the different models.  Among
the former, we point towards the different grid structures and
resolutions.  Our models were run on spherical grids with a radial
resolution that is finest at the centre and decreases towards larger
radii such that at several tens of kilometres, our grid cells are
larger than those of the Cartesian AMR models of
\cite{Mosta_et_al__2014__apjl__MagnetorotationalCore-collapseSupernovaeinThreeDimensions,Moesta_et_al__2018__apj__r-processNucleosynthesisfromThree-dimensionalMagnetorotationalCore-collapseSupernovae,Halevi_Moesta__2018__mnras__r-Processnucleosynthesisfromthree-dimensionaljet-drivencore-collapsesupernovaewithmagneticmisalignments,Kuroda_et_al__2020__apj__MagnetorotationalExplosionofaMassiveStarSupportedbyNeutrinoHeatinginGeneralRelativisticThreeDimensionalSimulations}.
We note that we apply a mesh coarsening scheme at the $z$-axis in
order to circumvent the timestep restriction of spherical coordinates.
This scheme ensures that all grid cells have approximately equal
widths in the three coordinate directions, which makes them similar to
those of a Cartesian mesh.  Concerning aspects of the physics of the
models, the tendency of stronger magnetic fields to produce jets less
affected by kinks modes, seen by
\cite{Moesta_et_al__2018__apj__r-processNucleosynthesisfromThree-dimensionalMagnetorotationalCore-collapseSupernovae,Halevi_Moesta__2018__mnras__r-Processnucleosynthesisfromthree-dimensionaljet-drivencore-collapsesupernovaewithmagneticmisalignments}
as well as in our models suggests a strong impact of the field
strength and geometry.  Furthermore, the structure of the progenitor
stars, different for the different works cited here and ours, will
affect the ram pressure the jets to overcome.  This, in turn, may have
in influence in the balance between kink and dynamic times and,
therefore, modify the growth of kink modes.

\section{Summary and conclusions}
\label{Sek:Concl}

We continued our previous investigations of the magnetorotational core
collapse of massive stars by performing a series of three-dimensional
simulations of possible progenitors of GRBs
\cite{Obergaulinger_Aloy__2017__mnras__Protomagnetarandblackholeformationinhigh-massstars};
\citetalias{Obergaulinger_Aloy__2020__mnras__MagnetorotationalCoreCollapseofPossibleGRBProgenitorsIExplosionMechanisms,Aloy_Obergaulinger_2020__mnras_PaperII}.
As initial model, we chose a star of zero-age main-sequence mass
$M_{\textsc{zams}} = 35 \, \msol$ evolved until the onset of core
collapse in spherical symmetry including a model for the magnetic
fields and rotation
\cite{Woosley_Heger__2006__apj__TheProgenitorStarsofGamma-RayBursts}.
To address the uncertainty of the geometry and the strength of the
magnetic field owing to the approximate nature of this model, we
computed four versions of the same progenitor star with different
field configurations.  One of them is based on the stellar-evolution
model and possesses a rather strong magnetic field with a strength of
up to $b_{\mathrm{max}} \approx \zehnh{1.3}{12} \, \mathrm{G}$ limited
to the convectively stable layers of the star (\mRO).  In another
simulation, \mRp, we explored the effect of an artificial increase of
the poloidal component, energetically subdominant in \mRO, by a global
factor of 3.  The remaining models are set up with a large-scale
dipolar field geometry normalized to two different central values,
\viz $b_{\mathrm{centre}} = \zehn{10} \, \mathrm{G}$ (\mRw) and
$b_{\mathrm{centre}} = \zehn{12} \, \mathrm{G}$ (\mRs).  All of them
correspond to axisymmetric models from
\citetalias{Obergaulinger_Aloy__2020__mnras__MagnetorotationalCoreCollapseofPossibleGRBProgenitorsIExplosionMechanisms}
and \citetalias{Aloy_Obergaulinger_2020__mnras_PaperII}.  Our
simulations were run with our numerical code combining special
relativistic MHD with a spectral two-moment neutrino transport
and including the relevant reactions between neutrinos and matter.

Our axisymmetric models confirmed the development of highly energetic,
strongly bipolar explosions driven by a combination of neutrino
heating and magnetic extraction of rotational energy from the core as
well as showing paths towards the formation of GRB progenitors driven
by proto-magnetars or collapsars.  The goal of our present study is to
scrutinize these possibilities in full three-dimensional geometry.  In
this effort, we complement previous work along similar lines done by
\cite{Scheidegger_et_al__2010__CQG__GW_from_SN_matter,Winteler_et_al__2012__apjl__MagnetorotationallyDrivenSupernovaeastheOriginofEarlyGalaxyr-processElements,Mosta_et_al__2014__apjl__MagnetorotationalCore-collapseSupernovaeinThreeDimensions,Moesta_et_al__2018__apj__r-processNucleosynthesisfromThree-dimensionalMagnetorotationalCore-collapseSupernovae,Kuroda_et_al__2020__apj__MagnetorotationalExplosionofaMassiveStarSupportedbyNeutrinoHeatinginGeneralRelativisticThreeDimensionalSimulations}
using a variety of physical approximations and numerical methods.
Among issues of the explosion mechanisms, a main question
emerging from these studies pertains to the development of
non-axisymmetric instabilities perturbing the polar outflows and
potentially disrupting them before they manage to break out of the
core.

We ran our simulations for a comparably long times of between $0.8$
and $1.5 \, \sek$ after bounce, which extends into a relatively long
phase after the four models develop an explosion.  Depending on the
initial magnetic field, several evolutionary paths are possible.

The relatively weakly magnetized \mRw produces shock revival within
about $250 \, \ms$ after bounce due to neutrino heating.  Before the
explosion, the shock wave gradually expands at low latitudes and
experiences a strong $m = 1$ spiral deformation, which is also visible
in the pattern of efficient neutrino heating and of a favourable ratio
of the timescales of advection through the gain layer and neutrino
heating.  This explosion mechanism leads to a moderately oblate
geometry of the shock wave as it propagates outwards.  Behind it,
large bubbles of hot gas expand in a stochastic geometry, in contrast
to the polar explosion of the axisymmetric version of the model.  At
late times, however, when the PNS has acquired a sufficiently strong
magnetic field, a very hot outflow of moderate magnetization emerges
from a polar region at the PNS surface and starts to catch up with the
more spherical shock wave. This outflow is highly variable and changes
its direction, sharing qualitatively some of the properties of the
jittering jets model
\citep{Papish_Soker__2011__mnras__Explodingcorecollapsesupernovaewithjitteringjets}. Whether
this is a generic feature of mildly magnetised pre-SN cores requires
further exploration with a larger grid of models with different masses
and rotational properties.

Compared to \mRw, the stronger fields of \mnRO compensate for a less
important neutrino heating such that the explosion time is very
similar.  The magnetic contribution to the explosion mechanism favours
a bipolar rather than equatorial or spherical explosion geometry.  The
model develops a pair of collimated, fast outflows along the
rotational axis that reach a radius of $r = \zehn{4} \, \km$ within
$0.5 \, \sek$ after they have been launched.  By the end of the
simulations, the diagnostic explosion energies of the two models level
off at comparable values around $\eej \approx \zehnh{5}{50} \, \erg$,
\ie.  The PNSs of both models retain high rotational energies of more
than $\zehnh{1.5}{52} \, \erg$ by the end of the simulations.  If
subsequently released by magnetic braking on longer timescales, this
energy reservoir would be sufficient to power a hypernova-like
explosion.  The simulations offer an indication of such a possibility
in the maximum of $\Erotpns$ reached after more than half a second and
the subsequent decline.  We note that the model displays an evolution
that is in terms of the shock propagation speed as well as the ejecta
morphology similar to that of
\cite{Mosta_et_al__2014__apjl__MagnetorotationalCore-collapseSupernovaeinThreeDimensions}
with its dual-lobe explosions.

The two models with the strongest fields, \mnRp and \mnRs, explode due
to magnetic fields and rotation alone.  They show only a short (\mnRp)
epoch of shock stagnation or none at all (\mnRs).  Around the axis of
the magnetic field, which is identical to the rotational axis, the
\Alfven waves pass faster through the gain layer than fluid elements
falling towards the PNS.  The corresponding strong magnetic field
accelerates the gas along the axis.  The resulting jets propagate very
rapidly, reaching distances of $r \approx \zehn{4} \, \km$ within
$0.4 \, \sek$.  The explosions energies are in excess of the canonical
value of $\zehn{51} \, \erg$ with a value of
$\eej \approx \zehnh{2}{51} \, \erg$ for \mRp and \mnRs exceeding
$\eej > \zehn{52} \, \erg$ without having converged to a final value
by the end of the simulation.  These two models have rotational
energies in the PNS significantly smaller than \modls{W} and \RO,
though still above $\zehn{52}\,\erg$ in \mRp, and
$\sim \zehnh{4}{51}\,\erg$ in \mRs. Hence, also in \mRp, the prospects
of a very energetic SN explosion are large.

Mass accretion onto the PNSs ceases in all models within at most
$0.7 \, \sek$ after bounce, which is not sufficient to increase their
masses beyond the limit for BH formation.  The ordering of final
masses is the inverse of the magnetic field with values between
$\MPNS \approx 2.16 \, \msol$ for \mRw and
$\MPNS \approx 1.75 \, \msol$ for \mRs.  The strong MHD explosions of
models \mnRp and \mnRs quench mass accretion most effectively and the
PNSs start gradually losing mass.  All PNSs possess high rotational
and magnetic energies.  In models \mnRp and \mnRw, the presence of
strong fields, in particular their poloidal components, cause the PNSs
to rotate slower than in the other two models.  \MRs experiences a
pronounced spin-down in parallel to the mass loss of the PNS and the
prolonged increase of the explosion energy.  The work done by the
field leads to a decrease of the magnetic energy, too.  The less
magnetically dominated PNSs of models \mnRO and \mnRw, on the other
hand, maintain or even increase the magnetic energies.  At the end of
the simulations, the surface-averaged fields of the PNS, ranging
between $\sim 7 \times 10^{13} \, \Gauss$ for the poloidal and
toroidal components of \mRp to $\sim 4.2 \times 10^{14} \, \Gauss$ for
the toroidal component of \Rw, are in the range of magnetar fields.
The poloidal and toroidal components tend to be of similar magnitude.

The strong rotation flattens the shapes of the PNSs to a strong
degree.  In \mRw, we observe highly asymmetric downflows impinging on
the PNS that slowly tilt the orientation of its rotation axis.  This
effect results in a complex topology of the magnetic field
characterized by loops aligned along different directions.  This
geometry differs strongly from that of the other models where the
combination of poloidal and toroidal components follows roughly the
pattern observed in axisymmetric models.

We summarize elements our results have in common with our axisymmetric
models and the simulations of other authors and where they differ from
them.  The times of the explosion and the mechanisms by which they are
initiated are similar to the axisymmetric versions of the models.  The
neutrino-driven explosion of \mRw, on the other hand, with its strong
$m = 1$ mode and the predominantly equatorial shock revival differs
from the bipolar explosion in 2D, as does the tilting
rotational axis of the PNS.  The explosion geometry of the other
models is much closer to the axisymmetric models.  The explosion
energies grow to lower values than in axisymmetry and tend to
stabilise within the time simulated here.  The exception to the latter
behaviour is \mRs with an ongoing rise of the explosion energy, albeit
slower than in 2D.  Since at the same time the PNS loses mass and
rotational energy, this evolution resembles that of the proto-magnetar
cases we had found in axisymmetry.  Unlike in 2D, where BH formation
is a common outcome of several of the models considered here (in
particular \modls{W} and \RO), mass accretion stops, at least for the
moment, in all 3D models before the PNS reaches a mass sufficient for
gravitational instability.  Longer simulations would be required to
check the possibility of the accretion of fallback material during
later epochs.

The explosion mechanisms --rotationally modified neutrino-driven and
MHD explosions-- as well as the explosion energies agree in general
with the results of other groups such as the ones of
\cite{Takiwaki_et_al__2016__mnras__ThreeDimensionalSimulationsofRapidlyRotatingCoreCollapseSupernovaeFindingaNeutrinoPoweredExplosionAidedbyNonAxisymmetricFlows,Summa_et_al__2018__apj__Rotation-supportedNeutrino-drivenSupernovaExplosionsinThreeDimensionsandtheCriticalLuminosityCondition}
for rotating stars without magnetic fields and of
\cite{Winteler_et_al__2012__apjl__MagnetorotationallyDrivenSupernovaeastheOriginofEarlyGalaxyr-processElements,Mosta_et_al__2014__apjl__MagnetorotationalCore-collapseSupernovaeinThreeDimensions,Kuroda_et_al__2020__apj__MagnetorotationalExplosionofaMassiveStarSupportedbyNeutrinoHeatinginGeneralRelativisticThreeDimensionalSimulations}
for magneto-rotational core collapse.

The magnetically driven outflows of models \mnRO, \mnRp, and \mnRs are
not subject to strong non-axisymmetric instabilities.  The
displacement of the barycentre of the magnetic field in the jets can
grow exponentially early on, but the growth is limited and does not
lead to a strong perturbation or a disruption of the outflows.

On the issue of the disruption of the MHD-driven outflows by
non-axisymmetric instabilities, our results are more in line with
\cite{Winteler_et_al__2012__apjl__MagnetorotationallyDrivenSupernovaeastheOriginofEarlyGalaxyr-processElements},
who did not observe such a behaviour, than with
\cite{Mosta_et_al__2014__apjl__MagnetorotationalCore-collapseSupernovaeinThreeDimensions}
and
\cite{Kuroda_et_al__2020__apj__MagnetorotationalExplosionofaMassiveStarSupportedbyNeutrinoHeatinginGeneralRelativisticThreeDimensionalSimulations},
whose simulations show strong kink modes.  However, the disagreement
may not be as large as the dichotomy of failed or successful jets
might suggest.  We find a growth of non-axisymmetric modes at times
and locations similar to the cases presented by
\cite{Mosta_et_al__2014__apjl__MagnetorotationalCore-collapseSupernovaeinThreeDimensions,Moesta_et_al__2018__apj__r-processNucleosynthesisfromThree-dimensionalMagnetorotationalCore-collapseSupernovae,Halevi_Moesta__2018__mnras__r-Processnucleosynthesisfromthree-dimensionaljet-drivencore-collapsesupernovaewithmagneticmisalignments,Kuroda_et_al__2020__apj__MagnetorotationalExplosionofaMassiveStarSupportedbyNeutrinoHeatinginGeneralRelativisticThreeDimensionalSimulations},
\viz the innermost few tens of km in the immediate post-bounce phase.
However, in our models the jets manage to overcome these perturbations
and propagate towards larger radii.  After this critical phase, they
are subject to only minor influence of kink modes.  We find
indications for a continuous dependence of the importance of the
instabilities on the field strength and, thus, the energetics and
speed of the jet with the weaker jets showing stronger distortions
than stronger ones.  This finding seems to be supported by
\cite{Mosta_et_al__2014__apjl__MagnetorotationalCore-collapseSupernovaeinThreeDimensions}
who mention a test simulation with a stronger field than the one in
their 3D model showing weaker kink modes and model B13 of
\cite{Moesta_et_al__2018__apj__r-processNucleosynthesisfromThree-dimensionalMagnetorotationalCore-collapseSupernovae,Halevi_Moesta__2018__mnras__r-Processnucleosynthesisfromthree-dimensionaljet-drivencore-collapsesupernovaewithmagneticmisalignments},
in which a stronger magnetic field produces more stable jets.
Possible reasons for the differences \wrt the cited studies include
the spherical grids of our models and the grid resolution as well as
the profiles of density, rotation, and pressure of the progenitors
that will affect the ram pressure against which the jets have to
propagate and, thus, the ability of kink modes to grow within
dynamical times.

Though a quantitative comparison is made difficult by the different
physical settings, the growth of the kink instabilities is compatible
with the analysis and the results for jet propagation of
\cite{Moll_et_al__2008__aap__Kink_instabilities_in_jets_from_rotating_magnetic_fields}
as well as with the results obtained by
\cite{Bromberg_Tchekhobskoy_2016MNRAS.456.1739} for collapsar jets.
According to the findings of
\citeauthor{Moll_et_al__2008__aap__Kink_instabilities_in_jets_from_rotating_magnetic_fields},
the typical length scales and the growth times of the kink modes are
given by the magnetic pitch of the helical field, larger for stronger
radial field, and the time an \Alfven wave requires for one full
revolution along the helical structure.  The jet can be stabilised if
it accelerates, expands laterally, or if differential rotation
generates a strong magnetic shear.  During later stages of the
evolution, when the jet has propagated beyond several $100 \, \km$,
the conditions are favourable for the growth of very long (hundreds to
thousands of km) modes on short times of several ms.  However, all the
aforementioned inhibiting effects are also present, reducing the
impact of the non-axisymmetric modes.  Among them, the morphology of
the magnetic field seems to play a prominent role.  The toroidal field
has only a small positive or even negative gradient with cylindrical
radius, which, according to
\cite{Begelman__1998__apj__InstabilityofToroidalMagneticFieldinJetsandPlerions},
may account for a stabilization of the outflows.

A further difference to previous work
\citep[e.g.,][]{Winteler_et_al__2012__apjl__MagnetorotationallyDrivenSupernovaeastheOriginofEarlyGalaxyr-processElements,Moesta_et_al__2018__apj__r-processNucleosynthesisfromThree-dimensionalMagnetorotationalCore-collapseSupernovae,Halevi_Moesta__2018__mnras__r-Processnucleosynthesisfromthree-dimensionaljet-drivencore-collapsesupernovaewithmagneticmisalignments}
lies in the proton-rich composition of our jets.  MHD jets are
commonly assumed to present favourable conditions for neutron-capture
nucleosynthesis, \ie, low $Y_e$ because their rapid expansion should
permit the gas to leave the region where strong neutrino fluxes can
modify the electron fraction.  Our models defy these expectations by
hosting jets whose beams consist of almost symmetric matter with $Y_e
\approx 0.5$ despite the high speeds of the gas.  Although the matter
forming the jet beams originates from the immediate vicinity of the
polar caps of the PNS and thus starts its propagation with a very low
$Y_e \approx 0.2$, it is quickly releptonized by the neutrinos at
distances smaller than two PNS radii, and local temperatures in excess
of 10\,GK.  The reason for this evolution is that the magnetic field,
which the gas has to follow, is at the jet base predominantly toroidal
and has only a small radial component.  As a consequence of this field
geometry, caused by the strong radial differential rotation profile
across the PNS surface, the gas orbits the rotational axis several
times before finally being ejected in the jets.  The time it is thus
exposed to intense neutrino radiation is sufficient to increase the
electron fraction to $Y_e \approx 0.5$.  This finding may put in
question the ability of the MHD jet formation mechanism to produce
conditions for r-process nucleosynthesis.  However, our results also
indicate that other components of the ejecta may be more neutron-rich,
in particular the cocoons of the jets or, as shown in the long-term
axisymmetric models of paper II, as well as in
\cite{Reichert_et_al__2021__mnras}, late proto-magnetar-like winds
driven by the magnetic field.  Furthermore, changes of the neutrino
emission, in particular of the ratio between $\nu_e$ and
$\bar{\nu}_e$, might lead to more favourable conditions for
neutron-capture nucleosynthesis.  To a large degree, we attribute the
differences between our models and other studies to our use of a M1
neutrino transport method as opposed to the leakage schemes used by
them that tend to yield more neutron-rich conditions.  Moreover, the
results of the other studies are based on an analysis of Lagrangian
tracer particles whereas we use the data on the Eulerian grid.  We
note that the releptonization of the gas entering the outflows via the
jet base occurs in the regions where our grid has the finest
resolution, minimizing additional diffusion \wrt a Lagrangian
advection method.  A possible contribution of the methodological
differences to the reported values of the electron fraction, in
particular for parcels ejected without passing through the jet base,
remains to be studied thoroughly.

We point out some limitations of our work.  Besides a higher grid
resolution, a wider scope of initial models, in particular a more
realistic magnetic field configuration  derived self-consistently from
multi-dimensional stellar models, would be highly desirable.  Despite
these limitations, our results strengthen the case for rapidly
rotating and strongly magnetized stars as progenitors of energetic, bipolar
CCSNe.  Furthermore, the final state of our models, containing PNSs
with high rotational energy and strong magnetic fields, as well as the
spin-down phase of the strongest magnetised model suggest the
possibility of a later transformation into a proto-magnetar-driven
GRB.  Further exploring this option would require much longer
simulation times, which is not feasible using the same numerical
methods.  Aspects that will be addressed in future research
are the production of heavy elements in these models and the
multi-messenger observables of gravitational waves and neutrinos.

\section{Acknowledgements}
\label{Sek:Ackno}
We thank the anonymous referee for their helpful questions and
comments.  This work has been supported by the Spanish Ministry of
Science, Education and Universities (PGC2018-095984-B-I00) and the
Valencian Community (PROMETEU/2019/071). MO acknowledges support from
the European Research Council under grant EUROPIUM-667912, and from
the the Deutsche Forschungsgemeinschaft (DFG, German Research
Foundation) -- Projektnummer 279384907 -- SFB 1245 as well as from the
Spanish Ministry of Science via the Ram{\'o}n y Cajal programme
(\miRyC).  The authors thankfully acknowledge the computer resources
and the technical support provided by grants AECT-2017-2-0006,
AECT-2017-3-0007, AECT-2018-1-0010, AECT-2018-2-0003,
AECT-2018-3-0010, and AECT-2019-1-0009 of the Spanish Supercomputing
Network on cluster \textit{MareNostrum} of the Barcelona
Supercomputing Centre - Centro Nacional de Supercomputaci\'on, on
clusters \textit{Tirant} and \textit{Lluisvives} of the Servei
d'Inform\`atica of the University of Valencia (financed by the FEDER
funds for Scientific Infrastructures; IDIFEDER-2018-063), and on
cluster \textit{Lichtenberg} of the Technical University of Darmstadt
(grant 906).

\vspace{-.5cm}
\section*{Data Availability}
The data underlying this article will be shared upon reasonable request to the corresponding authors.

\end{document}